\documentclass[10pt,aps,prl,superscriptaddress,twocolumn,citeautoscript]{revtex4-2}
\usepackage[american]{babel}
\usepackage{graphicx}
\usepackage{amsfonts}
\usepackage{amsmath}
\usepackage{amssymb}
\usepackage{bm}
\usepackage{textcomp}
\usepackage{verbatim}
\usepackage{xspace}
\usepackage{soul}
\usepackage{bbold}
\usepackage{mathtools}
\usepackage{dcolumn}
\usepackage{hyperref}
\usepackage[usenames]{color}
\usepackage{subfig}
\usepackage{physics}
\usepackage{units}
\usepackage[export]{adjustbox}
\usepackage{tablefootnote}
\usepackage[dvipsnames]{xcolor}
\usepackage{ragged2e}
\DeclareCaptionJustification{justified}{\justifying}
\usepackage[commandnameprefix=always]{changes}

\def\beq{\begin{equation}}
	\def\eeq{\end{equation}}

\newcommand{\out}[1]{{}}

\newcommand{\highlight}[1]{\textcolor{black}{{#1}}}

\newcommand{\highlighttwo}[1]{\textcolor{black}{{#1}}}

\renewcommand{\sout}[1]{\unskip}

\footnotetext{\textsuperscript{$\dagger$} These authors contributed equally to this work.}

\begin{document}


\title{Ab initio spin Hamiltonians and magnetism of Ce and Yb triangular-lattice compounds}

	\author{Leonid V. Pourovskii\textsuperscript{$\dagger$}}
    
    \email{leonid.poyurovskiy@polytechnique.edu}
	\affiliation{CPHT, CNRS, \'Ecole polytechnique, Institut Polytechnique de Paris, 91120 Palaiseau, France}
	\affiliation{Coll\`ege de France, Université PSL, 11 place Marcelin Berthelot, 75005 Paris, France}
	
	\author{Rafael D. Soares\textsuperscript{$\dagger$}}

	\affiliation{Max Planck Institute for the Physics of Complex Systems, N\"{o}thnitzer Stra{\ss}e 38, 01187 Dresden, Germany}
	
	\author{Alexander Wietek}
	\affiliation{Max Planck Institute for the Physics of Complex Systems, N\"{o}thnitzer Stra{\ss}e 38, 01187 Dresden, Germany}
	
	\begin{abstract}
We calculate the crystal-field splitting, ground-state Kramers doublet and intersite exchange interactions within the ground-state doublet manifold using an ab initio Hubbard-I based approach for a representative set of Ce and Yb triangular-lattice compounds. These include the putative quantum spin liquids (QSL) RbCeO$_2$ and YbZn$_2$GaO$_5$ and the antiferromagnets KCeO$_2$ and KCeS$_2$. The calculated nearest-neighbor (NN) couplings are antiferromagnetic and exhibit noticeable anisotropy. The next-nearest-neighbor (NNN) couplings are ferromagnetic in the Ce systems and dominated by classical dipole–dipole interactions in the Yb case. Solving the resulting effective spin-1/2 models by exact diagonalization up to $N=36$ sites, we predict ordered magnetic ground states for all systems, including the two QSL candidates. We explore the phase space of an anisotropic  NN + isotropic NNN triangular-lattice model finding that a significant antiferromagnetic NNN coupling is required to stabilize QSL phases, while the NN exchange anisotropy is detrimental to them. Our findings highlight a possibly important role of deviations from the perfect triangular model --  like atomic disorder –  in real triangular-lattice materials.
\end{abstract}
	
	
\maketitle

{\it Introduction.} Triangular lattice spin-$1/2$ antiferromagnets (AFM), with their inherent geometric frustration and quantum fluctuations, stand as prime candidates for harboring exotic states of quantum matter, such as quantum spin liquids (QSL) and topologically ordered phases. While the nearest-neighbor Heisenberg model on the triangular lattice has been shown to stabilize a $120^\circ$ magnetically ordered N\'eel ground-state~\cite{Huse1988,Bernu1992,Capriotti1999,White2007}, the inclusion of additional interactions has revealed a variety of both magnetically ordered and quantum disordered phases. Notably, enhanced second nearest-neighbor interactions can induce a quantum paramagnetic state governed by the physics of the gapless $U(1)$ Dirac spin liquid (DSL)~\cite{Hastings2000,Hermele2004,Hermele2005}, where distinct proposals have suggested a direct emergence of the DSL~\cite{Kaneko2014,Iqbal2016,Hu2019}, a valence bond solid~\cite{Wietek2024}, a gapped $\mathbf{Z}_2$ spin liquids~\cite{Zhu2015,Hu2015,Saadatmand2016,Jiang2022} and chiral spin liquids~\cite{Saadatmand2017,Gong2017,Wietek2017}, which are understood to be descendants of the parent DSL order~\cite{Hermele2004,Hermele2005,Song2019,Song2020}. While the precise nature of the paramagnetic regime is still under debate, it has become evident that the physics of the DSL and its descendants serves as an organizing principle of the phase diagrams of triangular lattice antiferromagnets~\cite{Wietek2024}. 


Geometrically perfect quasi-2D triangular lattices of magnetic ions are realized in numerous layered rare-earth oxides and chalcogenides. In particular, Kramers 3+ ions of $R$=Ce, Yb, Nd etc. in systems like  delafossites $ARX_2$ (where $A$ is an alkali metal, $X$=O, S, Se)~\cite{Bordelon2017,Bastien2020,Bordelon2021,Kulbakov2021,Ortiz2022,Grussler2023,Xie2024}, heptatantalates $R$Ta$_7$O$_{19}$~\cite{Arh2022},  YbMgGaO$_4$~\cite{Li2015,Shen2016,Li2016,Zhu2017b},  and YbZn$_2$GaO$_5$~\cite{Bag2024} have recently attracted a lot of interest. The crystal-field (CF) splitting in these systems is typically much larger than rather weak superexchange between the well localized 4$f$ shells of rare-earth ions. Consequently, admixture of the excited CF levels by superexchange can be neglected with the lowest Kramers doublet considered as an effective spin-1/2. However, strong spin-orbit (SO) entanglement in the  4$f$ ground state multiplet in conjunction with the weaker CF effect results in anisotropic charge and magnetic density of the Kramers states, which leads, in turn, to anisotropic superexchange and $g$-factors. Hence, the intersite coupling anisotropy that has been shown to strongly affect QSL formation  is naturally present in those compounds. A gapless QSL ground state has been suggested on the basis of intensive experimental investigation for NaYbO$_2$~\cite{Bordelon2017}, several Yb delafossite selenides and sulfides~\cite{Dai2021,Wu2022} as well as for YbZn$_2$GaO$_5$~\cite{Bag2024}.  

Low-energy Hamiltonians for those rare-earth triangular materials remain to date quite uncertain. In particular, due to lack of single crystals for majority of delafossites~\cite{Bag2024},  two-site anisotropies were assessed experimentally only in few cases~\cite{Xie2024prl,Xie2024}. 
Hence, isotropic NN and NNN interactions are often assumed in the analysis~\cite{Scheie2024}, which is further complicated by possible admixture of excited CF levels by magnetic field. Theoretically, ab initio calculations of exchange interactions in Yb delafossites on the basis of DFT electronic structure have so far predicted too small NNN couplings compared to experimental expectations~\cite{Villanova2023,Scheie2024}.


In this work, we tackle the problem of evaluating realistic spin Hamiltonians for triangular rare-earth systems using DFT+dynamical mean-field theory (DMFT)~\cite{Georges1996,Anisimov1997_1,Lichtenstein_LDApp} treating strongly localized rare-earth 4$f$ states in a quasiatomic (Hubbard-I) approximation~\cite{hubbard_1} to obtain the high-temperature paramagnetic electronic structure, as well as, the 4$f$-shell CF splitting and ground-state Kramers doublet. Intersite exchange interactions (IEI) are evaluated from the same paramagnetic electronic structure using the force-theorem in Hubbard-I (FT-HI) method~\cite{Pourovskii2016}.
 In contrast to the previous ab initio calculations deriving IEI in triangular rare-earth systems from DFT electronic structure  with metallic 4$f$ bands~\cite{Villanova2023}, within our framework  the IEI are obtained from realistic self-consistent DFT+HI electronic structure of the target compounds with localized 4$f$ states. A significant impact of DFT+DMFT charge-density self-consistency on the electronic structure and CF splitting is well documented~\cite{Pourovskii2007,Bhandary2016,Delange2017}. 
We apply this framework to derive realistic spin Hamiltonians for a representative set of Ce- and Yb- triangular compounds, namely: KCeO$_2$, KCeS$_2$, RbCeO$_2$, and YbZn$_2$GaO$_5$ (YZGO). These materials are either experimentally inferred to exhibit conventional magnetic orders  (KCeO$_2$~\cite{Bordelon2021} and KCeS$_2$~\cite{Bastien2020,Kulbakov2021}), a putative DSL (YZGO~\cite{Bag2024,wu2025}) or remain to be investigated in detail (RbCeO$_2$, for which no order was found in a preliminary study~\cite{Ortiz2022}, but a full characterization has not been carried out).
We then study the resulting Hamiltonians through exact diagonalization (ED) calculations~\cite{Weisse2008,wietek2025xdiagexactdiagonalizationquantum}, obtaining numerically exact ground- and low-energy eigenstates resolved by the lattice space group on clusters with sizes up to $36$ spins. This, then, allows us to analyze in detail the nature of the magnetic ground-state (GS) spin-spin correlations for each of the considered compounds. \highlight{Previous works have already investigated certain regions of the parameter space for these type of Hamiltonians, through classical Monte Carlo studies~\cite{PhysRevB.94.035107,PhysRevB.97.184413}, ED~\cite{Wu2021} and notably using DMRG~\cite{Zhu2017b,PhysRevB.95.165110,Zhu2018,PhysRevLett.134.196702}.}

{\it Ab initio method.}  We employ a charge self-consistent implementation \cite{Aichhorn2009,Aichhorn2011,Aichhorn2016} of DFT+DMFT based on the Wien2k linearized augmented-plane-wave (LAPW) full-potential code~\cite{Wien2k} and ”TRIQS” library~\cite{Parcollet2015}.
All calculations were carried out using the reported experimental lattice structures~\cite{Bordelon2021,Kulbakov2021,Ortiz2022,Bag2024}. The IEI are calculated by FT-HI approach~\cite{Pourovskii2016} using the MagInt code~\cite{magint} analogously to previous applications of this approach to correlated magnetic insulators, see Ref.~\cite{hidden_order_review} for a review. The FT-HI method includes all kinetic exchange contributions due to virtual hopping of 4$f$ electrons. See the Supplemental Material (SM)~\cite{supplmat} for more details on our ab initio calculations \highlight{and the effect of charge-density self-consistency in DFT+HI}.

\begin{figure}[!t]
 \centering
 \includegraphics[width=0.9\columnwidth]{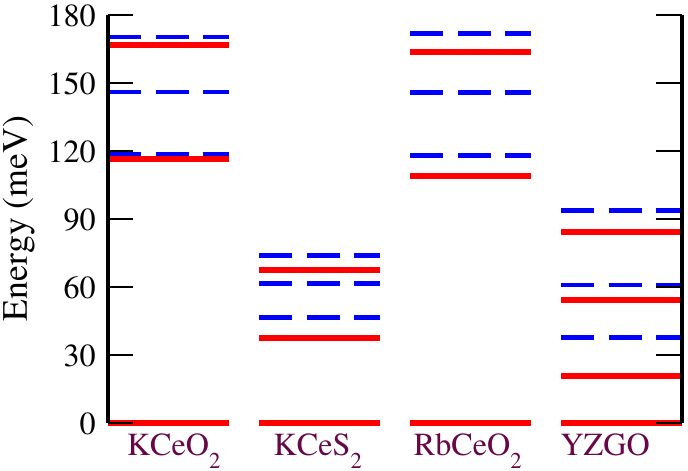}
\caption{Calculated CF level splitting. The thick red lines are calculated CF levels; the dashed blue lines are CF excitations measured in inelastic neutron scattering (INS) from Refs.~\cite{Bordelon2021,Bastien2020,Ortiz2022,Bag2024}. Note that  three excitations are seen in INS spectra for the Ce compounds. 
}
\label{fig:CF}
\end{figure}

\begin{table}[!bp]
	\begin{center}
		\begin{ruledtabular}
          \renewcommand{\arraystretch}{1.2}
			\begin{tabular}{l c c c c c}
            & & KCeO$_2$ & KCeS$_2$ & RbCeO$_2$ & YZGO \\
            \hline 
            $g_{ab}$ & This work & 1.43 & 2.00 & 1.54 & 3.24 \\
                 & Refs.  & 2.00$^a$ & 1.67$^b$, 2.47$^c$ & 1.46$^d$ & 3.04$^e$ \\
                 \hline
            $g_{c}$ & This work & 0.17 & -0.10 & 0.30 & 3.04 \\
             & Refs. & 0.29$^a$ & 0.58$^b$,0.65$^c$ & 0.01$^d$ & 3.44$^e$ \\
            \end{tabular}
		\end{ruledtabular}
	\end{center}
	\caption{Calculated values for the in-plane ($g_{ab}$) and out-of-plane ($g_{c}$) gyromagnetic tensor components. Other values are obtained by electron spin resonance (ESR), Refs.~\cite{Bordelon2021} (a) ESR and quantum chemistry calculations (QCC) \cite{Bastien2020} (b); QCC, Ref.~\cite{Kulbakov2021} (c); INS, Ref.~\cite{Ortiz2022} (d); magnetization measurements, Ref.~\cite{Bag2024} (e).}
   \label{tab:g_tensors} 
\end{table}

{\it Crystal-field splitting and ground state.} 
DFT+HI calculations converge to the expected insulating state with $f^1$ and $f^{13}$ GS occupancy of the Ce and Yb 4$f$ shells, respectively (see the SM for the calculated DFT+HI spectral functions). 
The calculated CF splitting  are shown in Fig.~\ref{fig:CF} and the corresponding Kramers doublets are listed in the SM~\cite{supplmat}.
We find a particularly large total CF splitting -- about 60\% of the SO gap (the latter is 0.3~eV in all Ce systems) -- for the Ce oxides. Correspondingly, there is a non-negligible admixture $\approx 0.17|7/2;\pm1/2\rangle$ of the SO excited multiplet to their GS Kramers doublets. The CF splitting in KCeS$_2$ is almost 3 times smaller leading to the corresponding reduction of the $J=7/2$ admixture.
No noticeable multiplet mixing is found for YZGO due to  its much larger SO gap of 1.37~eV. The dominating contribution of $|M=\pm1/2\rangle$ to the GS of all Ce systems leads to an in-plane magnetic anisotropy. In contrast,  the $|M=\pm7/2\rangle$ states provide the largest contribution to the GS doublet of YZGO. The corresponding $g$-tensor extracted from the calculated magnetic moment operators $M_{\alpha}=g_{\alpha}S^{\alpha}$, $\alpha=x,y,z$ within the GS Kramers doublet is listed in Table~\ref{tab:g_tensors}. Here we introduced the pseudo-spin $S=1/2$ to label the states within the GS doublet and the corresponding spin operators $S^{\alpha}$. One sees that the in-plane components $g_{ab}=g_{x}(=g_{y})$ 
largely dominate in all Ce systems. 
In KCeS$_2$, the out-of-plane component even becomes negative as a result of a near-perfect cancellation between the $J=$5/2 states, with the moment induced by a $J=$7/2 admixture. The $g$-tensor in YZGO is significantly larger and quasi-isotropic.

The experimental determination of the CF splitting and GS in the Ce compounds under consideration is complicated by the appearance of an additional, third, mode in their INS spectra~\cite{Bordelon2017,Ortiz2022,Bastien2020}, though only two CF excitations are expected in the case of a $J=5/2$ GS. This phenomenon, observed also in other Ce systems, remains so far without commonly accepted explanation, see Refs.~\cite{Kulbakov2021,Mouding2025,Smith2024}.
Taking this complication into account, our predictions for the CF levels energies in all compounds under consideration are in good agreement with experiment (Fig.~\ref{fig:CF}), albeit with some systematic underestimation of the splitting by the present theory. Similarly, a good agreement is found for the $g$-tensor (Table~\ref{tab:g_tensors}) in comparison with previous experimental estimates and quantum chemistry calculations. We note that no signal  was observed in the electronic spin resonance for the out-of-plane direction in KCeS$_2$ suggesting $g_{c}\ll g_{ab}$~\cite{Bastien2020}.

{\it Effective magnetic Hamiltonian.} 
The IEI within 
the GS Kramers 
manifold were calculated by the FT-HI method. Besides IEI, we also included the classical intersite dipole-dipole interaction, which can be important given rather small magnitudes of anisotropic IEI in the target materials, evaluated using the ab initio gyromagnetic tensors (Table~\ref{tab:g_tensors}).

The symmetry of the triangular rare-earth layer constrains the most general coupling within the $S=1/2$ manifold for the NN  $ij$ bond along $\mathbf{r}_0^{\mathrm{NN}}=[100]||x$ to \cite{Li2015,Zhu2018,Maksimov2019}:
\begin{equation}
\begin{aligned}
 H_{ij}& = \mathbf{S}_i^T\hat{J} \mathbf{S}_j= J(\Delta S_i^zS_j^z+S_i^xS_j^x+S_i^yS_j^y)+ \\
&  2J_{\pm\pm}(S_i^xS_j^x-S_i^yS_j^y)+J_{z\pm}(S_i^zS_j^y+S_i^yS_j^z),
\end{aligned}
\label{eq:HNN}
\end{equation}
where $\mathbf{S}^T_i=[S^x_i,S^y_i,S^z_i]$, $\Delta$ is the diagonal XXZ anisotropy, $J_{\pm\pm}$ is the diagonal XY anisotropy and $J_{z\pm}$ are off-diagonal terms. 
The coupling for the NNN bond along $\mathbf{r}_0^{\mathrm{NNN}}=[010]||y$ takes the same form of Eq.~\eqref{eq:HNN}; we will use the prime to distinguish the NNN terms ($J^{\prime}$,$J_{\pm\pm}^{\prime}$ etc.). The coupling for other NN and NNN bonds follows from the symmetry~\cite{Iaconis2018,Maksimov2019}, see SM~\cite{supplmat} for details.

\begin{table}[!tp]
	\begin{center}
		\begin{ruledtabular}
          \renewcommand{\arraystretch}{1.2}
			\begin{tabular}{l c c c c}
                 & KCeO$_2$ & KCeS$_2$ & RbCeO$_2$ & YZGO \\
                \hline
                $J$ (meV) &0.51 & 0.021 & 0.36 & 0.25 \\
                $\Delta$ & 1.03 & 1.90 & 1.05 & 1.03 \\
                $J_{\pm\pm}/J$ & 0 & 0.18  & 0 & -0.055 \\
                $J_{z\pm}/J$ & 0.10 & -1.45 &  -0.121 & 0.039 \\
                \hline
                $J^{\prime}/J$ &-0.055 & -0.023 & -0.063  &  0.003 \\
                $\Delta^{\prime}$ & 0.70 & -0.64  & 0.67 & 5.96 \\
                $J_{\pm\pm}^{\prime}/J^{\prime}$ & 0.009 & -0.62 &  -0.002 & 2.62\\
                $J_{z\pm}^{\prime}/J^{\prime}$ & 0.043 & 1.00  & -0.040 & 0.073 \\
                %
                \hline
                $J^{^{\mathrm{int}}}/J$ & 0.011 &  & 0.016 & 0.012 \\
                \hline
                $\theta$ (K) & -8.5  & -0.43$^{ab}$ & -6.2 & -3.9$^{ab}$,-1.9$^{c}$ \\
                $\theta_{\mathrm{Exp}}$ (K) & -7.7 & -2.8$^{ab}$  & -5.1  & -5.2$^{ab}$,-3.8$^{c}$ \\
            \end{tabular}
		\end{ruledtabular}
	\end{center}
	\caption{Calculated intersite coupling including the IEI and dipole-dipole contributions. Experimental Curie-Weiss temperatures $\theta_{\mathrm{Exp}}$  extracted from out-of-plane ($^{c}$), in-plane ($^{ab}$), or polycrystalline data are from Refs.~\cite{Bordelon2021,Bastien2020,Ortiz2022,Bag2024}.  }
   \label{tab:Js} 
\end{table}

The resulting intersite couplings are listed in Table~\ref{tab:Js}. All systems exhibit AFM NN $J$, as expected; we find $\Delta \gtrsim 1$ everywhere. The NNN interactions are ferroic in all Ce systems and antiferroic in YZGO. There is an essential difference between the Ce and Yb systems in the nature of NNN interactions: in the former, the dipole-dipole interactions are negligible compared to the exchange. In YZGO, due to small NNN IEI and large $g$ factors, the anisotropic dipole-dipole contribution dominates over an almost isotropic NNN exchange. As a result, the total NNN couplings in YZGO are very anisotropic ($\Delta\approx 6$) with $J_{xx}^{\prime}(=J^{\prime}+2J^{\prime}_{\pm\pm})=0.020J$, $J_{yy}^{\prime}(=J^{\prime}-2J^{\prime}_{\pm\pm})=-0.013J$, and $J_{yy}^{\prime}=\Delta J^{\prime}=0.019J$. Hence, though the NNN interactions in YZGO are small, they are not as negligible as may seem from the $J^{\prime}/J$ ratio listed in Table~\ref{tab:Js}. The effect of a strong NNN anisotropy on the phase stability on the triangular lattice  has not been studied so far. Finally, the coupling $J^{\mathrm{int}}$ for the shortest interlayer bond is about 1\%-1.5\% of $J$ in all oxide systems (we could not extract it in KCeS$_2$ due to the general smallness of IEI in this system).

\begin{figure}[t]
    \centering
    \includegraphics[width=\columnwidth]{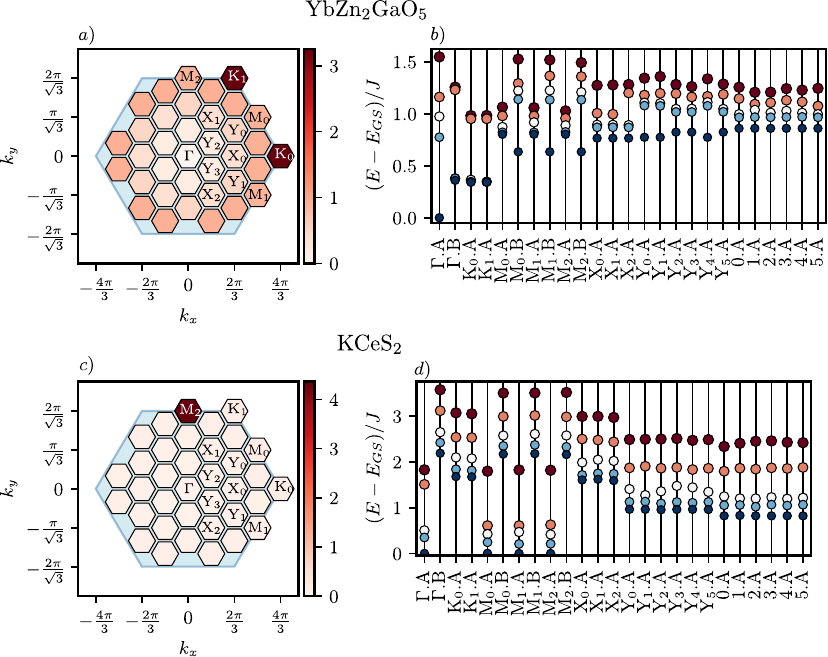}
    \caption{Spin structure factors and low-energy spectrum for the YZGO ($a,b$) and KCeS$_2$ ($c,d$) compounds in a $36$-site cluster. The isotropic structure factor $\mathcal{S}\left(\boldsymbol{k} \right)$ in $a)$ is peaked at the $K$ points, indicating $120^\circ$ order, consistent with low-energy excitations at $K_0.\text{A}$ and $K_1.\text{B}$. The peak at $M_2$ of $\mathcal{S}^{yy}(\boldsymbol{k})$ in $c)$ indicates stripe-$\perp$ order, consistent with low-energy excitations at the $M$ points in $d)$.}
    \label{fig:ybzno_ed}
\end{figure}
To assess the overall agreement in magnitude with experiments, we calculated high-temperature susceptibility by solving (\ref{eq:HNN}) in mean-field in an applied field using the McPhase package~\cite{Rotter2004} and an in-house module. 
For all oxide systems we find quite good agreement of the calculated Curie-Weiss temperatures $\theta$, extracted from the averaged  susceptibility $(2\chi_{ab}+\chi_{c})/3$, with available experimental data (Table~\ref{tab:Js}). In the case of KCeS$_2$, theoretical intersite couplings are severely underestimated.  This is also apparent from comparison of our result with the estimates of Ref.~\cite{Avdoshenko2022} that were obtained by quantum chemistry calculations and subsequently adjusted to agree with experimental INS. 4$f$ kinetic exchange captured by FT-HI does not, apparently, give the leading contribution to the IEI in  this sulfide.

{\it Low-energy spectra and ground-state correlations.} We now investigate the GS phase diagram of the effective magnetic Hamiltonians introduced in Eq.~\eqref{eq:HNN} with ED using the XDiag library~\cite{wietek2025xdiagexactdiagonalizationquantum,Wietek2018}. We obtain symmetry-resolved excitation spectra employing translational symmetries and the lattice inversion symmetry without spin rotation on cluster sizes $N=12, 32, 36$, see the SM~\cite{supplmat} for the employed lattice geometries. The irreducible representations (irreps) are represented by $\boldsymbol{k}.\rho$, where $\boldsymbol{k}$ labels the momentum and $\rho$ labels the inversion irrep, i.e. either $\rho=\text{A}$ for the even irrep or $\rho=\text{B}$ for odd irrep. We note that the additional point group symmetries are not employed, as an implementation of combined lattice-spin rotations is not readily accessible. 
We also obtained the GS static spin structure factors, 
\begin{equation}
\mathcal{S}^{\alpha\alpha}\left(\boldsymbol{k} \right) = \dfrac{1}{N}\sum_{n,m} e^{i \boldsymbol{k} \cdot \left(\boldsymbol{r}_n - \boldsymbol{r}_m \right)} \left\langle S^\alpha_{n} \cdot S^\alpha_{m} \right\rangle,   
\label{eq:su(2)_ssf}
\end{equation}
where $\alpha=x,y,z$, $N$ denotes the number of lattice sites and $\boldsymbol{r}_n$ the position of the $n^{\rm th}$ spin. Furthermore, we denote by $\mathcal{S}\left(\boldsymbol{k} \right) = \sum_{\alpha} \mathcal{S}^{\alpha\alpha}\left(\boldsymbol{k} \right)$, the isotropic spin structure factor. 

In panels $a)$ and $b)$ of Fig.~\ref{fig:ybzno_ed}, we show $\mathcal{S}\left(\boldsymbol{k} \right)$ and the low-lying energy spectrum for the model parameters of YZGO (see table~\ref{tab:Js}) on the $N=36$ cluster. Both the organization of the spectrum and the spin structure factor are consistent with the system realizing the $120^{\circ}$ AFM phase. The GS transforms under the $\Gamma.\text{A}$ irrep, while the first excited states (nearly degenerate) belong to the $\Gamma.\text{B}$ and $K.\text{A}$ irreps. These symmetry quantum numbers match the expected signatures of the $120^{\circ}$ AFM phase~\cite{PhysRevB.50.10048,wietek2017studyingcontinuoussymmetrybreaking}, see the SM~\cite{supplmat} for further details on how these quantum numbers are predicted. 

A key signature of a DSL is a gapless singlet mode at the $X_0$ point in the first Brillouin zone (FBZ), corresponding to the singlet monopole excitations~\cite{Song2019}. \highlight{Although in the present case we cannot identify these excitations as singlets due to the absence of spin SU(2) symmetry, they still occur at the same momentum.\sout{However,} In contrast,} the excitation energy at $X_0$ is significantly larger compared to the putative DSL of the $J_1$-$J_2$ model~\cite{Wietek2024}, rendering this scenario unlikely. Furthermore, $\mathcal{S}(\boldsymbol{k})$ exhibits a pronounced peak at crystal momentum $K$, in agreement with the known ordering wave vector of the $120^{\circ}$ AFM state~\cite{Bernu1992,PhysRevB.83.184401}. Analogously, we also observe spin correlations and a low-energy spectrum consistent with a $120^\circ$ state for the effective Hamiltonians of the KCeO$_2$ and RbCeO$_2$ compounds, see the SM~\cite{supplmat}.

\begin{figure*}[t]
    \centering
    \includegraphics{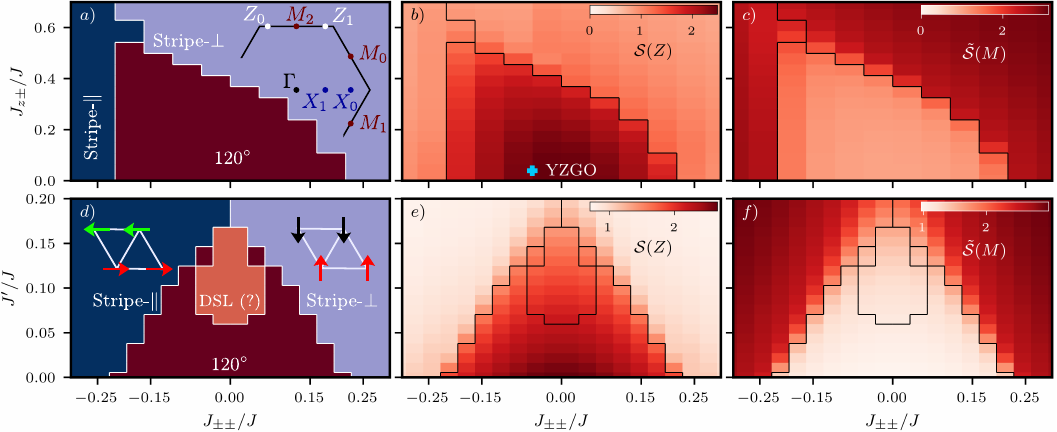}
    \caption{$a)$ and $d)$ - Approximate phase diagram of the effective magnetic Hamiltonian, inferred from the quantum number organization of the low-energy spectrum. Dark red marks the $120^\circ$ AFM phase, dark blue the stripy-$\parallel$ order, pale purple the stripy-$\perp$ order, and orange the regime with first non-zero momentum excitation in the $X_1.\text{A}$ irrep, suggesting a possible DSL phase. A typical classical ordering for each stripe phase is shown in the sketch in $d)$. The remaining panels show the static spin structure factor at selected momenta in the FBZ (see inset in $a$)): panel $b)$ and $e)$ at $Z$, while $c)$ and $f)$ show the average over the three inequivalent $M$ points. In $b)$ we identify the parameters compatible with the YZGO system. Parameters: $\Delta = 1.03J$, $J^\prime = 0.02J$ in $a)$–$c)$, and $J_{z\pm} = 0$ in $d)$–$f)$.}
    \label{fig:phase_diagram_n32}
\end{figure*}


On the other hand, the spectrum and correlations for the KCeS$_2$ system indicate a collinear stripy antiferromagnetic GS, as shown in panels $c)$ and $d)$ of Fig.~\ref{fig:ybzno_ed}. The low energy spectra reveals an approximate GS degeneracy between the $\Gamma$.A, and $M.\text{A}$ irreps, consistent with the stripy AFM state~\cite{wietek2017studyingcontinuoussymmetrybreaking}. Accordingly, we observe a pronounced peak in $\mathcal{S}^{yy}(\bm{k})$ and $\mathcal{S}^{zz}(\bm{k})$ at momentum $M_2$\footnote{The peak appears at only one $M$ point because slight numerical rounding in the ab-initio couplings weakly breaks the Hamiltonian’s $C_3$ symmetry.}. As noted~\cite{PhysRevB.94.035107}, the symmetries of Eq.~\eqref{eq:HNN} allow for two distinct stripe orders, differing by how they transform under the combined lattice-spin rotation around the triangular lattice axis directions. The first aligns the spins along the direction of the stripe, we call it stripe-$\parallel$ (stripe-parallel) (this is also named stripe-$x$). It is invariant under the $C_2$ axis rotations, as the spins align with it. In the second, the spins align perpendicular to the stripe direction, and thus transforms non-trivially under the $C_2$ axis rotation. We name it stripe-$\perp$ (stripe-perpendicular) (also known as stripe-$yz$). Thus, the (in)variance under this symmetry constitutes the distinguishing feature between the two states . A representative configuration of each stripe type is shown in Fig.~\ref{fig:phase_diagram_n32} $d)$. 


The stabilized stripe is oriented along the $x$-direction of the lattice, so, the peak in the $\mathcal{S}^{yy}(\bm{k})$ at $M_2$ indicates the realization of the stripe-$\perp$ state, see the SM for further details. Remarkably, this is in complete agreement with experiments on KCeS$_2$~\cite{Kulbakov2021}.


{\it Effective Hamiltonian phase diagram.} 
We now explore the extended phase diagram Eq.~\eqref{eq:HNN} and thereby the possibility of an emergent DSL at particular coupling parameters. Here, we consider a subset of the parameter space where the NN interactions can be fully anisotropic with $J_{\pm\pm} \neq 0$ and $J_{z\pm} \neq 0$, but the NNN interactions are restricted to isotropic Heisenberg interactions, $H_{ij} = J' \boldsymbol{S}_i \cdot \boldsymbol{S}_j$. This choice is compatible with the small NNN anisotropy observed in most materials, with the exception of the YZGO compound, cf. table~\ref{tab:Js}. However, in that case, the average strength of the NNN coupling, which can be estimated as $\text{Tr}\left[|\hat{J}|\right]$, is only 1.7~\% of the NN coupling. 

 We perform extensive ED calculations on the $32$-site cluster and use the $12$-site cluster to corroborate the phase diagram obtained, whose results and discussion are left to the SM~\cite{supplmat}. 
 

We obtain an approximate phase diagram (Fig.~\ref{fig:phase_diagram_n32} $a)$ and $d)$) based on the quantum numbers of the spectrum (check the SM). We begin by fixing $J' = 0.02J$ (a value compatible with the largest matrix elements of the NNN coupling $\hat{J'}$ in YZGO; see Table~\ref{tab:Js}) and study the phase diagram as a function $J_{\pm\pm}$ and $J_{z\pm}$. This corresponds to the panels ($a,b,c$) in the upper row of Fig.~\ref{fig:phase_diagram_n32}. In this case, the phase diagram consists of three distinct magnetically ordered states, a $120^\circ$ AFM region, a stripe-$\parallel$ phase, and a stripe-$\perp$ phase, which correspond to the dark red, dark blue, and pale purple regions shown in panel $a)$, respectively. 

As in the 32-site cluster, the $K$ momentum is not resolved, $\mathcal{S}(\boldsymbol{k})$ peaks at nearby wave vectors $Z_0$ and $Z_1$ in the $120^\circ$ AFM phase, as shown in Fig.~\ref{fig:phase_diagram_n32} $b)$. This reflects the frustrated realization of $120^\circ$ AFM order in this cluster. 

Upon tuning both $J_{\pm\pm}$ and $J_{z\pm}$, $\mathcal{S}(\boldsymbol{k})$ develops a peak at the $M$ points, signaling the onset of the stripe phase (see panel $c)$) We find that along the $J_{z\pm} = 0$ line, negative values of $J_{\pm\pm}/J$ favor a stripe-$\parallel$, whereas positive values stabilize a stripe-$\perp$ state. The coupling $J_{z\pm}$ further extends the stripe-$\perp$ region into negative $J_{\pm\pm}/J$ values, which agrees with previous studies. In certain parameter regimes, the peak in $\mathcal{S}(\boldsymbol{k})$ appears at both $M_0$ and $M_1$, while in others it shifts to $M_2$. This shift does not signal a transition between stripe phases, but originates from the absence of $C_3$ symmetry in the $32$-site cluster, which renders the three $M$ points inequivalent (unlike in the $12$-site cluster). Consequently, the lattice geometry lifts the degeneracy among the six stripe configurations. To account for this, in Fig.~\ref{fig:phase_diagram_n32} $c)$ we plot $\mathcal{S}(\boldsymbol{k})$ averaged over the three $M$ points,
\begin{equation}
   \tilde{\mathcal{S}}(M) = \frac{\mathcal{S}(M_0) + \mathcal{S}(M_1) + \mathcal{S}(M_2)}{3}.
\end{equation}
Distinguishing the two stripe types requires examining additional spin-spin correlators, which we do in the SM. We observe in panel $b)$ that the couplings determined for the YZGO system place it within the $120^\circ$ AFM region as expected.



We set $J_{z\pm} = 0$ and study the phase diagram as a function of $J'$ and $J_{\pm\pm}$, identifying four distinct regions, as shown in Fig.~\ref{fig:phase_diagram_n32} $d)$. For $J' \in [0,0.05]J$, the GS is in a $120^\circ$ AFM state, confirmed by the peak of $\mathcal{S}(\boldsymbol{k})$ at the $Z$ points (panel $e)$). This state persists up to $J_{\pm\pm} \sim 0.1 J$. Increasing $J_{\pm\pm}$ induces stripe order, whose orientation depends on the sign of $J_{\pm\pm}$: negative values yield stripe-$\parallel$, while positive values stripe-$\perp$ GS. This is corroborated by the peak of $\mathcal{S}(\boldsymbol{k})$ at the $M$ points as shown in panel $f)$.



For $J' \sim [0.05,0.16]J$ and $J_{\pm\pm} \sim [-0.1,0.1]J$, the first non-zero momentum excitation belongs to the $X_1.\text{A}$ irrep (see SM), corresponding to the orange region in panel $d)$. As discussed, the excitation at momentum $X$ is particularly noteworthy, as it is interpreted as a signature of the singlet-monopole of a DSL~\cite{Song2019,Song2020}, predicted to appear in the low-energy spectrum of the $J_1$–$J_2$ triangular-lattice antiferromagnet in the spin-liquid regime~\cite{Wietek2024}. An alternative explanation for such a low-lying excitation is a 12-site valence-bond solid, for which consistent correlations have also been observed in the spin-liquid regime~\cite{Wietek2024}. We find that this region is stabilized only for small values of $J_{\pm\pm} \sim \pm 0.1 J$, while larger values drive the system towards the stripe phase. This analysis indicates that the previously reported quantum paramagnetic phases in earlier DMRG studies~\cite{Zhu2018} (identified even in the absence of NNN coupling) might not correspond to a DSL phase. In the SM, we provide additional results of the energy spectra, which reveal that the $X.\text{A}$ states persist at higher energies, whereas the low-lying excitations belong to the tower of states characteristic of the $120^\circ$ AFM order.\highlighttwo{ While these results do not definitively exclude other paramagnetic states, the sharp $K$-point peak in the spin structure factor (computed also for an $N=36$ cluster at $J^\prime = 0$, $J_{z\pm} = 0.3J$, and $J_{\pm\pm} = -0.08J$) strongly supports $120^\circ$ AFM order. Although a DSL may exhibit similar quasi–long-range signatures, which finite-size ED cannot unambiguously distinguish from true long-range order, our spectral analysis offers little evidence for such a phase at this parameter point.}

{\it Conclusions -- } We have derived from first principles the crystal-field splitting and 
and intersite exchange interactions within the  Kramers GS manifold for a set of rare-earth triangular materials, including two systems (KCeO$_2$ and KCeS$_2$) that order magnetically, as well as two putative QSLs (RbCeO$_2$ and YZGO). The calculated CF splitting and CF GS $g$-factors are in good agreement with experiment. The mean-field Curie-Weiss temperature evaluated from the resulting low-energy Hamiltonian 
agrees well with experiment for all three oxide systems but is severely underestimated for KCeS$_2$. 

Solving the resulting low-energy Hamiltonian with ED, we find evidence for the \(120^{\circ}\) order in all systems apart from KCeS$_2$, where a stripe-$\perp$ state is instead stabilized, consistent with experimental observations. We find strong evidence for a magnetic $120^\circ$ order in the two DSL candidates. From the extended phase diagram of their Hamiltonian, we concluded that a stronger antiferromagnetic NNN coupling is required to realize the DSL phases in comparison to those present in the compounds.

We notice that a very recent neutron-scattering study~\cite{Zhao2025} reports significant site disorder in YZGO as evidenced by broadening  of crystal-field excitations and small random crystalline distortions. The effect of site disorder was discussed in the context of a well-studied similar system, YbMgGaO$_4$, where the resulting randomization of intersite exchange was predicted to mimic a QSL state~\cite{Zhu2017b}. Our results together with the observations of Ref.~\cite{Zhao2025} suggest that the same phenomenon may be at play in YZGO.

Our calculated intersite exchange interactions include all $f$-$f$  kinetic exchange contributions (in  particular, superexchange). However, in rare-earth semiconductors there can be another type of $f$-$f$ intersite exchange generated through 4$f$-5$d$ hopping and the inter-shell Hund's rule interaction~\cite{Mauger1986}. This mechanism could become important in the case of very weak superexchange like in KCeS$_2$, where our calculations apparently reproduce the structure of intersite exchange but not its magnitude. \highlight{Its impact on the NNN exchange $J'$ also needs to be evaluated, especially in the light of our present predictions for  $J'$ to be too small to stablize a DSL phase in  YZGO.} Further studies to evaluate the importance of such exchange mechanisms in triangular rare-earth compounds are worthwhile.

{\it Acknowledgments -- } The exact diagonalization calculations have been performed using the XDiag library~\cite{wietek2025xdiagexactdiagonalizationquantum}. L.V.P. is grateful to Sidhartha Dash, Dario Fiore Mosca, and Antoine Georges for useful discussions. L.V.P acknowledges the support by CNRS through the Tremplin@Physique 2025 program and  by the  CPHT computer team. A.W. acknowledges support by the German Research Foundation (DFG) through the Emmy Noether program (Grant No. 509755282).


\begin{thebibliography}{82}%
	\makeatletter
	\providecommand \@ifxundefined [1]{%
		\@ifx{#1\undefined}
	}%
	\providecommand \@ifnum [1]{%
		\ifnum #1\expandafter \@firstoftwo
		\else \expandafter \@secondoftwo
		\fi
	}%
	\providecommand \@ifx [1]{%
		\ifx #1\expandafter \@firstoftwo
		\else \expandafter \@secondoftwo
		\fi
	}%
	\providecommand \natexlab [1]{#1}%
	\providecommand \enquote  [1]{``#1''}%
	\providecommand \bibnamefont  [1]{#1}%
	\providecommand \bibfnamefont [1]{#1}%
	\providecommand \citenamefont [1]{#1}%
	\providecommand \href@noop [0]{\@secondoftwo}%
	\providecommand \href [0]{\begingroup \@sanitize@url \@href}%
	\providecommand \@href[1]{\@@startlink{#1}\@@href}%
	\providecommand \@@href[1]{\endgroup#1\@@endlink}%
	\providecommand \@sanitize@url [0]{\catcode `\\12\catcode `\$12\catcode
		`\&12\catcode `\#12\catcode `\^12\catcode `\_12\catcode `\%12\relax}%
	\providecommand \@@startlink[1]{}%
	\providecommand \@@endlink[0]{}%
	\providecommand \url  [0]{\begingroup\@sanitize@url \@url }%
	\providecommand \@url [1]{\endgroup\@href {#1}{\urlprefix }}%
	\providecommand \urlprefix  [0]{URL }%
	\providecommand \Eprint [0]{\href }%
	\providecommand \doibase [0]{https://doi.org/}%
	\providecommand \selectlanguage [0]{\@gobble}%
	\providecommand \bibinfo  [0]{\@secondoftwo}%
	\providecommand \bibfield  [0]{\@secondoftwo}%
	\providecommand \translation [1]{[#1]}%
	\providecommand \BibitemOpen [0]{}%
	\providecommand \bibitemStop [0]{}%
	\providecommand \bibitemNoStop [0]{.\EOS\space}%
	\providecommand \EOS [0]{\spacefactor3000\relax}%
	\providecommand \BibitemShut  [1]{\csname bibitem#1\endcsname}%
	\let\auto@bib@innerbib\@empty
	\bibitem [{\citenamefont {Huse}\ and\ \citenamefont {Elser}(1988)}]{Huse1988}%
	\BibitemOpen
	\bibfield  {author} {\bibinfo {author} {\bibfnamefont {D.~A.}\ \bibnamefont
			{Huse}}\ and\ \bibinfo {author} {\bibfnamefont {V.}~\bibnamefont {Elser}},\
	}\bibfield  {title} {\bibinfo {title} {Simple variational wave functions for
			two-dimensional heisenberg spin-\textonehalf{} antiferromagnets},\ }\href
	{https://doi.org/10.1103/PhysRevLett.60.2531} {\bibfield  {journal} {\bibinfo
			{journal} {Phys. Rev. Lett.}\ }\textbf {\bibinfo {volume} {60}},\ \bibinfo
		{pages} {2531} (\bibinfo {year} {1988})}\BibitemShut {NoStop}%
	\bibitem [{\citenamefont {Bernu}\ \emph {et~al.}(1992)\citenamefont {Bernu},
		\citenamefont {Lhuillier},\ and\ \citenamefont {Pierre}}]{Bernu1992}%
	\BibitemOpen
	\bibfield  {author} {\bibinfo {author} {\bibfnamefont {B.}~\bibnamefont
			{Bernu}}, \bibinfo {author} {\bibfnamefont {C.}~\bibnamefont {Lhuillier}},\
		and\ \bibinfo {author} {\bibfnamefont {L.}~\bibnamefont {Pierre}},\
	}\bibfield  {title} {\bibinfo {title} {Signature of n\'eel order in exact
			spectra of quantum antiferromagnets on finite lattices},\ }\href
	{https://doi.org/10.1103/PhysRevLett.69.2590} {\bibfield  {journal} {\bibinfo
			{journal} {Phys. Rev. Lett.}\ }\textbf {\bibinfo {volume} {69}},\ \bibinfo
		{pages} {2590} (\bibinfo {year} {1992})}\BibitemShut {NoStop}%
	\bibitem [{\citenamefont {Capriotti}\ \emph {et~al.}(1999)\citenamefont
		{Capriotti}, \citenamefont {Trumper},\ and\ \citenamefont
		{Sorella}}]{Capriotti1999}%
	\BibitemOpen
	\bibfield  {author} {\bibinfo {author} {\bibfnamefont {L.}~\bibnamefont
			{Capriotti}}, \bibinfo {author} {\bibfnamefont {A.~E.}\ \bibnamefont
			{Trumper}},\ and\ \bibinfo {author} {\bibfnamefont {S.}~\bibnamefont
			{Sorella}},\ }\bibfield  {title} {\bibinfo {title} {Long-range n\'eel order
			in the triangular heisenberg model},\ }\href
	{https://doi.org/10.1103/PhysRevLett.82.3899} {\bibfield  {journal} {\bibinfo
			{journal} {Phys. Rev. Lett.}\ }\textbf {\bibinfo {volume} {82}},\ \bibinfo
		{pages} {3899} (\bibinfo {year} {1999})}\BibitemShut {NoStop}%
	\bibitem [{\citenamefont {White}\ and\ \citenamefont
		{Chernyshev}(2007)}]{White2007}%
	\BibitemOpen
	\bibfield  {author} {\bibinfo {author} {\bibfnamefont {S.~R.}\ \bibnamefont
			{White}}\ and\ \bibinfo {author} {\bibfnamefont {A.~L.}\ \bibnamefont
			{Chernyshev}},\ }\bibfield  {title} {\bibinfo {title} {Ne\'el order in square
			and triangular lattice heisenberg models},\ }\href
	{https://doi.org/10.1103/PhysRevLett.99.127004} {\bibfield  {journal}
		{\bibinfo  {journal} {Phys. Rev. Lett.}\ }\textbf {\bibinfo {volume} {99}},\
		\bibinfo {pages} {127004} (\bibinfo {year} {2007})}\BibitemShut {NoStop}%
	\bibitem [{\citenamefont {Hastings}(2000)}]{Hastings2000}%
	\BibitemOpen
	\bibfield  {author} {\bibinfo {author} {\bibfnamefont {M.~B.}\ \bibnamefont
			{Hastings}},\ }\bibfield  {title} {\bibinfo {title} {{Dirac structure, RVB,
				and Goldstone modes in the kagom\'e antiferromagnet}},\ }\href
	{https://doi.org/10.1103/PhysRevB.63.014413} {\bibfield  {journal} {\bibinfo
			{journal} {Phys. Rev. B}\ }\textbf {\bibinfo {volume} {63}},\ \bibinfo
		{pages} {014413} (\bibinfo {year} {2000})}\BibitemShut {NoStop}%
	\bibitem [{\citenamefont {Hermele}\ \emph {et~al.}(2004)\citenamefont
		{Hermele}, \citenamefont {Senthil}, \citenamefont {Fisher}, \citenamefont
		{Lee}, \citenamefont {Nagaosa},\ and\ \citenamefont {Wen}}]{Hermele2004}%
	\BibitemOpen
	\bibfield  {author} {\bibinfo {author} {\bibfnamefont {M.}~\bibnamefont
			{Hermele}}, \bibinfo {author} {\bibfnamefont {T.}~\bibnamefont {Senthil}},
		\bibinfo {author} {\bibfnamefont {M.~P.~A.}\ \bibnamefont {Fisher}}, \bibinfo
		{author} {\bibfnamefont {P.~A.}\ \bibnamefont {Lee}}, \bibinfo {author}
		{\bibfnamefont {N.}~\bibnamefont {Nagaosa}},\ and\ \bibinfo {author}
		{\bibfnamefont {X.-G.}\ \bibnamefont {Wen}},\ }\bibfield  {title} {\bibinfo
		{title} {{Stability of $U(1)$ spin liquids in two dimensions}},\ }\href
	{https://doi.org/10.1103/PhysRevB.70.214437} {\bibfield  {journal} {\bibinfo
			{journal} {Phys. Rev. B}\ }\textbf {\bibinfo {volume} {70}},\ \bibinfo
		{pages} {214437} (\bibinfo {year} {2004})}\BibitemShut {NoStop}%
	\bibitem [{\citenamefont {Hermele}\ \emph {et~al.}(2005)\citenamefont
		{Hermele}, \citenamefont {Senthil},\ and\ \citenamefont
		{Fisher}}]{Hermele2005}%
	\BibitemOpen
	\bibfield  {author} {\bibinfo {author} {\bibfnamefont {M.}~\bibnamefont
			{Hermele}}, \bibinfo {author} {\bibfnamefont {T.}~\bibnamefont {Senthil}},\
		and\ \bibinfo {author} {\bibfnamefont {M.~P.~A.}\ \bibnamefont {Fisher}},\
	}\bibfield  {title} {\bibinfo {title} {Algebraic spin liquid as the mother of
			many competing orders},\ }\href {https://doi.org/10.1103/PhysRevB.72.104404}
	{\bibfield  {journal} {\bibinfo  {journal} {Phys. Rev. B}\ }\textbf {\bibinfo
			{volume} {72}},\ \bibinfo {pages} {104404} (\bibinfo {year}
		{2005})}\BibitemShut {NoStop}%
	\bibitem [{\citenamefont {Kaneko}\ \emph {et~al.}(2014)\citenamefont {Kaneko},
		\citenamefont {Morita},\ and\ \citenamefont {Imada}}]{Kaneko2014}%
	\BibitemOpen
	\bibfield  {author} {\bibinfo {author} {\bibfnamefont {R.}~\bibnamefont
			{Kaneko}}, \bibinfo {author} {\bibfnamefont {S.}~\bibnamefont {Morita}},\
		and\ \bibinfo {author} {\bibfnamefont {M.}~\bibnamefont {Imada}},\ }\bibfield
	{title} {\bibinfo {title} {Gapless spin-liquid phase in an extended spin 1/2
			triangular heisenberg model},\ }\href
	{https://doi.org/10.7566/jpsj.83.093707} {\bibfield  {journal} {\bibinfo
			{journal} {Journal of the Physical Society of Japan}\ }\textbf {\bibinfo
			{volume} {83}},\ \bibinfo {pages} {093707} (\bibinfo {year}
		{2014})}\BibitemShut {NoStop}%
	\bibitem [{\citenamefont {Iqbal}\ \emph {et~al.}(2016)\citenamefont {Iqbal},
		\citenamefont {Hu}, \citenamefont {Thomale}, \citenamefont {Poilblanc},\ and\
		\citenamefont {Becca}}]{Iqbal2016}%
	\BibitemOpen
	\bibfield  {author} {\bibinfo {author} {\bibfnamefont {Y.}~\bibnamefont
			{Iqbal}}, \bibinfo {author} {\bibfnamefont {W.-J.}\ \bibnamefont {Hu}},
		\bibinfo {author} {\bibfnamefont {R.}~\bibnamefont {Thomale}}, \bibinfo
		{author} {\bibfnamefont {D.}~\bibnamefont {Poilblanc}},\ and\ \bibinfo
		{author} {\bibfnamefont {F.}~\bibnamefont {Becca}},\ }\bibfield  {title}
	{\bibinfo {title} {Spin liquid nature in the heisenberg
			${J}_{1}\ensuremath{-}{J}_{2}$ triangular antiferromagnet},\ }\href
	{https://doi.org/10.1103/PhysRevB.93.144411} {\bibfield  {journal} {\bibinfo
			{journal} {Phys. Rev. B}\ }\textbf {\bibinfo {volume} {93}},\ \bibinfo
		{pages} {144411} (\bibinfo {year} {2016})}\BibitemShut {NoStop}%
	\bibitem [{\citenamefont {Hu}\ \emph {et~al.}(2019)\citenamefont {Hu},
		\citenamefont {Zhu}, \citenamefont {Eggert},\ and\ \citenamefont
		{He}}]{Hu2019}%
	\BibitemOpen
	\bibfield  {author} {\bibinfo {author} {\bibfnamefont {S.}~\bibnamefont
			{Hu}}, \bibinfo {author} {\bibfnamefont {W.}~\bibnamefont {Zhu}}, \bibinfo
		{author} {\bibfnamefont {S.}~\bibnamefont {Eggert}},\ and\ \bibinfo {author}
		{\bibfnamefont {Y.-C.}\ \bibnamefont {He}},\ }\bibfield  {title} {\bibinfo
		{title} {Dirac spin liquid on the spin-$1/2$ triangular {H}eisenberg
			antiferromagnet},\ }\href {https://doi.org/10.1103/PhysRevLett.123.207203}
	{\bibfield  {journal} {\bibinfo  {journal} {Phys. Rev. Lett.}\ }\textbf
		{\bibinfo {volume} {123}},\ \bibinfo {pages} {207203} (\bibinfo {year}
		{2019})}\BibitemShut {NoStop}%
	\bibitem [{\citenamefont {Wietek}\ \emph {et~al.}(2024)\citenamefont {Wietek},
		\citenamefont {Capponi},\ and\ \citenamefont {L\"auchli}}]{Wietek2024}%
	\BibitemOpen
	\bibfield  {author} {\bibinfo {author} {\bibfnamefont {A.}~\bibnamefont
			{Wietek}}, \bibinfo {author} {\bibfnamefont {S.}~\bibnamefont {Capponi}},\
		and\ \bibinfo {author} {\bibfnamefont {A.~M.}\ \bibnamefont {L\"auchli}},\
	}\bibfield  {title} {\bibinfo {title} {Quantum electrodynamics in $2+1$
			dimensions as the organizing principle of a triangular lattice
			antiferromagnet},\ }\href {https://doi.org/10.1103/PhysRevX.14.021010}
	{\bibfield  {journal} {\bibinfo  {journal} {Phys. Rev. X}\ }\textbf {\bibinfo
			{volume} {14}},\ \bibinfo {pages} {021010} (\bibinfo {year}
		{2024})}\BibitemShut {NoStop}%
	\bibitem [{\citenamefont {Zhu}\ and\ \citenamefont {White}(2015)}]{Zhu2015}%
	\BibitemOpen
	\bibfield  {author} {\bibinfo {author} {\bibfnamefont {Z.}~\bibnamefont
			{Zhu}}\ and\ \bibinfo {author} {\bibfnamefont {S.~R.}\ \bibnamefont
			{White}},\ }\bibfield  {title} {\bibinfo {title} {Spin liquid phase of the
			${S}=\frac{1}{2}\phantom{\rule{4.pt}{0ex}}{J}_{1}\ensuremath{-}{J}_{2}$
			{H}eisenberg model on the triangular lattice},\ }\href
	{https://doi.org/10.1103/PhysRevB.92.041105} {\bibfield  {journal} {\bibinfo
			{journal} {Phys. Rev. B}\ }\textbf {\bibinfo {volume} {92}},\ \bibinfo
		{pages} {041105} (\bibinfo {year} {2015})}\BibitemShut {NoStop}%
	\bibitem [{\citenamefont {Hu}\ \emph {et~al.}(2015)\citenamefont {Hu},
		\citenamefont {Gong}, \citenamefont {Zhu},\ and\ \citenamefont
		{Sheng}}]{Hu2015}%
	\BibitemOpen
	\bibfield  {author} {\bibinfo {author} {\bibfnamefont {W.-J.}\ \bibnamefont
			{Hu}}, \bibinfo {author} {\bibfnamefont {S.-S.}\ \bibnamefont {Gong}},
		\bibinfo {author} {\bibfnamefont {W.}~\bibnamefont {Zhu}},\ and\ \bibinfo
		{author} {\bibfnamefont {D.~N.}\ \bibnamefont {Sheng}},\ }\bibfield  {title}
	{\bibinfo {title} {Competing spin-liquid states in the spin-$\frac{1}{2}$
			{H}eisenberg model on the triangular lattice},\ }\href
	{https://doi.org/10.1103/PhysRevB.92.140403} {\bibfield  {journal} {\bibinfo
			{journal} {Phys. Rev. B}\ }\textbf {\bibinfo {volume} {92}},\ \bibinfo
		{pages} {140403} (\bibinfo {year} {2015})}\BibitemShut {NoStop}%
	\bibitem [{\citenamefont {Saadatmand}\ and\ \citenamefont
		{McCulloch}(2016)}]{Saadatmand2016}%
	\BibitemOpen
	\bibfield  {author} {\bibinfo {author} {\bibfnamefont {S.~N.}\ \bibnamefont
			{Saadatmand}}\ and\ \bibinfo {author} {\bibfnamefont {I.~P.}\ \bibnamefont
			{McCulloch}},\ }\bibfield  {title} {\bibinfo {title} {Symmetry
			fractionalization in the topological phase of the spin-$\frac{1}{2}$
			${J}_{1}\text{\ensuremath{-}}{J}_{2}$ triangular {H}eisenberg model},\ }\href
	{https://doi.org/10.1103/PhysRevB.94.121111} {\bibfield  {journal} {\bibinfo
			{journal} {Phys. Rev. B}\ }\textbf {\bibinfo {volume} {94}},\ \bibinfo
		{pages} {121111} (\bibinfo {year} {2016})}\BibitemShut {NoStop}%
	\bibitem [{\citenamefont {Jiang}\ and\ \citenamefont
		{Jiang}(2023)}]{Jiang2022}%
	\BibitemOpen
	\bibfield  {author} {\bibinfo {author} {\bibfnamefont {Y.-F.}\ \bibnamefont
			{Jiang}}\ and\ \bibinfo {author} {\bibfnamefont {H.-C.}\ \bibnamefont
			{Jiang}},\ }\bibfield  {title} {\bibinfo {title} {{Nature of quantum spin
				liquids of the $S=\frac{1}{2}$ Heisenberg antiferromagnet on the triangular
				lattice: A parallel DMRG study}},\ }\href
	{https://doi.org/10.1103/PhysRevB.107.L140411} {\bibfield  {journal}
		{\bibinfo  {journal} {Phys. Rev. B}\ }\textbf {\bibinfo {volume} {107}},\
		\bibinfo {pages} {L140411} (\bibinfo {year} {2023})}\BibitemShut {NoStop}%
	\bibitem [{\citenamefont {Saadatmand}\ and\ \citenamefont
		{McCulloch}(2017)}]{Saadatmand2017}%
	\BibitemOpen
	\bibfield  {author} {\bibinfo {author} {\bibfnamefont {S.~N.}\ \bibnamefont
			{Saadatmand}}\ and\ \bibinfo {author} {\bibfnamefont {I.~P.}\ \bibnamefont
			{McCulloch}},\ }\bibfield  {title} {\bibinfo {title} {Detection and
			characterization of symmetry-broken long-range orders in the
			spin-$\frac{1}{2}$ triangular {H}eisenberg model},\ }\href
	{https://doi.org/10.1103/PhysRevB.96.075117} {\bibfield  {journal} {\bibinfo
			{journal} {Phys. Rev. B}\ }\textbf {\bibinfo {volume} {96}},\ \bibinfo
		{pages} {075117} (\bibinfo {year} {2017})}\BibitemShut {NoStop}%
	\bibitem [{\citenamefont {Gong}\ \emph {et~al.}(2017)\citenamefont {Gong},
		\citenamefont {Zhu}, \citenamefont {Zhu}, \citenamefont {Sheng},\ and\
		\citenamefont {Yang}}]{Gong2017}%
	\BibitemOpen
	\bibfield  {author} {\bibinfo {author} {\bibfnamefont {S.-S.}\ \bibnamefont
			{Gong}}, \bibinfo {author} {\bibfnamefont {W.}~\bibnamefont {Zhu}}, \bibinfo
		{author} {\bibfnamefont {J.-X.}\ \bibnamefont {Zhu}}, \bibinfo {author}
		{\bibfnamefont {D.~N.}\ \bibnamefont {Sheng}},\ and\ \bibinfo {author}
		{\bibfnamefont {K.}~\bibnamefont {Yang}},\ }\bibfield  {title} {\bibinfo
		{title} {Global phase diagram and quantum spin liquids in a
			spin-$\frac{1}{2}$ triangular antiferromagnet},\ }\href
	{https://doi.org/10.1103/PhysRevB.96.075116} {\bibfield  {journal} {\bibinfo
			{journal} {Phys. Rev. B}\ }\textbf {\bibinfo {volume} {96}},\ \bibinfo
		{pages} {075116} (\bibinfo {year} {2017})}\BibitemShut {NoStop}%
	\bibitem [{\citenamefont {Wietek}\ and\ \citenamefont
		{L\"auchli}(2017)}]{Wietek2017}%
	\BibitemOpen
	\bibfield  {author} {\bibinfo {author} {\bibfnamefont {A.}~\bibnamefont
			{Wietek}}\ and\ \bibinfo {author} {\bibfnamefont {A.~M.}\ \bibnamefont
			{L\"auchli}},\ }\bibfield  {title} {\bibinfo {title} {Chiral spin liquid and
			quantum criticality in extended $s=\frac{1}{2}$ {H}eisenberg models on the
			triangular lattice},\ }\href {https://doi.org/10.1103/PhysRevB.95.035141}
	{\bibfield  {journal} {\bibinfo  {journal} {Phys. Rev. B}\ }\textbf {\bibinfo
			{volume} {95}},\ \bibinfo {pages} {035141} (\bibinfo {year}
		{2017})}\BibitemShut {NoStop}%
	\bibitem [{\citenamefont {Song}\ \emph {et~al.}(2019)\citenamefont {Song},
		\citenamefont {Wang}, \citenamefont {Vishwanath},\ and\ \citenamefont
		{He}}]{Song2019}%
	\BibitemOpen
	\bibfield  {author} {\bibinfo {author} {\bibfnamefont {X.-Y.}\ \bibnamefont
			{Song}}, \bibinfo {author} {\bibfnamefont {C.}~\bibnamefont {Wang}}, \bibinfo
		{author} {\bibfnamefont {A.}~\bibnamefont {Vishwanath}},\ and\ \bibinfo
		{author} {\bibfnamefont {Y.-C.}\ \bibnamefont {He}},\ }\bibfield  {title}
	{\bibinfo {title} {Unifying description of competing orders in
			two-dimensional quantum magnets},\ }\href
	{https://doi.org/10.1038/s41467-019-11727-3} {\bibfield  {journal} {\bibinfo
			{journal} {Nat. Commun.}\ }\textbf {\bibinfo {volume} {10}},\ \bibinfo
		{pages} {4254} (\bibinfo {year} {2019})}\BibitemShut {NoStop}%
	\bibitem [{\citenamefont {Song}\ \emph {et~al.}(2020)\citenamefont {Song},
		\citenamefont {He}, \citenamefont {Vishwanath},\ and\ \citenamefont
		{Wang}}]{Song2020}%
	\BibitemOpen
	\bibfield  {author} {\bibinfo {author} {\bibfnamefont {X.-Y.}\ \bibnamefont
			{Song}}, \bibinfo {author} {\bibfnamefont {Y.-C.}\ \bibnamefont {He}},
		\bibinfo {author} {\bibfnamefont {A.}~\bibnamefont {Vishwanath}},\ and\
		\bibinfo {author} {\bibfnamefont {C.}~\bibnamefont {Wang}},\ }\bibfield
	{title} {\bibinfo {title} {From spinon band topology to the symmetry quantum
			numbers of monopoles in {D}irac spin liquids},\ }\href
	{https://doi.org/10.1103/PhysRevX.10.011033} {\bibfield  {journal} {\bibinfo
			{journal} {Phys. Rev. X}\ }\textbf {\bibinfo {volume} {10}},\ \bibinfo
		{pages} {011033} (\bibinfo {year} {2020})}\BibitemShut {NoStop}%
	\bibitem [{\citenamefont {Bordelon}\ \emph {et~al.}(2019)\citenamefont
		{Bordelon}, \citenamefont {Kenney}, \citenamefont {Liu}, \citenamefont
		{Hogan}, \citenamefont {Posthuma}, \citenamefont {Kavand}, \citenamefont
		{Lyu}, \citenamefont {Sherwin}, \citenamefont {Butch}, \citenamefont {Brown},
		\citenamefont {Graf}, \citenamefont {Balents},\ and\ \citenamefont
		{Wilson}}]{Bordelon2017}%
	\BibitemOpen
	\bibfield  {author} {\bibinfo {author} {\bibfnamefont {M.~M.}\ \bibnamefont
			{Bordelon}}, \bibinfo {author} {\bibfnamefont {E.}~\bibnamefont {Kenney}},
		\bibinfo {author} {\bibfnamefont {C.}~\bibnamefont {Liu}}, \bibinfo {author}
		{\bibfnamefont {T.}~\bibnamefont {Hogan}}, \bibinfo {author} {\bibfnamefont
			{L.}~\bibnamefont {Posthuma}}, \bibinfo {author} {\bibfnamefont
			{M.}~\bibnamefont {Kavand}}, \bibinfo {author} {\bibfnamefont
			{Y.}~\bibnamefont {Lyu}}, \bibinfo {author} {\bibfnamefont {M.}~\bibnamefont
			{Sherwin}}, \bibinfo {author} {\bibfnamefont {N.~P.}\ \bibnamefont {Butch}},
		\bibinfo {author} {\bibfnamefont {C.}~\bibnamefont {Brown}}, \bibinfo
		{author} {\bibfnamefont {M.~J.}\ \bibnamefont {Graf}}, \bibinfo {author}
		{\bibfnamefont {L.}~\bibnamefont {Balents}},\ and\ \bibinfo {author}
		{\bibfnamefont {S.~D.}\ \bibnamefont {Wilson}},\ }\bibfield  {title}
	{\bibinfo {title} {Field-tunable quantum disordered ground state in the
			triangular-lattice antiferromagnet {NaYbO}$_2$},\ }\href
	{https://doi.org/10.1038/s41567-019-0594-5} {\bibfield  {journal} {\bibinfo
			{journal} {Nature Physics}\ }\textbf {\bibinfo {volume} {15}},\ \bibinfo
		{pages} {1058} (\bibinfo {year} {2019})}\BibitemShut {NoStop}%
	\bibitem [{\citenamefont {Bastien}\ \emph {et~al.}(2020)\citenamefont
		{Bastien}, \citenamefont {Rubrecht}, \citenamefont {Haeussler}, \citenamefont
		{Schlender}, \citenamefont {Zangeneh}, \citenamefont {Avdoshenko},
		\citenamefont {Sarkar}, \citenamefont {Alfonsov}, \citenamefont {Luther},
		\citenamefont {Onykiienko}, \citenamefont {Walker}, \citenamefont {Kühne},
		\citenamefont {Grinenko}, \citenamefont {Guguchia}, \citenamefont {Kataev},
		\citenamefont {Klauss}, \citenamefont {Hozoi}, \citenamefont {van~den Brink},
		\citenamefont {Inosov}, \citenamefont {Büchner}, \citenamefont {Wolter},\
		and\ \citenamefont {Doert}}]{Bastien2020}%
	\BibitemOpen
	\bibfield  {author} {\bibinfo {author} {\bibfnamefont {G.}~\bibnamefont
			{Bastien}}, \bibinfo {author} {\bibfnamefont {B.}~\bibnamefont {Rubrecht}},
		\bibinfo {author} {\bibfnamefont {E.}~\bibnamefont {Haeussler}}, \bibinfo
		{author} {\bibfnamefont {P.}~\bibnamefont {Schlender}}, \bibinfo {author}
		{\bibfnamefont {Z.}~\bibnamefont {Zangeneh}}, \bibinfo {author}
		{\bibfnamefont {S.}~\bibnamefont {Avdoshenko}}, \bibinfo {author}
		{\bibfnamefont {R.}~\bibnamefont {Sarkar}}, \bibinfo {author} {\bibfnamefont
			{A.}~\bibnamefont {Alfonsov}}, \bibinfo {author} {\bibfnamefont
			{S.}~\bibnamefont {Luther}}, \bibinfo {author} {\bibfnamefont {Y.~A.}\
			\bibnamefont {Onykiienko}}, \bibinfo {author} {\bibfnamefont {H.~C.}\
			\bibnamefont {Walker}}, \bibinfo {author} {\bibfnamefont {H.}~\bibnamefont
			{Kühne}}, \bibinfo {author} {\bibfnamefont {V.}~\bibnamefont {Grinenko}},
		\bibinfo {author} {\bibfnamefont {Z.}~\bibnamefont {Guguchia}}, \bibinfo
		{author} {\bibfnamefont {V.}~\bibnamefont {Kataev}}, \bibinfo {author}
		{\bibfnamefont {H.~H.}\ \bibnamefont {Klauss}}, \bibinfo {author}
		{\bibfnamefont {L.}~\bibnamefont {Hozoi}}, \bibinfo {author} {\bibfnamefont
			{J.}~\bibnamefont {van~den Brink}}, \bibinfo {author} {\bibfnamefont {D.~S.}\
			\bibnamefont {Inosov}}, \bibinfo {author} {\bibfnamefont {B.}~\bibnamefont
			{Büchner}}, \bibinfo {author} {\bibfnamefont {A.~U.~B.}\ \bibnamefont
			{Wolter}},\ and\ \bibinfo {author} {\bibfnamefont {T.}~\bibnamefont
			{Doert}},\ }\bibfield  {title} {\bibinfo {title} {{Long-range magnetic order
				in the ${\tilde S}=1/2$ triangular lattice antiferromagnet KCeS$_2$}},\
	}\href {https://doi.org/10.21468/SciPostPhys.9.3.041} {\bibfield  {journal}
		{\bibinfo  {journal} {SciPost Phys.}\ }\textbf {\bibinfo {volume} {9}},\
		\bibinfo {pages} {041} (\bibinfo {year} {2020})}\BibitemShut {NoStop}%
	\bibitem [{\citenamefont {Bordelon}\ \emph {et~al.}(2021)\citenamefont
		{Bordelon}, \citenamefont {Wang}, \citenamefont {Pajerowski}, \citenamefont
		{Banerjee}, \citenamefont {Sherwin}, \citenamefont {Brown}, \citenamefont
		{Eldeeb}, \citenamefont {Petersen}, \citenamefont {Hozoi}, \citenamefont
		{R\"o\ss{}ler}, \citenamefont {Mourigal},\ and\ \citenamefont
		{Wilson}}]{Bordelon2021}%
	\BibitemOpen
	\bibfield  {author} {\bibinfo {author} {\bibfnamefont {M.~M.}\ \bibnamefont
			{Bordelon}}, \bibinfo {author} {\bibfnamefont {X.}~\bibnamefont {Wang}},
		\bibinfo {author} {\bibfnamefont {D.~M.}\ \bibnamefont {Pajerowski}},
		\bibinfo {author} {\bibfnamefont {A.}~\bibnamefont {Banerjee}}, \bibinfo
		{author} {\bibfnamefont {M.}~\bibnamefont {Sherwin}}, \bibinfo {author}
		{\bibfnamefont {C.~M.}\ \bibnamefont {Brown}}, \bibinfo {author}
		{\bibfnamefont {M.~S.}\ \bibnamefont {Eldeeb}}, \bibinfo {author}
		{\bibfnamefont {T.}~\bibnamefont {Petersen}}, \bibinfo {author}
		{\bibfnamefont {L.}~\bibnamefont {Hozoi}}, \bibinfo {author} {\bibfnamefont
			{U.~K.}\ \bibnamefont {R\"o\ss{}ler}}, \bibinfo {author} {\bibfnamefont
			{M.}~\bibnamefont {Mourigal}},\ and\ \bibinfo {author} {\bibfnamefont
			{S.~D.}\ \bibnamefont {Wilson}},\ }\bibfield  {title} {\bibinfo {title}
		{Magnetic properties and signatures of moment ordering in the triangular
			lattice antiferromagnet {KCeO}$_{2}$},\ }\href
	{https://doi.org/10.1103/PhysRevB.104.094421} {\bibfield  {journal} {\bibinfo
			{journal} {Phys. Rev. B}\ }\textbf {\bibinfo {volume} {104}},\ \bibinfo
		{pages} {094421} (\bibinfo {year} {2021})}\BibitemShut {NoStop}%
	\bibitem [{\citenamefont {Kulbakov}\ \emph {et~al.}(2021)\citenamefont
		{Kulbakov}, \citenamefont {Avdoshenko}, \citenamefont {Puente-Orench},
		\citenamefont {Deeb}, \citenamefont {Doerr}, \citenamefont {Schlender},
		\citenamefont {Doert},\ and\ \citenamefont {Inosov}}]{Kulbakov2021}%
	\BibitemOpen
	\bibfield  {author} {\bibinfo {author} {\bibfnamefont {A.~A.}\ \bibnamefont
			{Kulbakov}}, \bibinfo {author} {\bibfnamefont {S.~M.}\ \bibnamefont
			{Avdoshenko}}, \bibinfo {author} {\bibfnamefont {I.}~\bibnamefont
			{Puente-Orench}}, \bibinfo {author} {\bibfnamefont {M.}~\bibnamefont {Deeb}},
		\bibinfo {author} {\bibfnamefont {M.}~\bibnamefont {Doerr}}, \bibinfo
		{author} {\bibfnamefont {P.}~\bibnamefont {Schlender}}, \bibinfo {author}
		{\bibfnamefont {T.}~\bibnamefont {Doert}},\ and\ \bibinfo {author}
		{\bibfnamefont {D.~S.}\ \bibnamefont {Inosov}},\ }\bibfield  {title}
	{\bibinfo {title} {Stripe-yz magnetic order in the triangular-lattice
			antiferromagnet {KCeS}$_2$},\ }\href
	{https://doi.org/10.1088/1361-648X/ac15d6} {\bibfield  {journal} {\bibinfo
			{journal} {Journal of Physics: Condensed Matter}\ }\textbf {\bibinfo {volume}
			{33}},\ \bibinfo {pages} {425802} (\bibinfo {year} {2021})}\BibitemShut
	{NoStop}%
	\bibitem [{\citenamefont {Ortiz}\ \emph {et~al.}(2022)\citenamefont {Ortiz},
		\citenamefont {Bordelon}, \citenamefont {Bhattacharyya}, \citenamefont
		{Pokharel}, \citenamefont {Sarte}, \citenamefont {Posthuma}, \citenamefont
		{Petersen}, \citenamefont {Eldeeb}, \citenamefont {Granroth}, \citenamefont
		{Dela~Cruz}, \citenamefont {Calder}, \citenamefont {Abernathy}, \citenamefont
		{Hozoi},\ and\ \citenamefont {Wilson}}]{Ortiz2022}%
	\BibitemOpen
	\bibfield  {author} {\bibinfo {author} {\bibfnamefont {B.~R.}\ \bibnamefont
			{Ortiz}}, \bibinfo {author} {\bibfnamefont {M.~M.}\ \bibnamefont {Bordelon}},
		\bibinfo {author} {\bibfnamefont {P.}~\bibnamefont {Bhattacharyya}}, \bibinfo
		{author} {\bibfnamefont {G.}~\bibnamefont {Pokharel}}, \bibinfo {author}
		{\bibfnamefont {P.~M.}\ \bibnamefont {Sarte}}, \bibinfo {author}
		{\bibfnamefont {L.}~\bibnamefont {Posthuma}}, \bibinfo {author}
		{\bibfnamefont {T.}~\bibnamefont {Petersen}}, \bibinfo {author}
		{\bibfnamefont {M.~S.}\ \bibnamefont {Eldeeb}}, \bibinfo {author}
		{\bibfnamefont {G.~E.}\ \bibnamefont {Granroth}}, \bibinfo {author}
		{\bibfnamefont {C.~R.}\ \bibnamefont {Dela~Cruz}}, \bibinfo {author}
		{\bibfnamefont {S.}~\bibnamefont {Calder}}, \bibinfo {author} {\bibfnamefont
			{D.~L.}\ \bibnamefont {Abernathy}}, \bibinfo {author} {\bibfnamefont
			{L.}~\bibnamefont {Hozoi}},\ and\ \bibinfo {author} {\bibfnamefont {S.~D.}\
			\bibnamefont {Wilson}},\ }\bibfield  {title} {\bibinfo {title} {Electronic
			and structural properties of {RbCe}${X}_{2}$ $({X}_{2}:$ {O}$_{2},$
			{S}$_{2}$, {SeS}, {Se}$_{2}$, {TeSe}, {Te}$_{2}$)},\ }\href
	{https://doi.org/10.1103/PhysRevMaterials.6.084402} {\bibfield  {journal}
		{\bibinfo  {journal} {Phys. Rev. Mater.}\ }\textbf {\bibinfo {volume} {6}},\
		\bibinfo {pages} {084402} (\bibinfo {year} {2022})}\BibitemShut {NoStop}%
	\bibitem [{\citenamefont {Gru\ss{}ler}\ \emph {et~al.}(2023)\citenamefont
		{Gru\ss{}ler}, \citenamefont {Hemmida}, \citenamefont {Bachus}, \citenamefont
		{Skourski}, \citenamefont {Krug~von Nidda}, \citenamefont {Gegenwart},\ and\
		\citenamefont {Tsirlin}}]{Grussler2023}%
	\BibitemOpen
	\bibfield  {author} {\bibinfo {author} {\bibfnamefont {F.}~\bibnamefont
			{Gru\ss{}ler}}, \bibinfo {author} {\bibfnamefont {M.}~\bibnamefont
			{Hemmida}}, \bibinfo {author} {\bibfnamefont {S.}~\bibnamefont {Bachus}},
		\bibinfo {author} {\bibfnamefont {Y.}~\bibnamefont {Skourski}}, \bibinfo
		{author} {\bibfnamefont {H.-A.}\ \bibnamefont {Krug~von Nidda}}, \bibinfo
		{author} {\bibfnamefont {P.}~\bibnamefont {Gegenwart}},\ and\ \bibinfo
		{author} {\bibfnamefont {A.~A.}\ \bibnamefont {Tsirlin}},\ }\bibfield
	{title} {\bibinfo {title} {Role of alkaline metal in the rare-earth
			triangular antiferromagnet {KYbO}$_{2}$},\ }\href
	{https://doi.org/10.1103/PhysRevB.107.224416} {\bibfield  {journal} {\bibinfo
			{journal} {Phys. Rev. B}\ }\textbf {\bibinfo {volume} {107}},\ \bibinfo
		{pages} {224416} (\bibinfo {year} {2023})}\BibitemShut {NoStop}%
	\bibitem [{\citenamefont {Xie}\ \emph {et~al.}(2024{\natexlab{a}})\citenamefont
		{Xie}, \citenamefont {Zhuo}, \citenamefont {Cai}, \citenamefont {Zhang},\
		and\ \citenamefont {Zhang}}]{Xie2024}%
	\BibitemOpen
	\bibfield  {author} {\bibinfo {author} {\bibfnamefont {M.}~\bibnamefont
			{Xie}}, \bibinfo {author} {\bibfnamefont {W.}~\bibnamefont {Zhuo}}, \bibinfo
		{author} {\bibfnamefont {Y.}~\bibnamefont {Cai}}, \bibinfo {author}
		{\bibfnamefont {Z.}~\bibnamefont {Zhang}},\ and\ \bibinfo {author}
		{\bibfnamefont {Q.}~\bibnamefont {Zhang}},\ }\bibfield  {title} {\bibinfo
		{title} {Rare-earth chalcogenides: An inspiring playground for exploring
			frustrated magnetism},\ }\href
	{https://doi.org/10.1088/0256-307X/41/11/117505} {\bibfield  {journal}
		{\bibinfo  {journal} {Chinese Physics Letters}\ }\textbf {\bibinfo {volume}
			{41}},\ \bibinfo {pages} {117505} (\bibinfo {year}
		{2024}{\natexlab{a}})}\BibitemShut {NoStop}%
	\bibitem [{\citenamefont {Arh}\ \emph {et~al.}(2022)\citenamefont {Arh},
		\citenamefont {Sana}, \citenamefont {Pregelj}, \citenamefont {Khuntia},
		\citenamefont {Jagli{\v c}i{\'c}}, \citenamefont {Le}, \citenamefont
		{Biswas}, \citenamefont {Manuel}, \citenamefont {Mangin-Thro}, \citenamefont
		{Ozarowski},\ and\ \citenamefont {Zorko}}]{Arh2022}%
	\BibitemOpen
	\bibfield  {author} {\bibinfo {author} {\bibfnamefont {T.}~\bibnamefont
			{Arh}}, \bibinfo {author} {\bibfnamefont {B.}~\bibnamefont {Sana}}, \bibinfo
		{author} {\bibfnamefont {M.}~\bibnamefont {Pregelj}}, \bibinfo {author}
		{\bibfnamefont {P.}~\bibnamefont {Khuntia}}, \bibinfo {author} {\bibfnamefont
			{Z.}~\bibnamefont {Jagli{\v c}i{\'c}}}, \bibinfo {author} {\bibfnamefont
			{M.~D.}\ \bibnamefont {Le}}, \bibinfo {author} {\bibfnamefont {P.~K.}\
			\bibnamefont {Biswas}}, \bibinfo {author} {\bibfnamefont {P.}~\bibnamefont
			{Manuel}}, \bibinfo {author} {\bibfnamefont {L.}~\bibnamefont {Mangin-Thro}},
		\bibinfo {author} {\bibfnamefont {A.}~\bibnamefont {Ozarowski}},\ and\
		\bibinfo {author} {\bibfnamefont {A.}~\bibnamefont {Zorko}},\ }\bibfield
	{title} {\bibinfo {title} {The ising triangular-lattice antiferromagnet
			neodymium heptatantalate as a quantum spin liquid candidate},\ }\href
	{https://doi.org/10.1038/s41563-021-01169-y} {\bibfield  {journal} {\bibinfo
			{journal} {Nature Materials}\ }\textbf {\bibinfo {volume} {21}},\ \bibinfo
		{pages} {416} (\bibinfo {year} {2022})}\BibitemShut {NoStop}%
	\bibitem [{\citenamefont {Li}\ \emph {et~al.}(2015)\citenamefont {Li},
		\citenamefont {Chen}, \citenamefont {Tong}, \citenamefont {Pi}, \citenamefont
		{Liu}, \citenamefont {Yang}, \citenamefont {Wang},\ and\ \citenamefont
		{Zhang}}]{Li2015}%
	\BibitemOpen
	\bibfield  {author} {\bibinfo {author} {\bibfnamefont {Y.}~\bibnamefont
			{Li}}, \bibinfo {author} {\bibfnamefont {G.}~\bibnamefont {Chen}}, \bibinfo
		{author} {\bibfnamefont {W.}~\bibnamefont {Tong}}, \bibinfo {author}
		{\bibfnamefont {L.}~\bibnamefont {Pi}}, \bibinfo {author} {\bibfnamefont
			{J.}~\bibnamefont {Liu}}, \bibinfo {author} {\bibfnamefont {Z.}~\bibnamefont
			{Yang}}, \bibinfo {author} {\bibfnamefont {X.}~\bibnamefont {Wang}},\ and\
		\bibinfo {author} {\bibfnamefont {Q.}~\bibnamefont {Zhang}},\ }\bibfield
	{title} {\bibinfo {title} {Rare-earth triangular lattice spin liquid: A
			single-crystal study of {YbMgGaO}$_{4}$},\ }\href
	{https://doi.org/10.1103/PhysRevLett.115.167203} {\bibfield  {journal}
		{\bibinfo  {journal} {Phys. Rev. Lett.}\ }\textbf {\bibinfo {volume} {115}},\
		\bibinfo {pages} {167203} (\bibinfo {year} {2015})}\BibitemShut {NoStop}%
	\bibitem [{\citenamefont {Shen}\ \emph {et~al.}(2016)\citenamefont {Shen},
		\citenamefont {Li}, \citenamefont {Wo}, \citenamefont {Li}, \citenamefont
		{Shen}, \citenamefont {Pan}, \citenamefont {Wang}, \citenamefont {Walker},
		\citenamefont {Steffens}, \citenamefont {Boehm}, \citenamefont {Hao},
		\citenamefont {Quintero-Castro}, \citenamefont {Harriger}, \citenamefont
		{Frontzek}, \citenamefont {Hao}, \citenamefont {Meng}, \citenamefont {Zhang},
		\citenamefont {Chen},\ and\ \citenamefont {Zhao}}]{Shen2016}%
	\BibitemOpen
	\bibfield  {author} {\bibinfo {author} {\bibfnamefont {Y.}~\bibnamefont
			{Shen}}, \bibinfo {author} {\bibfnamefont {Y.-D.}\ \bibnamefont {Li}},
		\bibinfo {author} {\bibfnamefont {H.}~\bibnamefont {Wo}}, \bibinfo {author}
		{\bibfnamefont {Y.}~\bibnamefont {Li}}, \bibinfo {author} {\bibfnamefont
			{S.}~\bibnamefont {Shen}}, \bibinfo {author} {\bibfnamefont {B.}~\bibnamefont
			{Pan}}, \bibinfo {author} {\bibfnamefont {Q.}~\bibnamefont {Wang}}, \bibinfo
		{author} {\bibfnamefont {H.~C.}\ \bibnamefont {Walker}}, \bibinfo {author}
		{\bibfnamefont {P.}~\bibnamefont {Steffens}}, \bibinfo {author}
		{\bibfnamefont {M.}~\bibnamefont {Boehm}}, \bibinfo {author} {\bibfnamefont
			{Y.}~\bibnamefont {Hao}}, \bibinfo {author} {\bibfnamefont {D.~L.}\
			\bibnamefont {Quintero-Castro}}, \bibinfo {author} {\bibfnamefont {L.~W.}\
			\bibnamefont {Harriger}}, \bibinfo {author} {\bibfnamefont {M.~D.}\
			\bibnamefont {Frontzek}}, \bibinfo {author} {\bibfnamefont {L.}~\bibnamefont
			{Hao}}, \bibinfo {author} {\bibfnamefont {S.}~\bibnamefont {Meng}}, \bibinfo
		{author} {\bibfnamefont {Q.}~\bibnamefont {Zhang}}, \bibinfo {author}
		{\bibfnamefont {G.}~\bibnamefont {Chen}},\ and\ \bibinfo {author}
		{\bibfnamefont {J.}~\bibnamefont {Zhao}},\ }\bibfield  {title} {\bibinfo
		{title} {Evidence for a spinon fermi surface in a triangular-lattice
			quantum-spin-liquid candidate},\ }\href {https://doi.org/10.1038/nature20614}
	{\bibfield  {journal} {\bibinfo  {journal} {Nature}\ }\textbf {\bibinfo
			{volume} {540}},\ \bibinfo {pages} {559} (\bibinfo {year}
		{2016})}\BibitemShut {NoStop}%
	\bibitem [{\citenamefont {Li}\ \emph {et~al.}(2016{\natexlab{a}})\citenamefont
		{Li}, \citenamefont {Adroja}, \citenamefont {Biswas}, \citenamefont {Baker},
		\citenamefont {Zhang}, \citenamefont {Liu}, \citenamefont {Tsirlin},
		\citenamefont {Gegenwart},\ and\ \citenamefont {Zhang}}]{Li2016}%
	\BibitemOpen
	\bibfield  {author} {\bibinfo {author} {\bibfnamefont {Y.}~\bibnamefont
			{Li}}, \bibinfo {author} {\bibfnamefont {D.}~\bibnamefont {Adroja}}, \bibinfo
		{author} {\bibfnamefont {P.~K.}\ \bibnamefont {Biswas}}, \bibinfo {author}
		{\bibfnamefont {P.~J.}\ \bibnamefont {Baker}}, \bibinfo {author}
		{\bibfnamefont {Q.}~\bibnamefont {Zhang}}, \bibinfo {author} {\bibfnamefont
			{J.}~\bibnamefont {Liu}}, \bibinfo {author} {\bibfnamefont {A.~A.}\
			\bibnamefont {Tsirlin}}, \bibinfo {author} {\bibfnamefont {P.}~\bibnamefont
			{Gegenwart}},\ and\ \bibinfo {author} {\bibfnamefont {Q.}~\bibnamefont
			{Zhang}},\ }\bibfield  {title} {\bibinfo {title} {Muon spin relaxation
			evidence for the u(1) quantum spin-liquid ground state in the triangular
			antiferromagnet {YbMgGaO}$_{4}$},\ }\href
	{https://doi.org/10.1103/PhysRevLett.117.097201} {\bibfield  {journal}
		{\bibinfo  {journal} {Phys. Rev. Lett.}\ }\textbf {\bibinfo {volume} {117}},\
		\bibinfo {pages} {097201} (\bibinfo {year} {2016}{\natexlab{a}})}\BibitemShut
	{NoStop}%
	\bibitem [{\citenamefont {Zhu}\ \emph {et~al.}(2017)\citenamefont {Zhu},
		\citenamefont {Maksimov}, \citenamefont {White},\ and\ \citenamefont
		{Chernyshev}}]{Zhu2017b}%
	\BibitemOpen
	\bibfield  {author} {\bibinfo {author} {\bibfnamefont {Z.}~\bibnamefont
			{Zhu}}, \bibinfo {author} {\bibfnamefont {P.~A.}\ \bibnamefont {Maksimov}},
		\bibinfo {author} {\bibfnamefont {S.~R.}\ \bibnamefont {White}},\ and\
		\bibinfo {author} {\bibfnamefont {A.~L.}\ \bibnamefont {Chernyshev}},\
	}\bibfield  {title} {\bibinfo {title} {Disorder-induced mimicry of a spin
			liquid in {YbMgGaO}$_{4}$},\ }\href
	{https://doi.org/10.1103/PhysRevLett.119.157201} {\bibfield  {journal}
		{\bibinfo  {journal} {Phys. Rev. Lett.}\ }\textbf {\bibinfo {volume} {119}},\
		\bibinfo {pages} {157201} (\bibinfo {year} {2017})}\BibitemShut {NoStop}%
	\bibitem [{\citenamefont {Bag}\ \emph {et~al.}(2024)\citenamefont {Bag},
		\citenamefont {Xu}, \citenamefont {Sherman}, \citenamefont {Yadav},
		\citenamefont {Kolesnikov}, \citenamefont {Podlesnyak}, \citenamefont {Choi},
		\citenamefont {da~Silva}, \citenamefont {Moore},\ and\ \citenamefont
		{Haravifard}}]{Bag2024}%
	\BibitemOpen
	\bibfield  {author} {\bibinfo {author} {\bibfnamefont {R.}~\bibnamefont
			{Bag}}, \bibinfo {author} {\bibfnamefont {S.}~\bibnamefont {Xu}}, \bibinfo
		{author} {\bibfnamefont {N.~E.}\ \bibnamefont {Sherman}}, \bibinfo {author}
		{\bibfnamefont {L.}~\bibnamefont {Yadav}}, \bibinfo {author} {\bibfnamefont
			{A.~I.}\ \bibnamefont {Kolesnikov}}, \bibinfo {author} {\bibfnamefont
			{A.~A.}\ \bibnamefont {Podlesnyak}}, \bibinfo {author} {\bibfnamefont
			{E.~S.}\ \bibnamefont {Choi}}, \bibinfo {author} {\bibfnamefont
			{I.}~\bibnamefont {da~Silva}}, \bibinfo {author} {\bibfnamefont {J.~E.}\
			\bibnamefont {Moore}},\ and\ \bibinfo {author} {\bibfnamefont
			{S.}~\bibnamefont {Haravifard}},\ }\bibfield  {title} {\bibinfo {title}
		{Evidence of dirac quantum spin liquid in {YbZn}$_{2}${GaO}$_{5}$},\ }\href
	{https://doi.org/10.1103/PhysRevLett.133.266703} {\bibfield  {journal}
		{\bibinfo  {journal} {Phys. Rev. Lett.}\ }\textbf {\bibinfo {volume} {133}},\
		\bibinfo {pages} {266703} (\bibinfo {year} {2024})}\BibitemShut {NoStop}%
	\bibitem [{\citenamefont {Dai}\ \emph {et~al.}(2021)\citenamefont {Dai},
		\citenamefont {Zhang}, \citenamefont {Xie}, \citenamefont {Duan},
		\citenamefont {Gao}, \citenamefont {Zhu}, \citenamefont {Feng}, \citenamefont
		{Tao}, \citenamefont {Huang}, \citenamefont {Cao}, \citenamefont
		{Podlesnyak}, \citenamefont {Granroth}, \citenamefont {Everett},
		\citenamefont {Neuefeind}, \citenamefont {Voneshen}, \citenamefont {Wang},
		\citenamefont {Tan}, \citenamefont {Morosan}, \citenamefont {Wang},
		\citenamefont {Lin}, \citenamefont {Shu}, \citenamefont {Chen}, \citenamefont
		{Guo}, \citenamefont {Lu},\ and\ \citenamefont {Dai}}]{Dai2021}%
	\BibitemOpen
	\bibfield  {author} {\bibinfo {author} {\bibfnamefont {P.-L.}\ \bibnamefont
			{Dai}}, \bibinfo {author} {\bibfnamefont {G.}~\bibnamefont {Zhang}}, \bibinfo
		{author} {\bibfnamefont {Y.}~\bibnamefont {Xie}}, \bibinfo {author}
		{\bibfnamefont {C.}~\bibnamefont {Duan}}, \bibinfo {author} {\bibfnamefont
			{Y.}~\bibnamefont {Gao}}, \bibinfo {author} {\bibfnamefont {Z.}~\bibnamefont
			{Zhu}}, \bibinfo {author} {\bibfnamefont {E.}~\bibnamefont {Feng}}, \bibinfo
		{author} {\bibfnamefont {Z.}~\bibnamefont {Tao}}, \bibinfo {author}
		{\bibfnamefont {C.-L.}\ \bibnamefont {Huang}}, \bibinfo {author}
		{\bibfnamefont {H.}~\bibnamefont {Cao}}, \bibinfo {author} {\bibfnamefont
			{A.}~\bibnamefont {Podlesnyak}}, \bibinfo {author} {\bibfnamefont {G.~E.}\
			\bibnamefont {Granroth}}, \bibinfo {author} {\bibfnamefont {M.~S.}\
			\bibnamefont {Everett}}, \bibinfo {author} {\bibfnamefont {J.~C.}\
			\bibnamefont {Neuefeind}}, \bibinfo {author} {\bibfnamefont {D.}~\bibnamefont
			{Voneshen}}, \bibinfo {author} {\bibfnamefont {S.}~\bibnamefont {Wang}},
		\bibinfo {author} {\bibfnamefont {G.}~\bibnamefont {Tan}}, \bibinfo {author}
		{\bibfnamefont {E.}~\bibnamefont {Morosan}}, \bibinfo {author} {\bibfnamefont
			{X.}~\bibnamefont {Wang}}, \bibinfo {author} {\bibfnamefont {H.-Q.}\
			\bibnamefont {Lin}}, \bibinfo {author} {\bibfnamefont {L.}~\bibnamefont
			{Shu}}, \bibinfo {author} {\bibfnamefont {G.}~\bibnamefont {Chen}}, \bibinfo
		{author} {\bibfnamefont {Y.}~\bibnamefont {Guo}}, \bibinfo {author}
		{\bibfnamefont {X.}~\bibnamefont {Lu}},\ and\ \bibinfo {author}
		{\bibfnamefont {P.}~\bibnamefont {Dai}},\ }\bibfield  {title} {\bibinfo
		{title} {Spinon fermi surface spin liquid in a triangular lattice
			antiferromagnet {NaYbSe}$_{2}$},\ }\href
	{https://doi.org/10.1103/PhysRevX.11.021044} {\bibfield  {journal} {\bibinfo
			{journal} {Phys. Rev. X}\ }\textbf {\bibinfo {volume} {11}},\ \bibinfo
		{pages} {021044} (\bibinfo {year} {2021})}\BibitemShut {NoStop}%
	\bibitem [{\citenamefont {Wu}\ \emph {et~al.}(2022)\citenamefont {Wu},
		\citenamefont {Li}, \citenamefont {Zhang}, \citenamefont {Liu}, \citenamefont
		{Gao}, \citenamefont {Feng}, \citenamefont {Deng}, \citenamefont {Ren},
		\citenamefont {Wang}, \citenamefont {Chen}, \citenamefont {Embs},
		\citenamefont {Zhu}, \citenamefont {Huang}, \citenamefont {Xiang},
		\citenamefont {Chen}, \citenamefont {Wu}, \citenamefont {Choi}, \citenamefont
		{Qu}, \citenamefont {Li}, \citenamefont {Wang}, \citenamefont {Zhou},
		\citenamefont {Su}, \citenamefont {Wang}, \citenamefont {Chen}, \citenamefont
		{Zhang},\ and\ \citenamefont {Ma}}]{Wu2022}%
	\BibitemOpen
	\bibfield  {author} {\bibinfo {author} {\bibfnamefont {J.}~\bibnamefont
			{Wu}}, \bibinfo {author} {\bibfnamefont {J.}~\bibnamefont {Li}}, \bibinfo
		{author} {\bibfnamefont {Z.}~\bibnamefont {Zhang}}, \bibinfo {author}
		{\bibfnamefont {C.}~\bibnamefont {Liu}}, \bibinfo {author} {\bibfnamefont
			{Y.~H.}\ \bibnamefont {Gao}}, \bibinfo {author} {\bibfnamefont
			{E.}~\bibnamefont {Feng}}, \bibinfo {author} {\bibfnamefont {G.}~\bibnamefont
			{Deng}}, \bibinfo {author} {\bibfnamefont {Q.}~\bibnamefont {Ren}}, \bibinfo
		{author} {\bibfnamefont {Z.}~\bibnamefont {Wang}}, \bibinfo {author}
		{\bibfnamefont {R.}~\bibnamefont {Chen}}, \bibinfo {author} {\bibfnamefont
			{J.}~\bibnamefont {Embs}}, \bibinfo {author} {\bibfnamefont {F.}~\bibnamefont
			{Zhu}}, \bibinfo {author} {\bibfnamefont {Q.}~\bibnamefont {Huang}}, \bibinfo
		{author} {\bibfnamefont {Z.}~\bibnamefont {Xiang}}, \bibinfo {author}
		{\bibfnamefont {L.}~\bibnamefont {Chen}}, \bibinfo {author} {\bibfnamefont
			{Y.}~\bibnamefont {Wu}}, \bibinfo {author} {\bibfnamefont {E.~S.}\
			\bibnamefont {Choi}}, \bibinfo {author} {\bibfnamefont {Z.}~\bibnamefont
			{Qu}}, \bibinfo {author} {\bibfnamefont {L.}~\bibnamefont {Li}}, \bibinfo
		{author} {\bibfnamefont {J.}~\bibnamefont {Wang}}, \bibinfo {author}
		{\bibfnamefont {H.}~\bibnamefont {Zhou}}, \bibinfo {author} {\bibfnamefont
			{Y.}~\bibnamefont {Su}}, \bibinfo {author} {\bibfnamefont {X.}~\bibnamefont
			{Wang}}, \bibinfo {author} {\bibfnamefont {G.}~\bibnamefont {Chen}}, \bibinfo
		{author} {\bibfnamefont {Q.}~\bibnamefont {Zhang}},\ and\ \bibinfo {author}
		{\bibfnamefont {J.}~\bibnamefont {Ma}},\ }\bibfield  {title} {\bibinfo
		{title} {Magnetic field effects on the quantum spin liquid behaviors of
			{NaYbS}$_2$},\ }\href {https://doi.org/10.1007/s44214-022-00011-z} {\bibfield
		{journal} {\bibinfo  {journal} {Quantum Frontiers}\ }\textbf {\bibinfo
			{volume} {1}},\ \bibinfo {pages} {13} (\bibinfo {year} {2022})}\BibitemShut
	{NoStop}%
	\bibitem [{\citenamefont {Xie}\ \emph {et~al.}(2024{\natexlab{b}})\citenamefont
		{Xie}, \citenamefont {Gozel}, \citenamefont {Xing}, \citenamefont {Zhao},
		\citenamefont {Avdoshenko}, \citenamefont {Wu}, \citenamefont {Sefat},
		\citenamefont {Chernyshev}, \citenamefont {L\"auchli}, \citenamefont
		{Podlesnyak},\ and\ \citenamefont {Nikitin}}]{Xie2024prl}%
	\BibitemOpen
	\bibfield  {author} {\bibinfo {author} {\bibfnamefont {T.}~\bibnamefont
			{Xie}}, \bibinfo {author} {\bibfnamefont {S.}~\bibnamefont {Gozel}}, \bibinfo
		{author} {\bibfnamefont {J.}~\bibnamefont {Xing}}, \bibinfo {author}
		{\bibfnamefont {N.}~\bibnamefont {Zhao}}, \bibinfo {author} {\bibfnamefont
			{S.~M.}\ \bibnamefont {Avdoshenko}}, \bibinfo {author} {\bibfnamefont
			{L.}~\bibnamefont {Wu}}, \bibinfo {author} {\bibfnamefont {A.~S.}\
			\bibnamefont {Sefat}}, \bibinfo {author} {\bibfnamefont {A.~L.}\ \bibnamefont
			{Chernyshev}}, \bibinfo {author} {\bibfnamefont {A.~M.}\ \bibnamefont
			{L\"auchli}}, \bibinfo {author} {\bibfnamefont {A.}~\bibnamefont
			{Podlesnyak}},\ and\ \bibinfo {author} {\bibfnamefont {S.~E.}\ \bibnamefont
			{Nikitin}},\ }\bibfield  {title} {\bibinfo {title} {Quantum spin dynamics due
			to strong kitaev interactions in the triangular-lattice antiferromagnet
			${\mathrm{cscese}}_{2}$},\ }\href
	{https://doi.org/10.1103/PhysRevLett.133.096703} {\bibfield  {journal}
		{\bibinfo  {journal} {Phys. Rev. Lett.}\ }\textbf {\bibinfo {volume} {133}},\
		\bibinfo {pages} {096703} (\bibinfo {year} {2024}{\natexlab{b}})}\BibitemShut
	{NoStop}%
	\bibitem [{\citenamefont {Scheie}\ \emph {et~al.}(2024)\citenamefont {Scheie},
		\citenamefont {Kamiya}, \citenamefont {Zhang}, \citenamefont {Lee},
		\citenamefont {Woods}, \citenamefont {Ajeesh}, \citenamefont {Gonzalez},
		\citenamefont {Bernu}, \citenamefont {Villanova}, \citenamefont {Xing},
		\citenamefont {Huang}, \citenamefont {Zhang}, \citenamefont {Ma},
		\citenamefont {Choi}, \citenamefont {Pajerowski}, \citenamefont {Zhou},
		\citenamefont {Sefat}, \citenamefont {Okamoto}, \citenamefont {Berlijn},
		\citenamefont {Messio}, \citenamefont {Movshovich}, \citenamefont {Batista},\
		and\ \citenamefont {Tennant}}]{Scheie2024}%
	\BibitemOpen
	\bibfield  {author} {\bibinfo {author} {\bibfnamefont {A.~O.}\ \bibnamefont
			{Scheie}}, \bibinfo {author} {\bibfnamefont {Y.}~\bibnamefont {Kamiya}},
		\bibinfo {author} {\bibfnamefont {H.}~\bibnamefont {Zhang}}, \bibinfo
		{author} {\bibfnamefont {S.}~\bibnamefont {Lee}}, \bibinfo {author}
		{\bibfnamefont {A.~J.}\ \bibnamefont {Woods}}, \bibinfo {author}
		{\bibfnamefont {M.~O.}\ \bibnamefont {Ajeesh}}, \bibinfo {author}
		{\bibfnamefont {M.~G.}\ \bibnamefont {Gonzalez}}, \bibinfo {author}
		{\bibfnamefont {B.}~\bibnamefont {Bernu}}, \bibinfo {author} {\bibfnamefont
			{J.~W.}\ \bibnamefont {Villanova}}, \bibinfo {author} {\bibfnamefont
			{J.}~\bibnamefont {Xing}}, \bibinfo {author} {\bibfnamefont {Q.}~\bibnamefont
			{Huang}}, \bibinfo {author} {\bibfnamefont {Q.}~\bibnamefont {Zhang}},
		\bibinfo {author} {\bibfnamefont {J.}~\bibnamefont {Ma}}, \bibinfo {author}
		{\bibfnamefont {E.~S.}\ \bibnamefont {Choi}}, \bibinfo {author}
		{\bibfnamefont {D.~M.}\ \bibnamefont {Pajerowski}}, \bibinfo {author}
		{\bibfnamefont {H.}~\bibnamefont {Zhou}}, \bibinfo {author} {\bibfnamefont
			{A.~S.}\ \bibnamefont {Sefat}}, \bibinfo {author} {\bibfnamefont
			{S.}~\bibnamefont {Okamoto}}, \bibinfo {author} {\bibfnamefont
			{T.}~\bibnamefont {Berlijn}}, \bibinfo {author} {\bibfnamefont
			{L.}~\bibnamefont {Messio}}, \bibinfo {author} {\bibfnamefont
			{R.}~\bibnamefont {Movshovich}}, \bibinfo {author} {\bibfnamefont {C.~D.}\
			\bibnamefont {Batista}},\ and\ \bibinfo {author} {\bibfnamefont {D.~A.}\
			\bibnamefont {Tennant}},\ }\bibfield  {title} {\bibinfo {title} {Nonlinear
			magnons and exchange hamiltonians of the delafossite proximate quantum spin
			liquid candidates {KYbSe}$_{2}$ and {NaYbSe}$_{2}$},\ }\href
	{https://doi.org/10.1103/PhysRevB.109.014425} {\bibfield  {journal} {\bibinfo
			{journal} {Phys. Rev. B}\ }\textbf {\bibinfo {volume} {109}},\ \bibinfo
		{pages} {014425} (\bibinfo {year} {2024})}\BibitemShut {NoStop}%
	\bibitem [{\citenamefont {Villanova}\ \emph {et~al.}(2023)\citenamefont
		{Villanova}, \citenamefont {Scheie}, \citenamefont {Tennant}, \citenamefont
		{Okamoto},\ and\ \citenamefont {Berlijn}}]{Villanova2023}%
	\BibitemOpen
	\bibfield  {author} {\bibinfo {author} {\bibfnamefont {J.~W.}\ \bibnamefont
			{Villanova}}, \bibinfo {author} {\bibfnamefont {A.~O.}\ \bibnamefont
			{Scheie}}, \bibinfo {author} {\bibfnamefont {D.~A.}\ \bibnamefont {Tennant}},
		\bibinfo {author} {\bibfnamefont {S.}~\bibnamefont {Okamoto}},\ and\ \bibinfo
		{author} {\bibfnamefont {T.}~\bibnamefont {Berlijn}},\ }\bibfield  {title}
	{\bibinfo {title} {First-principles derivation of magnetic interactions in
			the triangular quantum spin liquid candidates {KYbCh}$_{2}$ ({$Ch$}={S},
			{Se}, {Te}) and {$A$}{YbSe}$_{2}$ ({$A$}={Na}, {Rb})},\ }\href
	{https://doi.org/10.1103/PhysRevResearch.5.033050} {\bibfield  {journal}
		{\bibinfo  {journal} {Phys. Rev. Res.}\ }\textbf {\bibinfo {volume} {5}},\
		\bibinfo {pages} {033050} (\bibinfo {year} {2023})}\BibitemShut {NoStop}%
	\bibitem [{\citenamefont {Georges}\ \emph {et~al.}(1996)\citenamefont
		{Georges}, \citenamefont {Kotliar}, \citenamefont {Krauth},\ and\
		\citenamefont {Rozenberg}}]{Georges1996}%
	\BibitemOpen
	\bibfield  {author} {\bibinfo {author} {\bibfnamefont {A.}~\bibnamefont
			{Georges}}, \bibinfo {author} {\bibfnamefont {G.}~\bibnamefont {Kotliar}},
		\bibinfo {author} {\bibfnamefont {W.}~\bibnamefont {Krauth}},\ and\ \bibinfo
		{author} {\bibfnamefont {M.~J.}\ \bibnamefont {Rozenberg}},\ }\bibfield
	{title} {\bibinfo {title} {Dynamical mean-field theory of strongly correlated
			fermion systems and the limit of infinite dimensions},\ }\href@noop {}
	{\bibfield  {journal} {\bibinfo  {journal} {Rev. Mod. Phys.}\ }\textbf
		{\bibinfo {volume} {68}},\ \bibinfo {pages} {13} (\bibinfo {year}
		{1996})}\BibitemShut {NoStop}%
	\bibitem [{\citenamefont {Anisimov}\ \emph {et~al.}(1997)\citenamefont
		{Anisimov}, \citenamefont {Poteryaev}, \citenamefont {Korotin}, \citenamefont
		{Anokhin},\ and\ \citenamefont {Kotliar}}]{Anisimov1997_1}%
	\BibitemOpen
	\bibfield  {author} {\bibinfo {author} {\bibfnamefont {V.~I.}\ \bibnamefont
			{Anisimov}}, \bibinfo {author} {\bibfnamefont {A.~I.}\ \bibnamefont
			{Poteryaev}}, \bibinfo {author} {\bibfnamefont {M.~A.}\ \bibnamefont
			{Korotin}}, \bibinfo {author} {\bibfnamefont {A.~O.}\ \bibnamefont
			{Anokhin}},\ and\ \bibinfo {author} {\bibfnamefont {G.}~\bibnamefont
			{Kotliar}},\ }\bibfield  {title} {\bibinfo {title} {First-principles
			calculations of the electronic structure and spectra of strongly correlated
			systems: dynamical mean-field theory},\ }\href@noop {} {\bibfield  {journal}
		{\bibinfo  {journal} {Journal of Physics: Condensed Matter}\ }\textbf
		{\bibinfo {volume} {9}},\ \bibinfo {pages} {7359} (\bibinfo {year}
		{1997})}\BibitemShut {NoStop}%
	\bibitem [{\citenamefont {Lichtenstein}\ and\ \citenamefont
		{Katsnelson}(1998)}]{Lichtenstein_LDApp}%
	\BibitemOpen
	\bibfield  {author} {\bibinfo {author} {\bibfnamefont {A.~I.}\ \bibnamefont
			{Lichtenstein}}\ and\ \bibinfo {author} {\bibfnamefont {M.~I.}\ \bibnamefont
			{Katsnelson}},\ }\bibfield  {title} {\bibinfo {title} {Ab initio calculations
			of quasiparticle band structure in correlated systems: {LDA++} approach},\
	}\href@noop {} {\bibfield  {journal} {\bibinfo  {journal} {Phys. Rev. B}\
		}\textbf {\bibinfo {volume} {57}},\ \bibinfo {pages} {6884} (\bibinfo {year}
		{1998})}\BibitemShut {NoStop}%
	\bibitem [{\citenamefont {Hubbard}(1963)}]{hubbard_1}%
	\BibitemOpen
	\bibfield  {author} {\bibinfo {author} {\bibfnamefont {J.}~\bibnamefont
			{Hubbard}},\ }\bibfield  {title} {\bibinfo {title} {Electron correlations in
			narrow energy bands},\ }\href@noop {} {\bibfield  {journal} {\bibinfo
			{journal} {Proc. Roy. Soc. (London)}\ }\textbf {\bibinfo {volume} {A 276}},\
		\bibinfo {pages} {238} (\bibinfo {year} {1963})}\BibitemShut {NoStop}%
	\bibitem [{\citenamefont {Pourovskii}(2016)}]{Pourovskii2016}%
	\BibitemOpen
	\bibfield  {author} {\bibinfo {author} {\bibfnamefont {L.~V.}\ \bibnamefont
			{Pourovskii}},\ }\bibfield  {title} {\bibinfo {title} {Two-site fluctuations
			and multipolar intersite exchange interactions in strongly correlated
			systems},\ }\href {https://doi.org/10.1103/PhysRevB.94.115117} {\bibfield
		{journal} {\bibinfo  {journal} {Phys. Rev. B}\ }\textbf {\bibinfo {volume}
			{94}},\ \bibinfo {pages} {115117} (\bibinfo {year} {2016})}\BibitemShut
	{NoStop}%
	\bibitem [{\citenamefont {{Pourovskii}}\ \emph {et~al.}(2007)\citenamefont
		{{Pourovskii}}, \citenamefont {{Amadon}}, \citenamefont {{Biermann}},\ and\
		\citenamefont {{Georges}}}]{Pourovskii2007}%
	\BibitemOpen
	\bibfield  {author} {\bibinfo {author} {\bibfnamefont {L.~V.}\ \bibnamefont
			{{Pourovskii}}}, \bibinfo {author} {\bibfnamefont {B.}~\bibnamefont
			{{Amadon}}}, \bibinfo {author} {\bibfnamefont {S.}~\bibnamefont
			{{Biermann}}},\ and\ \bibinfo {author} {\bibfnamefont {A.}~\bibnamefont
			{{Georges}}},\ }\bibfield  {title} {\bibinfo {title} {{Self-consistency over
				the charge density in dynamical mean-field theory: A linear muffin-tin
				implementation and some physical implications}},\ }\href
	{https://doi.org/10.1103/PhysRevB.76.235101} {\bibfield  {journal} {\bibinfo
			{journal} {Phys. Rev. B}\ }\textbf {\bibinfo {volume} {76}},\ \bibinfo
		{pages} {235101} (\bibinfo {year} {2007})}\BibitemShut {NoStop}%
	\bibitem [{\citenamefont {Bhandary}\ \emph {et~al.}(2016)\citenamefont
		{Bhandary}, \citenamefont {Assmann}, \citenamefont {Aichhorn},\ and\
		\citenamefont {Held}}]{Bhandary2016}%
	\BibitemOpen
	\bibfield  {author} {\bibinfo {author} {\bibfnamefont {S.}~\bibnamefont
			{Bhandary}}, \bibinfo {author} {\bibfnamefont {E.}~\bibnamefont {Assmann}},
		\bibinfo {author} {\bibfnamefont {M.}~\bibnamefont {Aichhorn}},\ and\
		\bibinfo {author} {\bibfnamefont {K.}~\bibnamefont {Held}},\ }\bibfield
	{title} {\bibinfo {title} {Charge self-consistency in density functional
			theory combined with dynamical mean field theory: $k$-space reoccupation and
			orbital order},\ }\href {https://doi.org/10.1103/PhysRevB.94.155131}
	{\bibfield  {journal} {\bibinfo  {journal} {Phys. Rev. B}\ }\textbf {\bibinfo
			{volume} {94}},\ \bibinfo {pages} {155131} (\bibinfo {year}
		{2016})}\BibitemShut {NoStop}%
	\bibitem [{\citenamefont {Delange}\ \emph {et~al.}(2017)\citenamefont
		{Delange}, \citenamefont {Biermann}, \citenamefont {Miyake},\ and\
		\citenamefont {Pourovskii}}]{Delange2017}%
	\BibitemOpen
	\bibfield  {author} {\bibinfo {author} {\bibfnamefont {P.}~\bibnamefont
			{Delange}}, \bibinfo {author} {\bibfnamefont {S.}~\bibnamefont {Biermann}},
		\bibinfo {author} {\bibfnamefont {T.}~\bibnamefont {Miyake}},\ and\ \bibinfo
		{author} {\bibfnamefont {L.}~\bibnamefont {Pourovskii}},\ }\bibfield  {title}
	{\bibinfo {title} {Crystal-field splittings in rare-earth-based hard magnets:
			An ab initio approach},\ }\href {https://doi.org/10.1103/PhysRevB.96.155132}
	{\bibfield  {journal} {\bibinfo  {journal} {Phys. Rev. B}\ }\textbf {\bibinfo
			{volume} {96}},\ \bibinfo {pages} {155132} (\bibinfo {year}
		{2017})}\BibitemShut {NoStop}%
	\bibitem [{\citenamefont {Wu}\ \emph {et~al.}(2025)\citenamefont {Wu},
		\citenamefont {Pratt}, \citenamefont {Huddart}, \citenamefont {Chatterjee},
		\citenamefont {Goddard}, \citenamefont {Singleton}, \citenamefont
		{Prabhakaran},\ and\ \citenamefont {Blundell}}]{wu2025}%
	\BibitemOpen
	\bibfield  {author} {\bibinfo {author} {\bibfnamefont {H.~C.~H.}\
			\bibnamefont {Wu}}, \bibinfo {author} {\bibfnamefont {F.~L.}\ \bibnamefont
			{Pratt}}, \bibinfo {author} {\bibfnamefont {B.~M.}\ \bibnamefont {Huddart}},
		\bibinfo {author} {\bibfnamefont {D.}~\bibnamefont {Chatterjee}}, \bibinfo
		{author} {\bibfnamefont {P.~A.}\ \bibnamefont {Goddard}}, \bibinfo {author}
		{\bibfnamefont {J.}~\bibnamefont {Singleton}}, \bibinfo {author}
		{\bibfnamefont {D.}~\bibnamefont {Prabhakaran}},\ and\ \bibinfo {author}
		{\bibfnamefont {S.~J.}\ \bibnamefont {Blundell}},\ }\bibfield  {title}
	{\bibinfo {title} {Spin dynamics in the dirac {$U$}(1) spin liquid
			{YbZn}$_{2}${GaO}$_{5}$},\ }\href {https://doi.org/10.1103/l93v-f576}
	{\bibfield  {journal} {\bibinfo  {journal} {Phys. Rev. Lett.}\ }\textbf
		{\bibinfo {volume} {135}},\ \bibinfo {pages} {046704} (\bibinfo {year}
		{2025})}\BibitemShut {NoStop}%
	\bibitem [{\citenamefont {Wei{\ss}e}\ and\ \citenamefont
		{Fehske}(2008)}]{Weisse2008}%
	\BibitemOpen
	\bibfield  {author} {\bibinfo {author} {\bibfnamefont {A.}~\bibnamefont
			{Wei{\ss}e}}\ and\ \bibinfo {author} {\bibfnamefont {H.}~\bibnamefont
			{Fehske}},\ }\bibinfo {title} {Exact diagonalization techniques},\ in\ \href
	{https://doi.org/10.1007/978-3-540-74686-7_18} {\emph {\bibinfo {booktitle}
			{Computational Many-Particle Physics}}},\ \bibinfo {editor} {edited by\
		\bibinfo {editor} {\bibfnamefont {H.}~\bibnamefont {Fehske}}, \bibinfo
		{editor} {\bibfnamefont {R.}~\bibnamefont {Schneider}},\ and\ \bibinfo
		{editor} {\bibfnamefont {A.}~\bibnamefont {Wei{\ss}e}}}\ (\bibinfo
	{publisher} {Springer Berlin Heidelberg},\ \bibinfo {address} {Berlin,
		Heidelberg},\ \bibinfo {year} {2008})\ pp.\ \bibinfo {pages}
	{529--544}\BibitemShut {NoStop}%
	\bibitem [{\citenamefont {Wietek}\ \emph {et~al.}(2025)\citenamefont {Wietek},
		\citenamefont {Staszewski}, \citenamefont {Ulaga}, \citenamefont {Ebert},
		\citenamefont {Karlsson}, \citenamefont {Sarkar}, \citenamefont {Shackleton},
		\citenamefont {Sinha},\ and\ \citenamefont
		{Soares}}]{wietek2025xdiagexactdiagonalizationquantum}%
	\BibitemOpen
	\bibfield  {author} {\bibinfo {author} {\bibfnamefont {A.}~\bibnamefont
			{Wietek}}, \bibinfo {author} {\bibfnamefont {L.}~\bibnamefont {Staszewski}},
		\bibinfo {author} {\bibfnamefont {M.}~\bibnamefont {Ulaga}}, \bibinfo
		{author} {\bibfnamefont {P.~L.}\ \bibnamefont {Ebert}}, \bibinfo {author}
		{\bibfnamefont {H.}~\bibnamefont {Karlsson}}, \bibinfo {author}
		{\bibfnamefont {S.}~\bibnamefont {Sarkar}}, \bibinfo {author} {\bibfnamefont
			{H.}~\bibnamefont {Shackleton}}, \bibinfo {author} {\bibfnamefont
			{A.}~\bibnamefont {Sinha}},\ and\ \bibinfo {author} {\bibfnamefont {R.~D.}\
			\bibnamefont {Soares}},\ }\href {https://arxiv.org/abs/2505.02901} {\bibinfo
		{title} {Xdiag: Exact diagonalization for quantum many-body systems}}
	(\bibinfo {year} {2025}),\ \Eprint {https://arxiv.org/abs/2505.02901}
	{arXiv:2505.02901 [cond-mat.str-el]} \BibitemShut {NoStop}%
	\bibitem [{\citenamefont {Li}\ \emph {et~al.}(2016{\natexlab{b}})\citenamefont
		{Li}, \citenamefont {Wang},\ and\ \citenamefont {Chen}}]{PhysRevB.94.035107}%
	\BibitemOpen
	\bibfield  {author} {\bibinfo {author} {\bibfnamefont {Y.-D.}\ \bibnamefont
			{Li}}, \bibinfo {author} {\bibfnamefont {X.}~\bibnamefont {Wang}},\ and\
		\bibinfo {author} {\bibfnamefont {G.}~\bibnamefont {Chen}},\ }\bibfield
	{title} {\bibinfo {title} {Anisotropic spin model of strong
			spin-orbit-coupled triangular antiferromagnets},\ }\href
	{https://doi.org/10.1103/PhysRevB.94.035107} {\bibfield  {journal} {\bibinfo
			{journal} {Phys. Rev. B}\ }\textbf {\bibinfo {volume} {94}},\ \bibinfo
		{pages} {035107} (\bibinfo {year} {2016}{\natexlab{b}})}\BibitemShut
	{NoStop}%
	\bibitem [{\citenamefont {Parker}\ and\ \citenamefont
		{Balents}(2018)}]{PhysRevB.97.184413}%
	\BibitemOpen
	\bibfield  {author} {\bibinfo {author} {\bibfnamefont {E.}~\bibnamefont
			{Parker}}\ and\ \bibinfo {author} {\bibfnamefont {L.}~\bibnamefont
			{Balents}},\ }\bibfield  {title} {\bibinfo {title} {Finite-temperature
			behavior of a classical spin-orbit-coupled model for {YbMgGaO}$_{4}$ with and
			without bond disorder},\ }\href {https://doi.org/10.1103/PhysRevB.97.184413}
	{\bibfield  {journal} {\bibinfo  {journal} {Phys. Rev. B}\ }\textbf {\bibinfo
			{volume} {97}},\ \bibinfo {pages} {184413} (\bibinfo {year}
		{2018})}\BibitemShut {NoStop}%
	\bibitem [{\citenamefont {Wu}\ \emph {et~al.}(2021)\citenamefont {Wu},
		\citenamefont {Yao},\ and\ \citenamefont {Wu}}]{Wu2021}%
	\BibitemOpen
	\bibfield  {author} {\bibinfo {author} {\bibfnamefont {M.}~\bibnamefont
			{Wu}}, \bibinfo {author} {\bibfnamefont {D.-X.}\ \bibnamefont {Yao}},\ and\
		\bibinfo {author} {\bibfnamefont {H.-Q.}\ \bibnamefont {Wu}},\ }\bibfield
	{title} {\bibinfo {title} {Exact diagonalization study of the anisotropic
			heisenberg model related to {YbMgGaO}$_{4}$},\ }\href
	{https://doi.org/10.1103/PhysRevB.103.205122} {\bibfield  {journal} {\bibinfo
			{journal} {Phys. Rev. B}\ }\textbf {\bibinfo {volume} {103}},\ \bibinfo
		{pages} {205122} (\bibinfo {year} {2021})}\BibitemShut {NoStop}%
	\bibitem [{\citenamefont {Luo}\ \emph {et~al.}(2017)\citenamefont {Luo},
		\citenamefont {Hu}, \citenamefont {Xi}, \citenamefont {Zhao},\ and\
		\citenamefont {Wang}}]{PhysRevB.95.165110}%
	\BibitemOpen
	\bibfield  {author} {\bibinfo {author} {\bibfnamefont {Q.}~\bibnamefont
			{Luo}}, \bibinfo {author} {\bibfnamefont {S.}~\bibnamefont {Hu}}, \bibinfo
		{author} {\bibfnamefont {B.}~\bibnamefont {Xi}}, \bibinfo {author}
		{\bibfnamefont {J.}~\bibnamefont {Zhao}},\ and\ \bibinfo {author}
		{\bibfnamefont {X.}~\bibnamefont {Wang}},\ }\bibfield  {title} {\bibinfo
		{title} {Ground-state phase diagram of an anisotropic spin-$\frac{1}{2}$
			model on the triangular lattice},\ }\href
	{https://doi.org/10.1103/PhysRevB.95.165110} {\bibfield  {journal} {\bibinfo
			{journal} {Phys. Rev. B}\ }\textbf {\bibinfo {volume} {95}},\ \bibinfo
		{pages} {165110} (\bibinfo {year} {2017})}\BibitemShut {NoStop}%
	\bibitem [{\citenamefont {Zhu}\ \emph {et~al.}(2018)\citenamefont {Zhu},
		\citenamefont {Maksimov}, \citenamefont {White},\ and\ \citenamefont
		{Chernyshev}}]{Zhu2018}%
	\BibitemOpen
	\bibfield  {author} {\bibinfo {author} {\bibfnamefont {Z.}~\bibnamefont
			{Zhu}}, \bibinfo {author} {\bibfnamefont {P.~A.}\ \bibnamefont {Maksimov}},
		\bibinfo {author} {\bibfnamefont {S.~R.}\ \bibnamefont {White}},\ and\
		\bibinfo {author} {\bibfnamefont {A.~L.}\ \bibnamefont {Chernyshev}},\
	}\bibfield  {title} {\bibinfo {title} {Topography of spin liquids on a
			triangular lattice},\ }\href {https://doi.org/10.1103/PhysRevLett.120.207203}
	{\bibfield  {journal} {\bibinfo  {journal} {Phys. Rev. Lett.}\ }\textbf
		{\bibinfo {volume} {120}},\ \bibinfo {pages} {207203} (\bibinfo {year}
		{2018})}\BibitemShut {NoStop}%
	\bibitem [{\citenamefont {Gallegos}\ \emph {et~al.}(2025)\citenamefont
		{Gallegos}, \citenamefont {Jiang}, \citenamefont {White},\ and\ \citenamefont
		{Chernyshev}}]{PhysRevLett.134.196702}%
	\BibitemOpen
	\bibfield  {author} {\bibinfo {author} {\bibfnamefont {C.~A.}\ \bibnamefont
			{Gallegos}}, \bibinfo {author} {\bibfnamefont {S.}~\bibnamefont {Jiang}},
		\bibinfo {author} {\bibfnamefont {S.~R.}\ \bibnamefont {White}},\ and\
		\bibinfo {author} {\bibfnamefont {A.~L.}\ \bibnamefont {Chernyshev}},\
	}\bibfield  {title} {\bibinfo {title} {Phase diagram of the easy-axis
			triangular-lattice ${J}_{1}\text{\ensuremath{-}}{J}_{2}$ model},\ }\href
	{https://doi.org/10.1103/PhysRevLett.134.196702} {\bibfield  {journal}
		{\bibinfo  {journal} {Phys. Rev. Lett.}\ }\textbf {\bibinfo {volume} {134}},\
		\bibinfo {pages} {196702} (\bibinfo {year} {2025})}\BibitemShut {NoStop}%
	\bibitem [{\citenamefont {Aichhorn}\ \emph {et~al.}(2009)\citenamefont
		{Aichhorn}, \citenamefont {Pourovskii}, \citenamefont {Vildosola},
		\citenamefont {Ferrero}, \citenamefont {Parcollet}, \citenamefont {Miyake},
		\citenamefont {Georges},\ and\ \citenamefont {Biermann}}]{Aichhorn2009}%
	\BibitemOpen
	\bibfield  {author} {\bibinfo {author} {\bibfnamefont {M.}~\bibnamefont
			{Aichhorn}}, \bibinfo {author} {\bibfnamefont {L.~V.}\ \bibnamefont
			{Pourovskii}}, \bibinfo {author} {\bibfnamefont {V.}~\bibnamefont
			{Vildosola}}, \bibinfo {author} {\bibfnamefont {M.}~\bibnamefont {Ferrero}},
		\bibinfo {author} {\bibfnamefont {O.}~\bibnamefont {Parcollet}}, \bibinfo
		{author} {\bibfnamefont {T.}~\bibnamefont {Miyake}}, \bibinfo {author}
		{\bibfnamefont {A.}~\bibnamefont {Georges}},\ and\ \bibinfo {author}
		{\bibfnamefont {S.}~\bibnamefont {Biermann}},\ }\bibfield  {title} {\bibinfo
		{title} {Dynamical mean-field theory within an augmented plane-wave
			framework: Assessing electronic correlations in the iron pnictide
			{LaFeAsO}},\ }\href@noop {} {\bibfield  {journal} {\bibinfo  {journal} {Phys.
				Rev. B}\ }\textbf {\bibinfo {volume} {80}},\ \bibinfo {pages} {085101}
		(\bibinfo {year} {2009})}\BibitemShut {NoStop}%
	\bibitem [{\citenamefont {Aichhorn}\ \emph {et~al.}(2011)\citenamefont
		{Aichhorn}, \citenamefont {Pourovskii},\ and\ \citenamefont
		{Georges}}]{Aichhorn2011}%
	\BibitemOpen
	\bibfield  {author} {\bibinfo {author} {\bibfnamefont {M.}~\bibnamefont
			{Aichhorn}}, \bibinfo {author} {\bibfnamefont {L.~V.}\ \bibnamefont
			{Pourovskii}},\ and\ \bibinfo {author} {\bibfnamefont {A.}~\bibnamefont
			{Georges}},\ }\bibfield  {title} {\bibinfo {title} {Importance of electronic
			correlations for structural and magnetic properties of the iron pnictide
			superconductor {LaFeAsO}},\ }\href@noop {} {\bibfield  {journal} {\bibinfo
			{journal} {Phys. Rev. B}\ }\textbf {\bibinfo {volume} {84}},\ \bibinfo
		{pages} {054529} (\bibinfo {year} {2011})}\BibitemShut {NoStop}%
	\bibitem [{\citenamefont {Aichhorn}\ \emph {et~al.}(2016)\citenamefont
		{Aichhorn}, \citenamefont {Pourovskii}, \citenamefont {Seth}, \citenamefont
		{Vildosola}, \citenamefont {Zingl}, \citenamefont {Peil}, \citenamefont
		{Deng}, \citenamefont {Mravlje}, \citenamefont {Kraberger}, \citenamefont
		{Martins} \emph {et~al.}}]{Aichhorn2016}%
	\BibitemOpen
	\bibfield  {author} {\bibinfo {author} {\bibfnamefont {M.}~\bibnamefont
			{Aichhorn}}, \bibinfo {author} {\bibfnamefont {L.~V.}\ \bibnamefont
			{Pourovskii}}, \bibinfo {author} {\bibfnamefont {P.}~\bibnamefont {Seth}},
		\bibinfo {author} {\bibfnamefont {V.}~\bibnamefont {Vildosola}}, \bibinfo
		{author} {\bibfnamefont {M.}~\bibnamefont {Zingl}}, \bibinfo {author}
		{\bibfnamefont {O.~E.}\ \bibnamefont {Peil}}, \bibinfo {author}
		{\bibfnamefont {X.}~\bibnamefont {Deng}}, \bibinfo {author} {\bibfnamefont
			{J.}~\bibnamefont {Mravlje}}, \bibinfo {author} {\bibfnamefont {G.~J.}\
			\bibnamefont {Kraberger}}, \bibinfo {author} {\bibfnamefont {C.}~\bibnamefont
			{Martins}}, \emph {et~al.},\ }\bibfield  {title} {\bibinfo {title}
		{{TRIQS/DFTTools: A TRIQS application for ab initio calculations of
				correlated materials}},\ }\href@noop {} {\bibfield  {journal} {\bibinfo
			{journal} {Computer Physics Communications}\ }\textbf {\bibinfo {volume}
			{204}},\ \bibinfo {pages} {200} (\bibinfo {year} {2016})}\BibitemShut
	{NoStop}%
	\bibitem [{\citenamefont {Blaha}\ \emph {et~al.}(2018)\citenamefont {Blaha},
		\citenamefont {Schwarz}, \citenamefont {Madsen}, \citenamefont {Kvasnicka},
		\citenamefont {Luitz}, \citenamefont {Laskowski}, \citenamefont {Tran},\ and\
		\citenamefont {Marks}}]{Wien2k}%
	\BibitemOpen
	\bibfield  {author} {\bibinfo {author} {\bibfnamefont {P.}~\bibnamefont
			{Blaha}}, \bibinfo {author} {\bibfnamefont {K.}~\bibnamefont {Schwarz}},
		\bibinfo {author} {\bibfnamefont {G.}~\bibnamefont {Madsen}}, \bibinfo
		{author} {\bibfnamefont {D.}~\bibnamefont {Kvasnicka}}, \bibinfo {author}
		{\bibfnamefont {J.}~\bibnamefont {Luitz}}, \bibinfo {author} {\bibfnamefont
			{R.}~\bibnamefont {Laskowski}}, \bibinfo {author} {\bibfnamefont
			{F.}~\bibnamefont {Tran}},\ and\ \bibinfo {author} {\bibfnamefont {L.~D.}\
			\bibnamefont {Marks}},\ }\href@noop {} {\emph {\bibinfo {title} {WIEN2k, An
				augmented Plane Wave + Local Orbitals Program for Calculating Crystal
				Properties}}}\ (\bibinfo  {publisher} {Karlheinz Schwarz, Techn. Universität
		Wien, Austria,ISBN 3-9501031-1-2},\ \bibinfo {year} {2018})\BibitemShut
	{NoStop}%
	\bibitem [{\citenamefont {Parcollet}\ \emph {et~al.}(2015)\citenamefont
		{Parcollet}, \citenamefont {Ferrero}, \citenamefont {Ayral}, \citenamefont
		{Hafermann}, \citenamefont {Krivenko}, \citenamefont {Messio},\ and\
		\citenamefont {Seth}}]{Parcollet2015}%
	\BibitemOpen
	\bibfield  {author} {\bibinfo {author} {\bibfnamefont {O.}~\bibnamefont
			{Parcollet}}, \bibinfo {author} {\bibfnamefont {M.}~\bibnamefont {Ferrero}},
		\bibinfo {author} {\bibfnamefont {T.}~\bibnamefont {Ayral}}, \bibinfo
		{author} {\bibfnamefont {H.}~\bibnamefont {Hafermann}}, \bibinfo {author}
		{\bibfnamefont {I.}~\bibnamefont {Krivenko}}, \bibinfo {author}
		{\bibfnamefont {L.}~\bibnamefont {Messio}},\ and\ \bibinfo {author}
		{\bibfnamefont {P.}~\bibnamefont {Seth}},\ }\bibfield  {title} {\bibinfo
		{title} {Triqs: A toolbox for research on interacting quantum systems},\
	}\href {https://doi.org/https://doi.org/10.1016/j.cpc.2015.04.023} {\bibfield
		{journal} {\bibinfo  {journal} {Computer Physics Communications}\ }\textbf
		{\bibinfo {volume} {196}},\ \bibinfo {pages} {398} (\bibinfo {year}
		{2015})}\BibitemShut {NoStop}%
	\bibitem [{\citenamefont {Pourovskii}\ and\ \citenamefont
		{Fiore~Mosca}()}]{magint}%
	\BibitemOpen
	\bibfield  {author} {\bibinfo {author} {\bibfnamefont {L.~V.}\ \bibnamefont
			{Pourovskii}}\ and\ \bibinfo {author} {\bibfnamefont {D.}~\bibnamefont
			{Fiore~Mosca}},\ }\href@noop {} {\bibinfo {title} {{MagInt}}},\ \bibinfo
	{howpublished} {\url{https://github.com/MagInteract/MagInt}}\BibitemShut
	{NoStop}%
	\bibitem [{\citenamefont {Pourovskii}\ \emph {et~al.}(2025)\citenamefont
		{Pourovskii}, \citenamefont {Fiore~Mosca}, \citenamefont {Celiberti},
		\citenamefont {Khmelevskyi}, \citenamefont {Paramekanti},\ and\ \citenamefont
		{Franchini}}]{hidden_order_review}%
	\BibitemOpen
	\bibfield  {author} {\bibinfo {author} {\bibfnamefont {L.~V.}\ \bibnamefont
			{Pourovskii}}, \bibinfo {author} {\bibfnamefont {D.}~\bibnamefont
			{Fiore~Mosca}}, \bibinfo {author} {\bibfnamefont {L.}~\bibnamefont
			{Celiberti}}, \bibinfo {author} {\bibfnamefont {S.}~\bibnamefont
			{Khmelevskyi}}, \bibinfo {author} {\bibfnamefont {A.}~\bibnamefont
			{Paramekanti}},\ and\ \bibinfo {author} {\bibfnamefont {C.}~\bibnamefont
			{Franchini}},\ }\bibfield  {title} {\bibinfo {title} {Hidden orders in
			spin--orbit-entangled correlated insulators},\ }\href
	{https://doi.org/10.1038/s41578-025-00824-z} {\bibfield  {journal} {\bibinfo
			{journal} {Nature Reviews Materials}\ }\textbf {\bibinfo {volume} {10}},\
		\bibinfo {pages} {674} (\bibinfo {year} {2025})}\BibitemShut {NoStop}%
	\bibitem [{sup()}]{supplmat}%
	\BibitemOpen
	\href@noop {} {\bibinfo {title} {See {Supplemental Material}, where we detail
			the numerical methods and provide additional results.}}\BibitemShut {Stop}%
	\bibitem [{\citenamefont {Moulding}\ \emph {et~al.}(2025)\citenamefont
		{Moulding}, \citenamefont {Shimizu}, \citenamefont {Pawbake}, \citenamefont
		{Gao}, \citenamefont {Ramakrishnan}, \citenamefont {Garbarino}, \citenamefont
		{Caroca-Canales}, \citenamefont {Debray}, \citenamefont {Faugeras},
		\citenamefont {Geibel}, \citenamefont {Yanase},\ and\ \citenamefont
		{Méasson}}]{Mouding2025}%
	\BibitemOpen
	\bibfield  {author} {\bibinfo {author} {\bibfnamefont {O.}~\bibnamefont
			{Moulding}}, \bibinfo {author} {\bibfnamefont {M.}~\bibnamefont {Shimizu}},
		\bibinfo {author} {\bibfnamefont {A.}~\bibnamefont {Pawbake}}, \bibinfo
		{author} {\bibfnamefont {Y.}~\bibnamefont {Gao}}, \bibinfo {author}
		{\bibfnamefont {S.}~\bibnamefont {Ramakrishnan}}, \bibinfo {author}
		{\bibfnamefont {G.}~\bibnamefont {Garbarino}}, \bibinfo {author}
		{\bibfnamefont {N.}~\bibnamefont {Caroca-Canales}}, \bibinfo {author}
		{\bibfnamefont {J.}~\bibnamefont {Debray}}, \bibinfo {author} {\bibfnamefont
			{C.}~\bibnamefont {Faugeras}}, \bibinfo {author} {\bibfnamefont
			{C.}~\bibnamefont {Geibel}}, \bibinfo {author} {\bibfnamefont
			{Y.}~\bibnamefont {Yanase}},\ and\ \bibinfo {author} {\bibfnamefont {M.-A.}\
			\bibnamefont {Méasson}},\ }\href {https://arxiv.org/abs/2505.03249}
	{\bibinfo {title} {Multilayer crystal field states from locally broken
			centrosymmetry}} (\bibinfo {year} {2025}),\ \Eprint
	{https://arxiv.org/abs/2505.03249} {arXiv:2505.03249 [cond-mat.str-el]}
	\BibitemShut {NoStop}%
	\bibitem [{\citenamefont {Smith}\ \emph {et~al.}(2024)\citenamefont {Smith},
		\citenamefont {Lhotel}, \citenamefont {Petit},\ and\ \citenamefont
		{Gaulin}}]{Smith2024}%
	\BibitemOpen
	\bibfield  {author} {\bibinfo {author} {\bibfnamefont {E.~M.}\ \bibnamefont
			{Smith}}, \bibinfo {author} {\bibfnamefont {E.}~\bibnamefont {Lhotel}},
		\bibinfo {author} {\bibfnamefont {S.}~\bibnamefont {Petit}},\ and\ \bibinfo
		{author} {\bibfnamefont {B.~D.}\ \bibnamefont {Gaulin}},\ }\bibfield  {title}
	{\bibinfo {title} {Experimental insights into quantum spin ice physics in
			dipole–octupole pyrochlore magnets},\ }\bibfield  {journal} {\bibinfo
		{journal} {Annu. Rev. Condens. Matter Phys.}\ }\href
	{https://doi.org/https://doi.org/10.1146/annurev-conmatphys-041124-015101}
	{https://doi.org/10.1146/annurev-conmatphys-041124-015101} (\bibinfo {year}
	{2024})\BibitemShut {NoStop}%
	\bibitem [{\citenamefont {Maksimov}\ \emph {et~al.}(2019)\citenamefont
		{Maksimov}, \citenamefont {Zhu}, \citenamefont {White},\ and\ \citenamefont
		{Chernyshev}}]{Maksimov2019}%
	\BibitemOpen
	\bibfield  {author} {\bibinfo {author} {\bibfnamefont {P.~A.}\ \bibnamefont
			{Maksimov}}, \bibinfo {author} {\bibfnamefont {Z.}~\bibnamefont {Zhu}},
		\bibinfo {author} {\bibfnamefont {S.~R.}\ \bibnamefont {White}},\ and\
		\bibinfo {author} {\bibfnamefont {A.~L.}\ \bibnamefont {Chernyshev}},\
	}\bibfield  {title} {\bibinfo {title} {Anisotropic-exchange magnets on a
			triangular lattice: Spin waves, accidental degeneracies, and dual spin
			liquids},\ }\href {https://doi.org/10.1103/PhysRevX.9.021017} {\bibfield
		{journal} {\bibinfo  {journal} {Phys. Rev. X}\ }\textbf {\bibinfo {volume}
			{9}},\ \bibinfo {pages} {021017} (\bibinfo {year} {2019})}\BibitemShut
	{NoStop}%
	\bibitem [{\citenamefont {Iaconis}\ \emph {et~al.}(2018)\citenamefont
		{Iaconis}, \citenamefont {Liu}, \citenamefont {Halász},\ and\ \citenamefont
		{Balents}}]{Iaconis2018}%
	\BibitemOpen
	\bibfield  {author} {\bibinfo {author} {\bibfnamefont {J.}~\bibnamefont
			{Iaconis}}, \bibinfo {author} {\bibfnamefont {C.}~\bibnamefont {Liu}},
		\bibinfo {author} {\bibfnamefont {G.~B.}\ \bibnamefont {Halász}},\ and\
		\bibinfo {author} {\bibfnamefont {L.}~\bibnamefont {Balents}},\ }\bibfield
	{title} {\bibinfo {title} {{Spin Liquid versus Spin Orbit Coupling on the
				Triangular Lattice}},\ }\href {https://doi.org/10.21468/SciPostPhys.4.1.003}
	{\bibfield  {journal} {\bibinfo  {journal} {SciPost Phys.}\ }\textbf
		{\bibinfo {volume} {4}},\ \bibinfo {pages} {003} (\bibinfo {year}
		{2018})}\BibitemShut {NoStop}%
	\bibitem [{\citenamefont {Rotter}(2004)}]{Rotter2004}%
	\BibitemOpen
	\bibfield  {author} {\bibinfo {author} {\bibfnamefont {M.}~\bibnamefont
			{Rotter}},\ }\bibfield  {title} {\bibinfo {title} {Using mcphase to calculate
			magnetic phase diagrams of rare earth compounds},\ }\href
	{https://doi.org/https://doi.org/10.1016/j.jmmm.2003.12.1394} {\bibfield
		{journal} {\bibinfo  {journal} {Journal of Magnetism and Magnetic Materials}\
		}\textbf {\bibinfo {volume} {272-276}},\ \bibinfo {pages} {E481} (\bibinfo
		{year} {2004})},\ \bibinfo {note} {proceedings of the International
		Conference on Magnetism (ICM 2003)}\BibitemShut {NoStop}%
	\bibitem [{\citenamefont {Avdoshenko}\ \emph {et~al.}(2022)\citenamefont
		{Avdoshenko}, \citenamefont {Kulbakov}, \citenamefont {H\"au\ss{}ler},
		\citenamefont {Schlender}, \citenamefont {Doert}, \citenamefont {Ollivier},\
		and\ \citenamefont {Inosov}}]{Avdoshenko2022}%
	\BibitemOpen
	\bibfield  {author} {\bibinfo {author} {\bibfnamefont {S.~M.}\ \bibnamefont
			{Avdoshenko}}, \bibinfo {author} {\bibfnamefont {A.~A.}\ \bibnamefont
			{Kulbakov}}, \bibinfo {author} {\bibfnamefont {E.}~\bibnamefont
			{H\"au\ss{}ler}}, \bibinfo {author} {\bibfnamefont {P.}~\bibnamefont
			{Schlender}}, \bibinfo {author} {\bibfnamefont {T.}~\bibnamefont {Doert}},
		\bibinfo {author} {\bibfnamefont {J.}~\bibnamefont {Ollivier}},\ and\
		\bibinfo {author} {\bibfnamefont {D.~S.}\ \bibnamefont {Inosov}},\ }\bibfield
	{title} {\bibinfo {title} {Spin-wave dynamics in the {KCeS}$_{2}$
			delafossite: A theoretical description of powder inelastic neutron-scattering
			data},\ }\href {https://doi.org/10.1103/PhysRevB.106.214431} {\bibfield
		{journal} {\bibinfo  {journal} {Phys. Rev. B}\ }\textbf {\bibinfo {volume}
			{106}},\ \bibinfo {pages} {214431} (\bibinfo {year} {2022})}\BibitemShut
	{NoStop}%
	\bibitem [{\citenamefont {Wietek}\ and\ \citenamefont
		{L\"auchli}(2018)}]{Wietek2018}%
	\BibitemOpen
	\bibfield  {author} {\bibinfo {author} {\bibfnamefont {A.}~\bibnamefont
			{Wietek}}\ and\ \bibinfo {author} {\bibfnamefont {A.~M.}\ \bibnamefont
			{L\"auchli}},\ }\bibfield  {title} {\bibinfo {title} {Sublattice coding
			algorithm and distributed memory parallelization for large-scale exact
			diagonalizations of quantum many-body systems},\ }\href
	{https://doi.org/10.1103/PhysRevE.98.033309} {\bibfield  {journal} {\bibinfo
			{journal} {Phys. Rev. E}\ }\textbf {\bibinfo {volume} {98}},\ \bibinfo
		{pages} {033309} (\bibinfo {year} {2018})}\BibitemShut {NoStop}%
	\bibitem [{\citenamefont {Bernu}\ \emph {et~al.}(1994)\citenamefont {Bernu},
		\citenamefont {Lecheminant}, \citenamefont {Lhuillier},\ and\ \citenamefont
		{Pierre}}]{PhysRevB.50.10048}%
	\BibitemOpen
	\bibfield  {author} {\bibinfo {author} {\bibfnamefont {B.}~\bibnamefont
			{Bernu}}, \bibinfo {author} {\bibfnamefont {P.}~\bibnamefont {Lecheminant}},
		\bibinfo {author} {\bibfnamefont {C.}~\bibnamefont {Lhuillier}},\ and\
		\bibinfo {author} {\bibfnamefont {L.}~\bibnamefont {Pierre}},\ }\bibfield
	{title} {\bibinfo {title} {Exact spectra, spin susceptibilities, and order
			parameter of the quantum heisenberg antiferromagnet on the triangular
			lattice},\ }\href {https://doi.org/10.1103/PhysRevB.50.10048} {\bibfield
		{journal} {\bibinfo  {journal} {Phys. Rev. B}\ }\textbf {\bibinfo {volume}
			{50}},\ \bibinfo {pages} {10048} (\bibinfo {year} {1994})}\BibitemShut
	{NoStop}%
	\bibitem [{\citenamefont {Wietek}\ \emph {et~al.}(2017)\citenamefont {Wietek},
		\citenamefont {Schuler},\ and\ \citenamefont
		{Läuchli}}]{wietek2017studyingcontinuoussymmetrybreaking}%
	\BibitemOpen
	\bibfield  {author} {\bibinfo {author} {\bibfnamefont {A.}~\bibnamefont
			{Wietek}}, \bibinfo {author} {\bibfnamefont {M.}~\bibnamefont {Schuler}},\
		and\ \bibinfo {author} {\bibfnamefont {A.~M.}\ \bibnamefont {Läuchli}},\
	}\href {https://arxiv.org/abs/1704.08622} {\bibinfo {title} {Studying
			continuous symmetry breaking using energy level spectroscopy}} (\bibinfo
	{year} {2017}),\ \Eprint {https://arxiv.org/abs/1704.08622} {arXiv:1704.08622
		[cond-mat.str-el]} \BibitemShut {NoStop}%
	\bibitem [{\citenamefont {Messio}\ \emph {et~al.}(2011)\citenamefont {Messio},
		\citenamefont {Lhuillier},\ and\ \citenamefont
		{Misguich}}]{PhysRevB.83.184401}%
	\BibitemOpen
	\bibfield  {author} {\bibinfo {author} {\bibfnamefont {L.}~\bibnamefont
			{Messio}}, \bibinfo {author} {\bibfnamefont {C.}~\bibnamefont {Lhuillier}},\
		and\ \bibinfo {author} {\bibfnamefont {G.}~\bibnamefont {Misguich}},\
	}\bibfield  {title} {\bibinfo {title} {Lattice symmetries and regular
			magnetic orders in classical frustrated antiferromagnets},\ }\href
	{https://doi.org/10.1103/PhysRevB.83.184401} {\bibfield  {journal} {\bibinfo
			{journal} {Phys. Rev. B}\ }\textbf {\bibinfo {volume} {83}},\ \bibinfo
		{pages} {184401} (\bibinfo {year} {2011})}\BibitemShut {NoStop}%
	\bibitem [{Note()}]{Note}%
	\BibitemOpen
	\bibinfo {note} {The peak appears at only one $M$ point because slight
		numerical rounding in the ab-initio couplings weakly breaks the
		Hamiltonian’s $C_3$ symmetry.}\BibitemShut {Stop}%
	\bibitem [{\citenamefont {Zhao}\ \emph {et~al.}(2025)\citenamefont {Zhao},
		\citenamefont {Chen}, \citenamefont {Stone}, \citenamefont {Zhang},
		\citenamefont {Sarkis}, \citenamefont {Koohpayeh},\ and\ \citenamefont
		{Broholm}}]{Zhao2025}%
	\BibitemOpen
	\bibfield  {author} {\bibinfo {author} {\bibfnamefont {L.}~\bibnamefont
			{Zhao}}, \bibinfo {author} {\bibfnamefont {T.}~\bibnamefont {Chen}}, \bibinfo
		{author} {\bibfnamefont {M.~B.}\ \bibnamefont {Stone}}, \bibinfo {author}
		{\bibfnamefont {Q.}~\bibnamefont {Zhang}}, \bibinfo {author} {\bibfnamefont
			{C.~L.}\ \bibnamefont {Sarkis}}, \bibinfo {author} {\bibfnamefont {S.~M.}\
			\bibnamefont {Koohpayeh}},\ and\ \bibinfo {author} {\bibfnamefont
			{C.}~\bibnamefont {Broholm}},\ }\href {https://arxiv.org/abs/2507.12592}
	{\bibinfo {title} {Quenched disorder in the triangular lattice
			antiferromagnet {YbZn$_2$GaO$_5$}}} (\bibinfo {year} {2025}),\ \Eprint
	{https://arxiv.org/abs/2507.12592} {arXiv:2507.12592 [cond-mat.str-el]}
	\BibitemShut {NoStop}%
	\bibitem [{\citenamefont {Mauger}\ and\ \citenamefont
		{Godart}(1986)}]{Mauger1986}%
	\BibitemOpen
	\bibfield  {author} {\bibinfo {author} {\bibfnamefont {A.}~\bibnamefont
			{Mauger}}\ and\ \bibinfo {author} {\bibfnamefont {C.}~\bibnamefont
			{Godart}},\ }\bibfield  {title} {\bibinfo {title} {The magnetic, optical, and
			transport properties of representatives of a class of magnetic
			semiconductors: The europium chalcogenides},\ }\href
	{https://doi.org/https://doi.org/10.1016/0370-1573(86)90139-0} {\bibfield
		{journal} {\bibinfo  {journal} {Physics Reports}\ }\textbf {\bibinfo {volume}
			{141}},\ \bibinfo {pages} {51} (\bibinfo {year} {1986})}\BibitemShut
	{NoStop}%
	\bibitem [{\citenamefont {Qiao}\ \emph {et~al.}(2025)\citenamefont {Qiao},
		\citenamefont {Lin}, \citenamefont {Khmelevskyi}, \citenamefont {Pourovskii},
		\citenamefont {Xu}, \citenamefont {He}, \citenamefont {Yu}, \citenamefont
		{Cao}, \citenamefont {Li},\ and\ \citenamefont {Xing}}]{Qiao2025}%
	\BibitemOpen
	\bibfield  {author} {\bibinfo {author} {\bibfnamefont {Z.}~\bibnamefont
			{Qiao}}, \bibinfo {author} {\bibfnamefont {K.}~\bibnamefont {Lin}}, \bibinfo
		{author} {\bibfnamefont {S.}~\bibnamefont {Khmelevskyi}}, \bibinfo {author}
		{\bibfnamefont {L.~V.}\ \bibnamefont {Pourovskii}}, \bibinfo {author}
		{\bibfnamefont {H.}~\bibnamefont {Xu}}, \bibinfo {author} {\bibfnamefont
			{M.}~\bibnamefont {He}}, \bibinfo {author} {\bibfnamefont {C.}~\bibnamefont
			{Yu}}, \bibinfo {author} {\bibfnamefont {Y.}~\bibnamefont {Cao}}, \bibinfo
		{author} {\bibfnamefont {Q.}~\bibnamefont {Li}},\ and\ \bibinfo {author}
		{\bibfnamefont {X.}~\bibnamefont {Xing}},\ }\bibfield  {title} {\bibinfo
		{title} {Uncovering hidden orders within deformable materials: The case of
			dysprosium},\ }\href {https://doi.org/10.1021/jacs.5c06003} {\bibfield
		{journal} {\bibinfo  {journal} {Journal of the American Chemical Society}\
		}\textbf {\bibinfo {volume} {147}},\ \bibinfo {pages} {23026} (\bibinfo
		{year} {2025})}\BibitemShut {NoStop}%
	\bibitem [{\citenamefont {Pourovskii}\ \emph {et~al.}(2020)\citenamefont
		{Pourovskii}, \citenamefont {Boust}, \citenamefont {Ballou}, \citenamefont
		{Eslava},\ and\ \citenamefont {Givord}}]{Pourovskii2020}%
	\BibitemOpen
	\bibfield  {author} {\bibinfo {author} {\bibfnamefont {L.~V.}\ \bibnamefont
			{Pourovskii}}, \bibinfo {author} {\bibfnamefont {J.}~\bibnamefont {Boust}},
		\bibinfo {author} {\bibfnamefont {R.}~\bibnamefont {Ballou}}, \bibinfo
		{author} {\bibfnamefont {G.~G.}\ \bibnamefont {Eslava}},\ and\ \bibinfo
		{author} {\bibfnamefont {D.}~\bibnamefont {Givord}},\ }\bibfield  {title}
	{\bibinfo {title} {Higher-order crystal field and rare-earth magnetism in
			rare-earth--{Co}$_{5}$ intermetallics},\ }\href
	{https://doi.org/10.1103/PhysRevB.101.214433} {\bibfield  {journal} {\bibinfo
			{journal} {Phys. Rev. B}\ }\textbf {\bibinfo {volume} {101}},\ \bibinfo
		{pages} {214433} (\bibinfo {year} {2020})}\BibitemShut {NoStop}%
	\bibitem [{\citenamefont {Freeman}\ and\ \citenamefont
		{Watson}(1962)}]{Freeman1965}%
	\BibitemOpen
	\bibfield  {author} {\bibinfo {author} {\bibfnamefont {A.~J.}\ \bibnamefont
			{Freeman}}\ and\ \bibinfo {author} {\bibfnamefont {R.~E.}\ \bibnamefont
			{Watson}},\ }\bibfield  {title} {\bibinfo {title} {Theoretical investigation
			of some magnetic and spectroscopic properties of rare-earth ions},\ }\href
	{https://doi.org/10.1103/PhysRev.127.2058} {\bibfield  {journal} {\bibinfo
			{journal} {Phys. Rev.}\ }\textbf {\bibinfo {volume} {127}},\ \bibinfo {pages}
		{2058} (\bibinfo {year} {1962})}\BibitemShut {NoStop}%
	\bibitem [{\citenamefont {Carnall}\ \emph {et~al.}(1989)\citenamefont
		{Carnall}, \citenamefont {Goodman}, \citenamefont {Rajnak},\ and\
		\citenamefont {Rana}}]{Carnall1989}%
	\BibitemOpen
	\bibfield  {author} {\bibinfo {author} {\bibfnamefont {W.~T.}\ \bibnamefont
			{Carnall}}, \bibinfo {author} {\bibfnamefont {G.~L.}\ \bibnamefont
			{Goodman}}, \bibinfo {author} {\bibfnamefont {K.}~\bibnamefont {Rajnak}},\
		and\ \bibinfo {author} {\bibfnamefont {R.~S.}\ \bibnamefont {Rana}},\
	}\bibfield  {title} {\bibinfo {title} {A systematic analysis of the spectra
			of the lanthanides doped into single crystal laf$_3$},\ }\href@noop {}
	{\bibfield  {journal} {\bibinfo  {journal} {The Journal of Chemical Physics}\
		}\textbf {\bibinfo {volume} {90}},\ \bibinfo {pages} {3443} (\bibinfo {year}
		{1989})}\BibitemShut {NoStop}%
	\bibitem [{\citenamefont {Larson}\ \emph {et~al.}(2007)\citenamefont {Larson},
		\citenamefont {Lambrecht}, \citenamefont {Chantis},\ and\ \citenamefont {van
			Schilfgaarde}}]{Larson2007}%
	\BibitemOpen
	\bibfield  {author} {\bibinfo {author} {\bibfnamefont {P.}~\bibnamefont
			{Larson}}, \bibinfo {author} {\bibfnamefont {W.~R.~L.}\ \bibnamefont
			{Lambrecht}}, \bibinfo {author} {\bibfnamefont {A.}~\bibnamefont {Chantis}},\
		and\ \bibinfo {author} {\bibfnamefont {M.}~\bibnamefont {van Schilfgaarde}},\
	}\bibfield  {title} {\bibinfo {title} {Electronic structure of rare-earth
			nitrides using the {LSDA+U} approach: Importance of allowing $4f$ orbitals to
			break the cubic crystal symmetry},\ }\href
	{https://doi.org/10.1103/PhysRevB.75.045114} {\bibfield  {journal} {\bibinfo
			{journal} {Phys. Rev. B}\ }\textbf {\bibinfo {volume} {75}},\ \bibinfo
		{pages} {045114} (\bibinfo {year} {2007})}\BibitemShut {NoStop}%
	\bibitem [{\citenamefont {Galler}\ and\ \citenamefont
		{Pourovskii}(2022)}]{Galler2022}%
	\BibitemOpen
	\bibfield  {author} {\bibinfo {author} {\bibfnamefont {A.}~\bibnamefont
			{Galler}}\ and\ \bibinfo {author} {\bibfnamefont {L.~V.}\ \bibnamefont
			{Pourovskii}},\ }\bibfield  {title} {\bibinfo {title} {Electronic structure
			of rare-earth mononitrides: quasiatomic excitations and semiconducting
			bands},\ }\href {https://doi.org/10.1088/1367-2630/ac6317} {\bibfield
		{journal} {\bibinfo  {journal} {New Journal of Physics}\ }\textbf {\bibinfo
			{volume} {24}},\ \bibinfo {pages} {043039} (\bibinfo {year}
		{2022})}\BibitemShut {NoStop}%
\end{thebibliography}
%

\end{document}


\title{Supplemental Material for 'Ab initio spin Hamiltonians and magnetism of Ce and Yb triangular-lattice compounds'}

\author{Leonid V. Pourovskii}

\affiliation{CPHT, CNRS, \'Ecole polytechnique, Institut Polytechnique de Paris, 91120 Palaiseau, France}
\affiliation{Coll\`ege de France, Université PSL, 11 place Marcelin Berthelot, 75005 Paris, France}

\author{Rafael D. Soares}

\affiliation{Max Planck Institute for the Physics of Complex Systems, N\"{o}thnitzer Stra{\ss}e 38, 01187 Dresden, Germany}

\author{Alexander Wietek}
\affiliation{Max Planck Institute for the Physics of Complex Systems, N\"{o}thnitzer Stra{\ss}e 38, 01187 Dresden, Germany}


\date{\today}	
\maketitle

\section{Ab initio methodology and calculational details}

Our DFT+DMFT calculations employ the quasi-atomic Hubbard-I(HI) approximation to solve the quantum impurity problem for the RE 4$f$ shells; the method is abbreviated below as DFT+HI. We use the Wien2k LAPW code as the DFT part of DFT+HI. 
The atomic sphere radii $R$ in the Ce delafossites $A$Ce$X_2$ are set to $R_{\mathrm{Ce}}$=2.32, 2.5, 2.34, $R_A$=2.5, 2.5, 2.5 and $R_X$=2, 2.31, 2.01~a.u. for KCeO$_2$, KCeS$_2$, and RbCeO$_2$, respectively. In YZGO the atomic sphere radii of Yb, Ga, Zn, and O are set to 2.3, 1.92, 1.95, and 1.67~a.u, respectively. We employ 1000 (200) $\mathbf{k}$-points in the full Brillouin zone and the LAPW basis cutoff $R_{\mathrm{MIN}}K_{\mathrm{MAX}}$=8 (7) for the Ce delafossites (YZGO), respectively. The spin-orbit coupling is included through the standard second-variation approach. The LDA is used as the exchange-correlation potential in DFT. 

The projective Wannier orbitals representing RE 4$f$ are constructed in accordance with Ref.~\cite{Aichhorn2009} using a narrow energy window enclosing mainly the 4$f$ Kohn-Sham (KS) bands. Due to different positions of the 4$f$ KS band (which shifts downwards in energy along the RE series) in the Ce (4$f^1$) and Yb (4f$^{13}$) systems, the approach to constructing this Wannier basis somewhat differs between those two cases.   

In the case of Ce delafossites, the Ce 4$f$ KS band manifold is separated from the ligand $p$-band by a wide gap  and it only slightly overlaps with the Ce 5$d$ band. The lower boundary of the energy window is thus chosen to be within the $pf$ gap; the upper boundary is placed at 1.5 eV above the KS Fermi level $E_{F}^{KS}$, right above the top of the 4$f$ band. 

In the case of YZGO, the KS 4$f$ band is found to strongly overlap with the top of the O~2$p$ band.   In the course of DFT+HI self-consistency, the Yb 4$f$ band shifts downwards in energy deeper into the O~2$p$ band manifold. This results in the 4$f$ Hubbard bands in DFT+HI spectral function being too low in energy stabilizing the 4$f^{14}$ totally filled shell  configuration  in DFT+HI. A similar problem of the 4$f$ KS band being too low was observed in previous studies of heavy RE with the present approach, in particular, for the Dy metal~\cite{Qiao2025}. In contrast, in the Ce delafossites, similarly to other light RE compounds studied with the present technique, see e.~g.~\cite{Delange2017,Pourovskii2020}, the KS 4$f$ band remains pinned at $E_F^{KS}$, i.~e., its position in pure DFT electronic structure. 

In order to correct for this problem in YZGO, we thus apply, within Wien2k, a uniform shift to the 4$f$ band position to  keep its centerweight position aligned to that in pure DFT (0.9~eV below $E_F^{KS}$) in the course of DFT+HI iterations.  To disentangle 4$f$ from other contributions to hybridized 2$p$-4$f$ bands, we extend the Wannier basis to include  ligand O 2$p$ and itinerant Zn 3$d$ states. It has been shown~\cite{Pourovskii2020} that employing a narrow energy window to construct 4$f$ Wannier orbitals is crucial to capture the hybridization contribution to CF. Since this narrow window excludes most of O 2$p$ and Zn 3$d$ states, we employ two energy windows. The 4$f$ Yb orbitals  are formed from applying the initial projection to the bands in the range [-1.7:1.7] eV around the centerweight of KS 4$f$ band, which fully encloses this band. The O 2$p$ and Zn 3$d$ Wannier orbitals are formed from projection in the large window of [-9.5:1.5]~eV around $E_F^{KS}$ that fully includes both corresponding band manifolds. The resulting  full set of Wannier orbitals including Yb 4$f$, O 2$p$ and Zn 3$d$  is then orthonormalized using the standard prescription of the projective construction approach~\cite{Aichhorn2009} to obtain a proper Wannier basis.  

We parametrize the 4$f$-shell rotationally-invariant on-site Coulomb repulsion by $U=F^0$ and the Hund's rule coupling $J_H$ using the values $U=$6(8)~eV and $J_H=$0.7(1.1)~eV for the Ce(Yb) 4$f$ shells, respectively. The values of $U$ and $J_H$ together with the standard assumptions on the ratios of the Slater parameters ($F^4/F^2$=0.668, $F^6/F^2$=0.45~\cite{Freeman1965}) fully determine the vertex. The chosen $J_H$ values are  consistent with  experimental data on RE 3+ ions in insulators~\cite{Carnall1989}, the values of $U$ are within the generally accepted range and increasing from Ce to Yb in agreement with theoretical estimates for the RE series~\cite{Larson2007,Galler2022}.  We employed the fully localized limit for the DMFT  double-counting correction, which was calculated using the nominal $f^1$ ($f^{13}$)  occupancy in  the Ce (Yb) cases as shown to be appropriate for the DFT+HI method~\cite{Pourovskii2007}. To avoid the DFT self-interaction error impacting the CF parameters we employ averaging over the ground-state multiplet in our self-consistent calculations in accordance with Ref.~\cite{Delange2017}. Namely, the Boltzmann weights for all levels within the ground-state multiplet  $^{2}F_{5/2}$ ($^2F_{7/2}$) of the Ce (Yb) 4$f$ shell are averaged in the course of self-consistent DFT+HI calculations. 

Having converged DFT+HI calculations, we compute the intersite exchange interactions between the CF ground-state Kramer's doublets  using the FT-HI method. This method extract the low-energy exchange Hamiltonian from the paramagnetic electronic structure by evaluating the  response of DFT+DMFT grand potential to small fluctuations around the paramagnetic state on two neighboring sites, see Ref.~\cite{Pourovskii2016} for details. To form a proper (pseudo-)spin-1/2, the phases of the Kramer's doublet states are chosen to have the proper symmetry under the time reversal $T$, $T|+1/2\rangle=|-1/2\rangle$. 

\begin{figure*}[tbh]
	\centering
	\includegraphics[width=0.7\columnwidth]{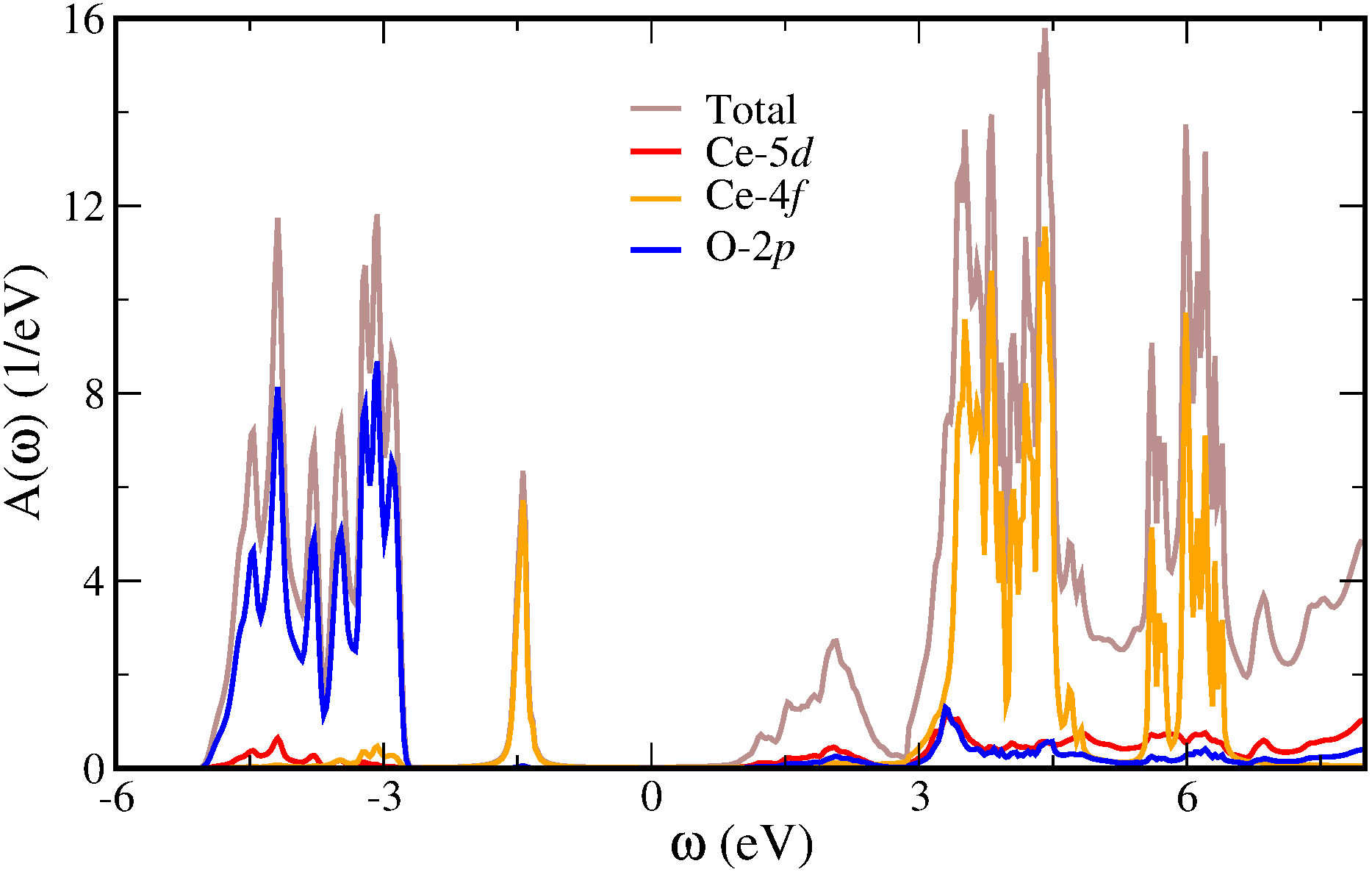}
	\includegraphics[width=0.7\columnwidth]{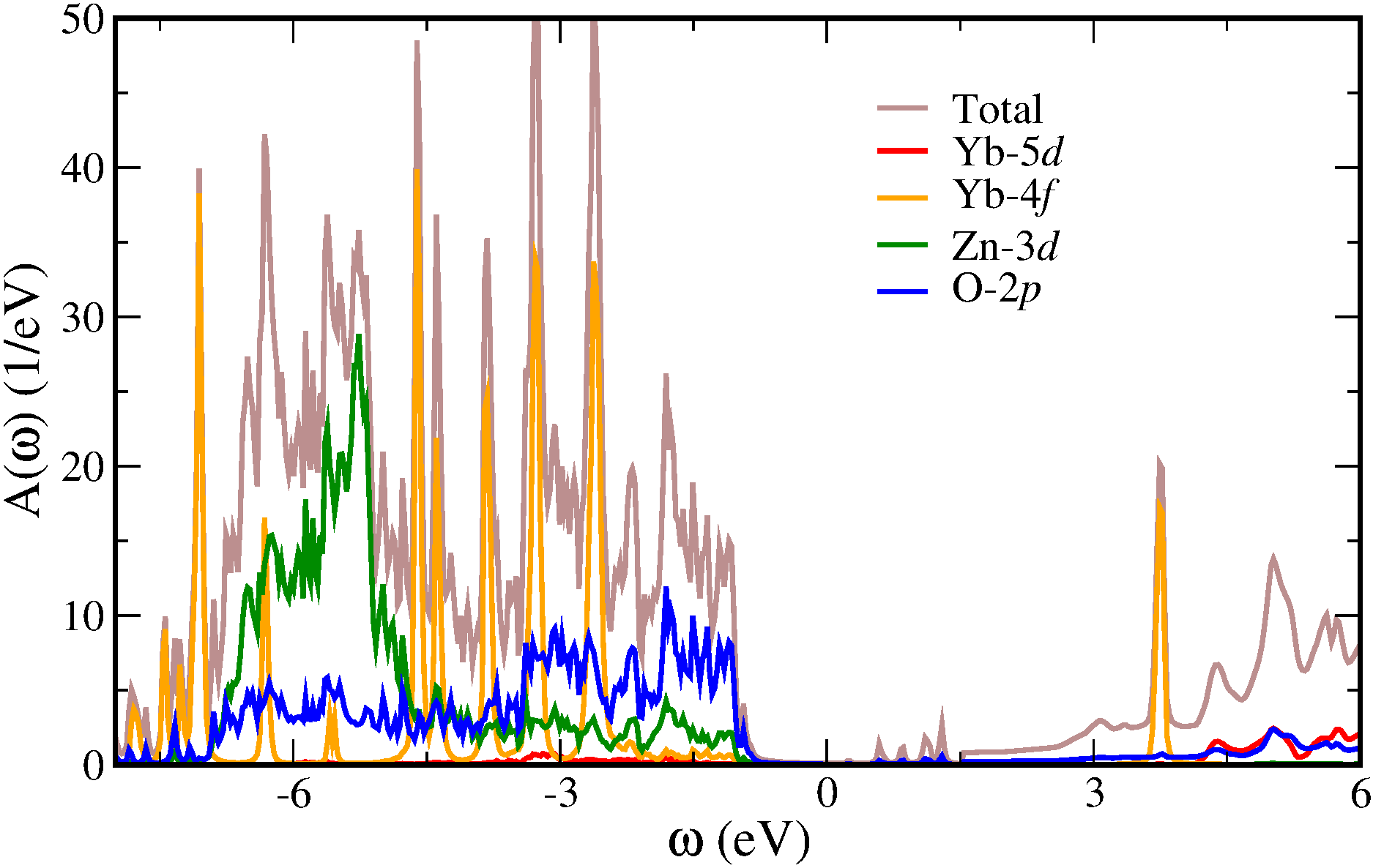}
	\caption{DFT+Hubbard-I spectral functions of KCeO$_2$ (top panel) and YbZn$_2$GaO$_5$ (bottom panel).}
	\label{fig:hubI_dos}
\end{figure*}

\section{DFT+Hubbard-I paramagnetic electronic structure}

In Fig.~\ref{fig:hubI_dos} we display the electronic structure of the paramagnetic phase in Ce delafossites (exemplified by KCeO$_2$) and YZGO calculated by the DFT+Hubbard-I approach. In the case of KCeO$_2$, the Ce 4$f$ lower Hubbard band is located in the wide semiconducting gap between O~2$p$ and Ce~5$d$. In the case of the Yb compounds, the 4$f$ lower Hubbard band is within the O~2$p$ band  and also overlaps with the filled Zn 3$d$ states, while the 4$f$ upper Hubbard band is located about 3.5~eV above the bottom of the conduction band (the latter is mainly formed by Yb 5$d$ states hybridized with O 2$p$).

\section{Crystal field levels and wavefunctions}

The  crystal-field levels and wavefunctions obtained by DFT+HI self-consistent calculations are listed in Table~\ref{tab:CF_WF}. The wavefunctions are expressed in the basis of $|J;M\rangle$ eigenstates  of the total angular momentum for the 4$f^1$ (the Ce compounds)  or 4$f^{13}$ (the Yb compound) shell. The quantization axis $z$ is along the direction orthogonal to the triangular RE  plane; the $y$ axis is along  one of the RE-RE nearest-neighbor bonds.

\begin{table}[hb]
	\begin{center}
		\begin{ruledtabular}
			\renewcommand{\arraystretch}{1.3}
			\begin{tabular}{ c   c  }

				\textbf{Energy (meV)} & \textbf{Wavefunction} \\
					\hline
					& \textbf{KCeO$_2$}  \\ 
					\hline
				0 & $0.853|5/2;\mp 1/2\rangle\pm 0.470|5/2;\pm 5/2\rangle \pm 0.177|7/2;\mp 1/2\rangle-0.134|7/2; \mp 7/2\rangle+0.041|7/2;\pm 5/2\rangle$  \\
							
				117 & $0.993|5/2;\pm 3/2\rangle\mp 0.116|7/2;\pm 3/2\rangle+0.039|7/2;\mp 3/2\rangle$  \\
						
				167 & $0.828|5/2;\pm 5/2\rangle \mp 0.484|5/2; \mp 1/2\rangle+0.243|7/2; \mp 1/2\rangle\pm 0.145|7/2;\mp 7/2\rangle$  \\
					\hline
			& \textbf{KCeS$_2$}  \\ 
				\hline
				     0 & $0.895|5/2;\mp 1/2\rangle \pm 0.444|5/2;\pm 5/2\rangle-0.045|7/2;\mp 7/2\rangle$  \\

				38 & $0.998|5/2;\pm 3/2\rangle\mp 0.055|7/2;\pm 3/2\rangle+0.032|7/2;\mp 3/2\rangle$  \\

				68 & $0.890|5/2;\pm 5/2\rangle\mp 0.441|5/2;\mp 1/2\rangle+0.087|7/2;\mp 1/2\rangle\pm 0.074|7/2;\mp 7/2\rangle$  \\
									\hline
				& \textbf{RbCeO$_2$}  \\ 
				\hline
				     0& $0.872|5/2;\mp 1/2\rangle\pm 0.439|5/2;\pm 5/2\rangle\pm 0.169|7/2;\mp 1/2\rangle-0.132|7/2;\mp 7/2\rangle+0.039|7/2;\pm 5/2\rangle$  \\
				
				109 & $0.993|5/2;\pm 3/2\rangle\mp 0.111|7/2;\pm 3/2\rangle + 0.040|7/2;\mp 3/2\rangle$  \\
				
				164 & $0.851|5/2;\pm 5/2\rangle\mp 0.453|5/2;\mp 1/2\rangle+0.229|7/2;\mp 1/2\rangle\pm 0.131|7/2;\mp 7/2\rangle$  \\
					\hline
					&\textbf{YbZn$_2$GaO$_5$}\\ 
				\hline
     0 & $0.725|7/2;\mp 7/2\rangle-0.501|7/2;\pm 5/2\rangle\mp 0.471|7/2;\mp 1/2\rangle$  \\

21 & $0.831|7/2;\mp 5/2\rangle+0.555|7/2;\pm 7/2\rangle \pm  0.033|7/2;\pm 1/2\rangle$  \\

54 & $|7/2;\pm 3/2\rangle$  \\

84 & $0.881|7/2;\mp 1/2\rangle \pm 0.408|7/2;\mp 7/2\rangle\mp 0.238|7/2;\pm 5/2\rangle$  \\

			\end{tabular}
		\end{ruledtabular}
	\end{center}
	\caption{Calculated crystal-field  energies and wavefunctions in all four compounds.  }
	\label{tab:CF_WF} 
\end{table}


\section{Point-group symmetry of the effective Hamiltonian}\label{sec:symm}

The intersite coupling along the NN bond $[100]||x$ and NNN bond $[010]||y$ is defined by Eq.~(1) of the main text. The intersite coupling for other NN and NNN bonds follows from the symmetry of 
the relevant point group that is generated by the threefold rotation about the $z$ axis, $C_3$, two-fold rotation $C_2$ around each axis $\mathbf{a}_1, \mathbf{a}_2, \mathbf{a}_3$, of the triangular lattice, and inversion $I$. The spin operator vector $\mathbf{S}$ transforms as a usual 3D vector upon proper rotations~\cite{Iaconis2018,Maksimov2019}.
By applying the $C_3$ and $C_3^2$ rotations about the $z$ axis, one obtains the couplings for the corresponding bonds as $\hat{J}_{\mathbf{r}}=\mathcal{R}\hat{J}_{\mathbf{r}_0}\mathcal{R}^T$, where $\mathbf{r}=\mathcal{R}\mathbf{r}_0$ and $\mathcal{R}$ is the corresponding rotation matrix. The couplings for the bonds related by inversion are identical, since inversion does not affect the spins. The symmetry of Eq.~(1) is not imposed in FT-HI calculations, in which all matrix elements of the intersite exchange Hamiltonian are evaluated independently, nevertheless it appears in the calculated IEI.

\section{One-shot DFT+HI calculations of KCeO$_2$}

In order to evaluate the impact of charge density self-consistency on the resulting CF splitting and IEI we carried out a test DFT+HI calculation for KCeO$_2$ in the "one-shot" mode, i.~e., using the converged DFT charge density and then applied the FT-HI method to calculate IEI. The resulting CF splitting and wavefunctions are displayed in the Table~\ref{tab:CF_non-sc}.  One may notice that the magnitude of CF splitting is severely underestimated compared to full self-consistent calculations (Table~\ref{tab:CF_WF}) and experiment~\cite{Bordelon2021}. The in-plane nature of the CF GS with a dominating $|5/2;\pm 1/2\rangle$ contribution is qualitatively captured, but the $g$-tensor anisotropy ($g_{ab}=$1.41, $g_c$=0.66) is significantly underestimated compared to the predictions of self-consistent DFT+HI and experimental estimates, see Table~I of the main text.

The KCeO$_2$ IEI calculated from one-shot DFT+HI agree reasonably with the full calculations (Table~II of the main text) with $J=$0.47~meV, $\Delta$=0.80, $J'/J=-0.04$ as well the anisotropic couplings $J_{\pm\pm}/J$
and $J_{z\pm}/J$ equal to 0.013 and 0.05, respectively. The XXZ anisotropy $\Delta$ is thus noticeably reduced, but anisotropic couplings remain small.  KCeO$_2$ lies deep in the $120^{\circ}$-AFM region of the calculated phase diagram (see Fig.~\ref{fig:12-site-results}), and these moderate changes of the IEI would not affect our predictions for its GS magnetic order.

\begin{table}[htb]
	\begin{center}
		\begin{ruledtabular}
			\renewcommand{\arraystretch}{1.3}
			\begin{tabular}{ c   c  }

				\textbf{Energy (meV)} & \textbf{Wavefunction} \\
					\hline
     0 & $0.800|5/2;\pm 1/2\rangle\mp 0.584|5/2;\mp 5/2\rangle \mp 0.112|7/2; \pm 1/2\rangle-0.067|7/2;\pm 7/2\rangle+0.046|7/2;\mp 5/2\rangle$  \\

    45 & $|5/2;\pm3/2\rangle$  \\

    79 & $0.790|5/2;\mp5/2\rangle\pm 0.578|5/2;\pm 1/2\rangle\mp 0.158|7/2;\pm 7/2\rangle+0.118|7/2;\pm 1/2\rangle\pm 0.054|7/2;\mp 5/2\rangle$  \\
			\end{tabular}
		\end{ruledtabular}
	\end{center}
	\caption{Crystal-field  energies and wavefunctions of KCeO$_2$ calculated with "one-shot" DFT+HI (i.~e., from DFT Kohn-Sham charge density).}
	\label{tab:CF_non-sc} 
\end{table}


\newpage

\section{Simulation clusters used in the ED calculations}
\begin{figure}[tbh]
	\centering
	\includegraphics{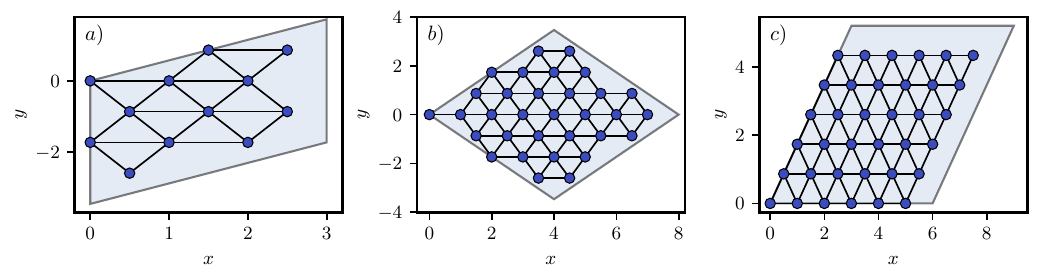}
	\caption{Geometry of the simulation clusters with (a) $N = 12$, (b) $N = 32$, and (c) $N = 36$ spins. The shaded area indicates the simulation box, with periodic boundary directions taken along the edges.}
	\label{fig:simulation_clusters}
\end{figure}

The simulation clusters that we use in our calculations are the $N=12,32$ and $36$ sites shown in Fig.~\ref{fig:simulation_clusters}. The shaded region corresponds to the simulation box with periodic boundary conditions taken along each of the edges. The $N=12$ and $36$ site clusters are compatible with the full point-group symmetries of the infinite triangular lattice, which is the full dihedral group of order 12 ($D_6$). On the other hand, the point group of the $N=32$ site cluster corresponds to the full dihedral group of order 4 ($D_2$). As such, the $12$ and $36$ site clusters are compatible with all the symmetries of the anisotropic Hamiltonian (see Sec.~\ref{sec:symm}), while the $32$ site is not. 


\section{Quantum Numbers of the symmetry broken states} 
\begin{figure}[tb]
	\centering
	\includegraphics[width=\columnwidth]{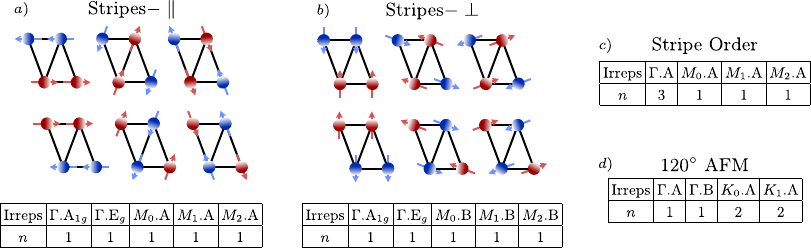}
	\caption{Scheme of the six degenerate classical ground states for $a)$ stripes-$\parallel$ and $b)$ stripes-$\perp$, together with the multiplicities ($n$) of the corresponding irreducible representations under the full $D_{3d}$ point-group symmetry of the Hamiltonian. $c)-d)$ Multiplicities of the irreducible representations for both stripe-$\parallel$ and stripe-$\perp$ (table $c)$) and for the $120^\circ$ AFM phase (table $d)$)  when the Hamiltonian possesses only the $C_{2}$ lattice symmetry.}
	\label{tab:stripy_ground-state_c2}
\end{figure}

In the thermodynamic limit, the ground state of magnetically ordered systems spontaneously breaks the symmetries of the Hamiltonian. This breaking is reflected in the organization of quantum numbers in the low-energy spectrum, forming the so-called Anderson Tower of States. Here we predict their quantum numbers using group representation theory, following Ref.~\cite{wietek2017studyingcontinuoussymmetrybreaking}. \highlight{Concretely, for each type of symmetry-breaking order, we identify the set of degenerate ground-states states, $\{|\phi_i\rangle\}$. For both the stripe and \(120^\circ\) AFM order, the number of degenerate ground states is finite, since these orders break only discrete symmetries. We study the action of the symmetry group on the manifold spanned by these states, which defines a representation of the group. This representation is then decomposed into a direct sum of irreducible representations, $\rho$, with multiplicities, $n_\rho$, determined from the character formula,
\begin{equation}
    n_{\rho} = \frac{1}{|G|} \sum_{g \in G} \chi_\rho^\ast(g) \, \mathrm{Tr}\!\left[\Lambda(g)\right],
\end{equation}
where $G$ is the symmetry group, $g$ a group element, and $\chi_\rho(g)$ its the character of the irrep. The $\Lambda (g)$ can be see as matrices defined by
\begin{equation}
    \Lambda(g)_{ij} = \langle \phi_i | \mathcal{O}_g | \phi_j \rangle,
\end{equation}
with $\mathcal{O}_g$ the symmetry operator associated with $g$ that acts on the Hilbert Space.}

\begin{figure}
    \centering
    \includegraphics[width=\columnwidth]{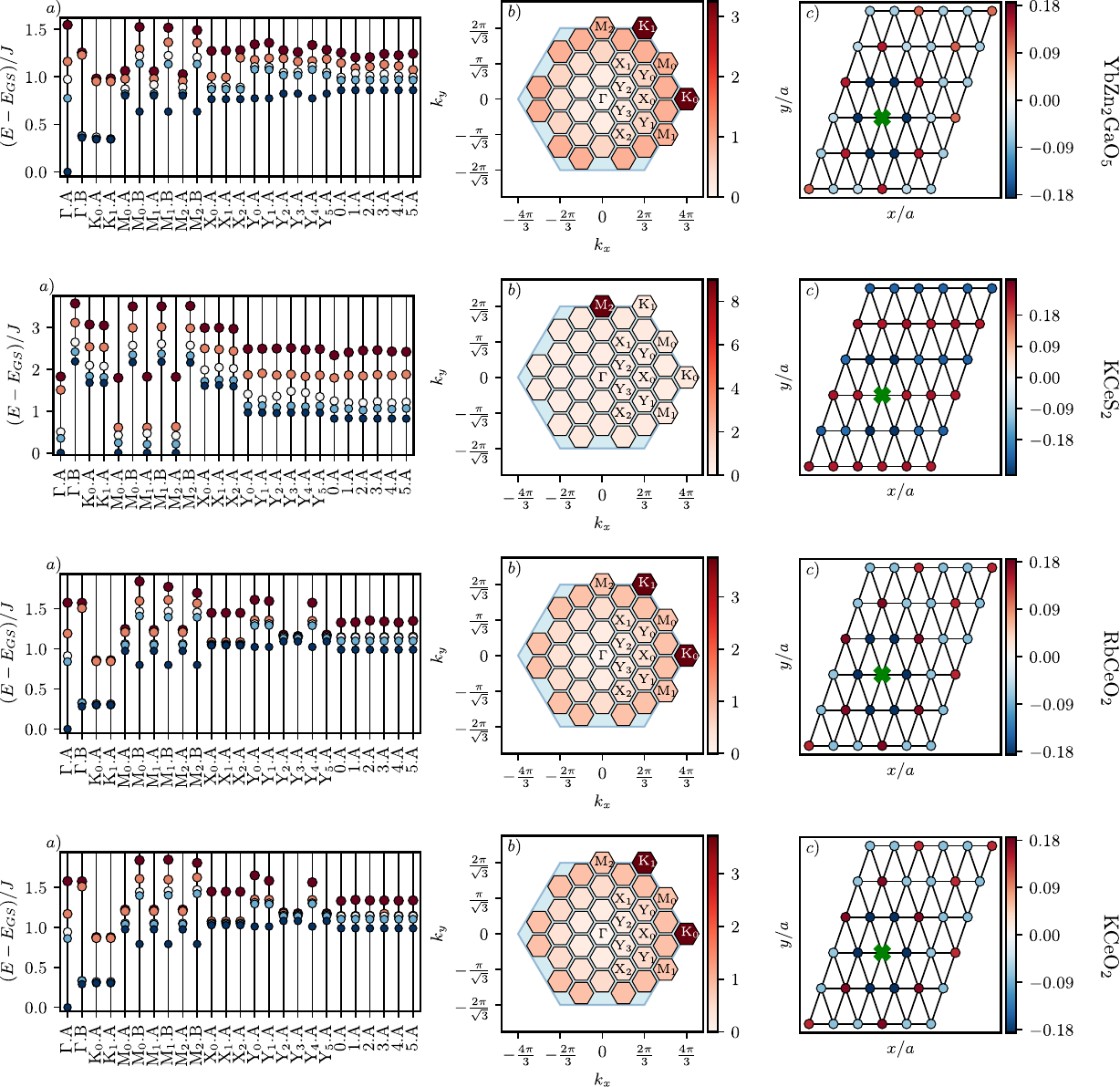}
    \caption{$a)$ Low-energy spectrum, $b)$ $\mathcal{S}(\boldsymbol{k})$ and $c)$ spin correlations $\langle \boldsymbol{S}_{\boldsymbol{r}_j} \cdot \boldsymbol{S}_{\boldsymbol{r}_0} \rangle$ in the GS of the $36$-site cluster effective magnetic Hamiltonians for the different compounds considered in this work (each row corresponds to a different compound). The spin correlations are computed with respect to the reference site $r_{0}$ marked by the green cross.}
     \label{fig:rbce_kceo_ed}
\end{figure}

We perform this analysis for two types of stripe-ordered states, stripe-$\parallel$ and stripe-$\perp$. The corresponding \highlight{degenerate} classical product states are shown in panels $a)$ and $b)$ of Fig.~\ref{tab:stripy_ground-state_c2}, respectively. \highlight{First, we consider the action of the full $D_{3d}$ point-group (the symmetry group relevant for the effective Hamiltonian as discussed in Sec.~\ref{sec:symm}), and compute the multiplicities of the irreps, shown in panels $a)$ and $b)$ of Fig.~\ref{tab:stripy_ground-state_c2}. From the action of the symmetry group on the prototypical states, we observe that} the two stripe orders differ in how they transform under combined lattice-spin $C_2$ rotation around the axis $\mathbf{a}_1$, $\mathbf{a}_2$, and $\mathbf{a}_3$ of the triangular lattice. The stripe-$\parallel$ states transform under the trivial irrep, while the stripe-$\perp$ states transform under the odd irrep. As a result, the quantum numbers in panels $a)$ and $b)$ differ in the point-group irrep. \highlight{\sout{The multiplicities of the irreps in the degenerate ground-state manifold are calculated considering the point-group $D_{3d}$.}}

However, in our ED calculations, we only use the translational and $C_2$ lattice rotation symmetries (ignoring spin rotations). Consequently, we do not resolve the quantum numbers of the entire $D_{3d}$ point-group shown in panel $a)$ and $b)$. \highlight{By computing the multiplicities of the irreps taking the point-group to be $C_2$, we find that \sout{Under the used symmetries in the ED calculations,} the two types of stripe order have \sout{identical} the same signatures in the spectra}, as shown in panel $c)$. Nevertheless, we can identify the transition between the two stripe types through a level crossing in the spectrum, specifically of the states that transform under the irreps with momentum $M$.

We have also carried out this analysis for the $120^\circ$ AFM state, again using only translational and $C_2$ lattice rotation symmetries. The corresponding quantum numbers are shown in panel $d)$.


\section{ED results effective Hamiltonians of $\text{KCeO}_2$ and $\text{RbCeO}_2$}


Here, we discuss the low-energy spectra and GS correlations on the effective Hamiltonians derived for the RbCeO$_2$ and KCeO$_2$ materials (consult Eq.~1  Tab.~III of the main text). These results are shown in the third and fourth row of Fig.~\ref{fig:rbce_kceo_ed}. For comparison, we show in the same figure the results for the YZGO and KCeS$_2$ compounds that were previously discussed in the main text. 

\begin{figure}
    \centering
    \includegraphics{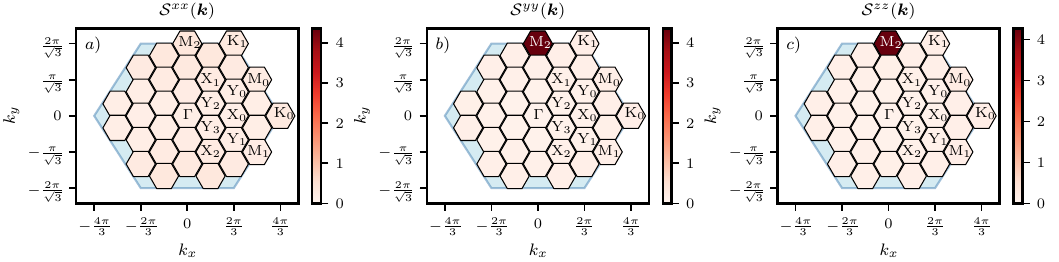}
    \caption{ a) $\mathcal{S}^{xx}(\boldsymbol{k})$, b) $\mathcal{S}^{yy}(\boldsymbol{k})$ and c) $\mathcal{S}^{zz}(\boldsymbol{k})$ in the 36-site cluster for the effective Hamiltonian GS of the KCeS$_2$ compound.}
     \label{fig:kces2_compound_sff}
\end{figure}

Similarly to the YZGO material, we observe an organization of the low-energy excitations and spin correlations pointing to the $120^\circ$ AFM order. First, in the low-energy excitations, we observe that the GS lies in the $\Gamma.A$ irrep, with the first excited state corresponding to the $\Gamma.\text{B}$ being almost degenerate with a pair of excitations at momentum $K$, as shown in panel $a)$ for both RbCeO$_2$ and KCeO$_2$ materials. These excitations in this symmetry sectors are compatible with those expected for the $120^\circ$ AFM phase (see Fig.~\ref{tab:stripy_ground-state_c2} d)). Secondly, $\mathcal{S}(\boldsymbol{k})$ peaks at the $K$ wave-vectors (shown in panels $b)$) . The real-space spin correlations [panels $c$] exhibit the sign-changing pattern characteristic of the three-sublattice structure of the $120^\circ$ order, reflecting the alternating spin orientations imposed by the triangular lattice geometry and frustrated antiferromagnetic interactions. This pattern is found in all compounds except KCeS$_2$, which instead shows GS correlations characteristic of stripe order. Along the $\hat{x}$ direction the spins align ferromagnetically, while antiferromagnetic correlations remain along $\hat{y}$, consistent with the strong peak of $\mathcal{S}(\boldsymbol{k})$ at momentum $M_2$. To determine whether the system realizes the stripe-$\parallel$ or stripe-$\perp$ phase, we analyze $\mathcal{S}^{xx} (\boldsymbol{k})$, $\mathcal{S}^{yy} (\boldsymbol{k})$, and $\mathcal{S}^{zz} (\boldsymbol{k})$, shown in panels $a)$–$c)$ of Fig.~\ref{fig:kces2_compound_sff}. All three components peak at $M_2$, confirming stripe order oriented along $[100]\parallel x$. However, the much stronger correlations in $\mathcal{S}^{yy}(M_2)$ and $\mathcal{S}^{zz}(M_2)$ indicate that the GS realizes the stripe-$\perp$ configuration.

\clearpage

\section{Low-energy spectrum for the $N=32$ cluster and Further Results on the Spin Structure Factor}

\begin{figure}[t]
    \centering
    \includegraphics{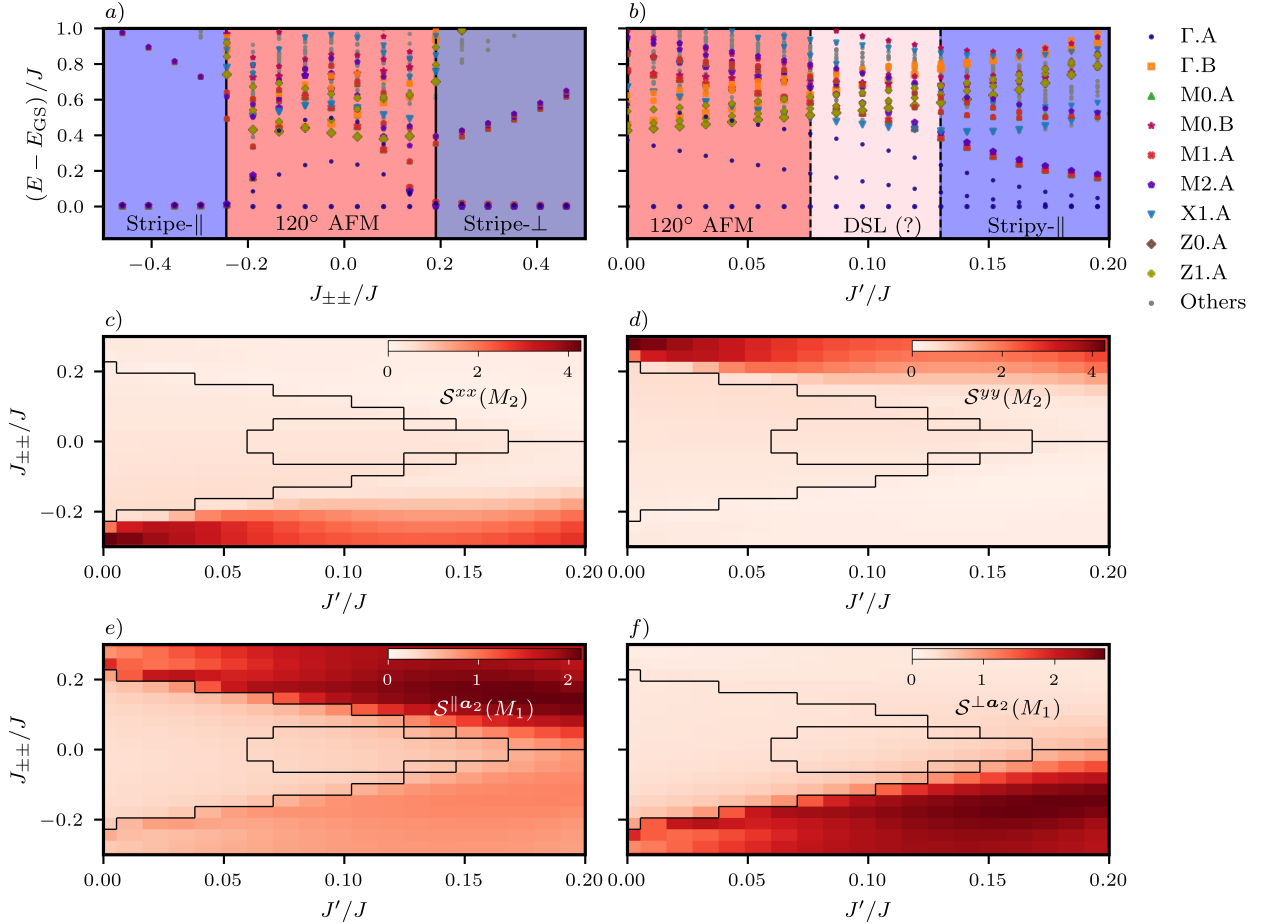}
    \caption{Low-energy spectra and static spin structure factors of the effective magnetic Hamiltonian for the 32-site cluster. $a)$ - Spectrum as a function of $J_{\pm\pm}/J$ for $J' = 0.022J$ and $J_{z\pm} = 0.173J$. $b)$- Spectrum as a function of $J'/J$ for $J_{\pm\pm} = -0.0488J$ and $J_{z\pm} = 0$. The colored regions indicate phases: blue – stripe-$\parallel$, dark blue – stripe-$\perp$, red – $120^\circ$ AFM, pale red – putative DSL. $c)–f)$ -  Static spin structure factors: (c) along $x$ and (d) along $y$ at momenta $M_2$; (e) parallel and (f) perpendicular to $\boldsymbol{a}_2$ at momenta $M_1$.}
    \label{fig:32_site_energy_spectra}
\end{figure}

Here, we present a detailed analysis of the low-energy spectrum of the effective magnetic Hamiltonian on the $N = 32$ site cluster used in the main text to construct the extended phase diagram. In addition, we discuss the static spin structure factor, which allow us to distinguish between the stripe-$\parallel$ and stripe-$\perp$ phases.

We begin by discussing the spectra. Fig.~3 of the main text shows the dependence of the low-energy levels on different symmetry sectors as a function of the model parameters. Panel (a) displays the dependence on $J_{\pm\pm}$ at fixed $J_{z\pm} = 0.173J$, taken from Fig.~3(a), while panel (b) shows the dependence on $J^\prime$ at fixed $J_{\pm\pm} = -0.0488J$, taken from Fig.~3(d). The different magnetic phases can be identified from the organization of the quantum numbers. The $120^\circ$ AFM region is signaled by a first excited state in the $\Gamma.\text{A}$ irrep together with two degenerate excited states in the $Z_0.\text{A}$ and $Z_1.\text{A}$ irreps. The transition to either stripe phase is marked by the closure of the gap in the $\Gamma.\text{A}$ irrep, corresponding to the phase boundaries in the main text. This criterion agrees remarkably well with the magnetic order inferred from the spin structure factors, as shown in the main text. Deep in the stripe phase, one also expects three degenerate states in the $M_0.\text{A}$, $M_1.\text{A}$, and $M_2.\text{A}$ irreps, becoming quasi-degenerate with the ground state in the $\Gamma.\text{A}$ irrep, as clearly seen in Fig.~3(a). The putative DSL regime is indicated by a first excited state with a finite momentum equal to $X$. This regime is shown in Fig.~\ref{fig:32_site_energy_spectra}(b), where the $X_1.\text{A}$ level becomes the lowest excitation with a finite momentum as $J^\prime/J$ increases. We emphasize that this criterion alone does not establish the existence of the DSL phase nor determine its precise boundaries. We represent the transition to and from this region by a dashed line to emphasize this point. 


As discussed above, the two types of stripe order cannot be distinguished solely from the quantum numbers of the spectra within the symmetries accessible to our ED calculations. However, they can be differentiated by analyzing the spin structure factor with spin components projected along the stripe direction and rotated by $90^\circ$ from it. Due to the symmetries of the $32$-site cluster, the six stripe configurations sketched in Fig.~\ref{tab:stripy_ground-state_c2} are no longer degenerate. We find that in some regions of the phase diagram the ground state exhibits a characteristic wave vector at the $M_2$ point, while in others the peak appears at both $M_0$ and $M_1$ wave vectors. This distinction does not yet separate stripe-$\parallel$ from stripe-$\perp$; rather, it indicates which specific stripe arrangement (within each stripe type) is energetically preferred for this cluster geometry.  

So for the parameters where the order is characterized by the $M_2$ wave vector, we inspect $\mathcal{S}^{xx}(\text{M}_2)$ and $\mathcal{S}^{yy}(\text{M}_2)$, shown in Figs.~\ref{fig:32_site_energy_spectra}(c) and (d). In the stripe-$\parallel$ phase, the spins preferably align along the stripe direction (the $x$-axis), leading to $\mathcal{S}^{xx}(M_2) > \mathcal{S}^{yy}(M_2)$. In contrast, in the stripe-$\perp$ phase the spins are rotated by $90^\circ$, resulting in $\mathcal{S}^{yy}(M_2) > \mathcal{S}^{xx}(M_2)$. 
From this criterion, we identify the stripe-$\parallel$ region for $J_{\pm\pm}>0$ and the stripe-$\perp$ region for $J_{\pm\pm}<0$.  

Nevertheless, there are regions within the stripe-ordered phases where neither $\mathcal{S}^{xx}(M_2)$ nor $\mathcal{S}^{yy}(M_2)$ show a peak. These correspond to regions where the GS develops stripe order with characteristic wave vectors at $M_1$ (or $M_0$). In this case, we analyze instead
\begin{equation}
\begin{aligned}
\mathcal{S}^{\parallel \boldsymbol{a}_2} \left( \text{M}_1 \right) &= \dfrac{1}{N}\sum_{i,j} e^{i M_1 (\boldsymbol{r}_i-\boldsymbol{r}_j)} \left\langle \left(\boldsymbol{S}_i\cdot \hat{\boldsymbol{a}}_2 \right) \left(\boldsymbol{S}_j\cdot \hat{\boldsymbol{a}}_2 \right) \right\rangle,\\ \mathcal{S}^{\perp \boldsymbol{a}_2} \left( \text{M}_1 \right) &= \dfrac{1}{N}\sum_{i,j} e^{i M_1 (\boldsymbol{r}_i-\boldsymbol{r}_j)} \left\langle \left(\boldsymbol{S}_i\cdot [1-\hat{\boldsymbol{a}}_2] \right) \left(\boldsymbol{S}_j\cdot [1-\hat{\boldsymbol{a}}_2] \right) \right\rangle,
\end{aligned}
\end{equation}
where $\hat{\boldsymbol{a}}_2$ is the unit vector in the $\boldsymbol{a}_2$ lattice direction. These are shown in Figs.~\ref{fig:32_site_energy_spectra}(e) and (f) as a function of $J_{\pm\pm}/J$ and $J^\prime/J$. From these we obtain the same result: stripe-$\parallel$ is stabilized for \highlight{$J_{\pm\pm}<0$}, while stripe-$\perp$ is stabilized for \highlight{$J_{\pm\pm}>0$}. Moreover, the transition between them, occurring around $J_{\pm\pm}=0$, agrees remarkably well with the one identified from the crossing of the ground state and the first excited state in the irreps with momentum $M$. The behavior is identical when analyzing the dependence on $J_{\pm\pm}$ and $J_{z\pm}$ with $J^\prime$ fixed.

\begin{figure}[t]
    \centering
    \includegraphics{n_32_energy_levels_with_quantum_numbers_second_scan.png}
    \caption{Low-energy spectra of the effective magnetic Hamiltonian for the 32-site cluster. $a)$ - Spectrum as a function of $J_{\pm\pm}/J$ for $J_{z\pm} = 0.390J$. $b)$- Spectrum as a function of $J_{z\pm}/J$ for $J_{\pm\pm} = -1.140J$. The colored regions indicate phases: blue – stripe-$\parallel$, dark blue – stripe-$\perp$, red – $120^\circ$ AFM, pale red – putative DSL. $c)–f)$. Other parameters: $J' = 0.022J$.}
    \label{fig:32_site_energy_spectra2}
\end{figure}

\section{Revisiting of the Spin Liquid Phase at Small $J^\prime$}

\begin{figure}[t]
    \centering
    \includegraphics[width=\columnwidth]{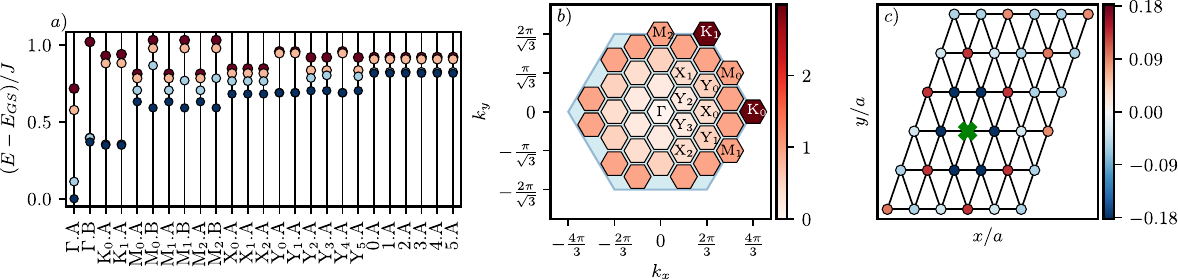}
    \caption{$a)$ Low-energy spectrum, $b)$ $\mathcal{S}(\boldsymbol{k})$ and $c)$ spin correlations $\langle \boldsymbol{S}_{\boldsymbol{r}_j} \cdot \boldsymbol{S}_{\boldsymbol{r}_0} \rangle$ in the GS of the $36$-site cluster effective magnetic Hamiltonians for $J^\prime=0.0J$, $J_{z\pm}=0.3J$ and $J_{\pm \pm} =-0.08J$. The spin correlations are computed with respect to the reference site $r_{0}$ marked by the green cross.}
    \label{fig:36_j2_zero_region}
\end{figure}

\highlightsecond{Previous results obtained using DMRG calculations~\cite{Zhu2018} suggest the existence of a broad quantum spin-liquid phase for $J^\prime=0$ with finite $J_{\pm\pm}$ and $J_{z\pm}$. This phase is proposed to be adiabatically connected to the quantum spin liquid found in the $J_1-J_2$ model (where $J_{\pm\pm}=J_{z\pm}=0$) around $J^\prime\sim 0.12J$. Here, we investigate this region in greater detail.} 

\highlightsecond{In Fig.~\ref{fig:32_site_energy_spectra2}, we present the low-energy spectra for fixed $J^\prime = 0.022J$ as a function of $J_{\pm\pm}$ (panel $a)$) and $J_{z\pm}$ (panel $b)$). We observe that in both cases the lowest excitations correspond to the towers of states associated with either the $120^\circ$ AFM or stripy orders, whereas the excitation in the X1.\text{A} sector appears at significantly higher energies. Based on this, we do not identify a DSL phase for this value of the parameters.}

\highlightsecond{We have also calculated the energy spectra and SSF in the $N=36$ cluster at $J^\prime=0$, $J_{z\pm}=0.3J$, and $J_{\pm\pm}=-0.08J$. While previous DMRG calculations suggested this to be a quantum paramagnetic region~\cite{Zhu2018}, our ED results point to $120^\circ$ magnetic order. In Fig.~\ref{fig:36_j2_zero_region}, we present the energy spectra for this cluster, resolved in the different momentum and point group sectors. The low-energy spectrum is compatible with the tower of states for $120^\circ$ AFM order, with the energy levels at momentum $X$ having a significantly higher energy than those at momenta $K$ and $M$. Moreover, the SSF exhibits strong peaks at the $K$ points, corresponding to the ordering vector of the $120^\circ$ AFM phase. These results contrast with the claim that this region corresponds to a quantum paramagnetic phase. However, even though our criterion does not identify a DSL for this parameter point, we cannot definitively rule out its existence given the available system sizes. In particular, the DSL could still develop a similar peak due to the presence of $120^\circ$ AFM algebraic correlations, which are not strictly distinguishable from long-range magnetic order when using ED. Thus, even though we find no evidence to identify the regime at $J^\prime = 0$, $J_{z\pm} = 0.3J$, and $J_{\pm\pm} = -0.08J$ as a DSL, finite-size effects likely preclude a definitive conclusion.}

\section{Low-energy spectrum and ground-state correlations for the $N=12$ cluster}

\begin{figure*}[t]
    \centering
    \includegraphics[width=\columnwidth]{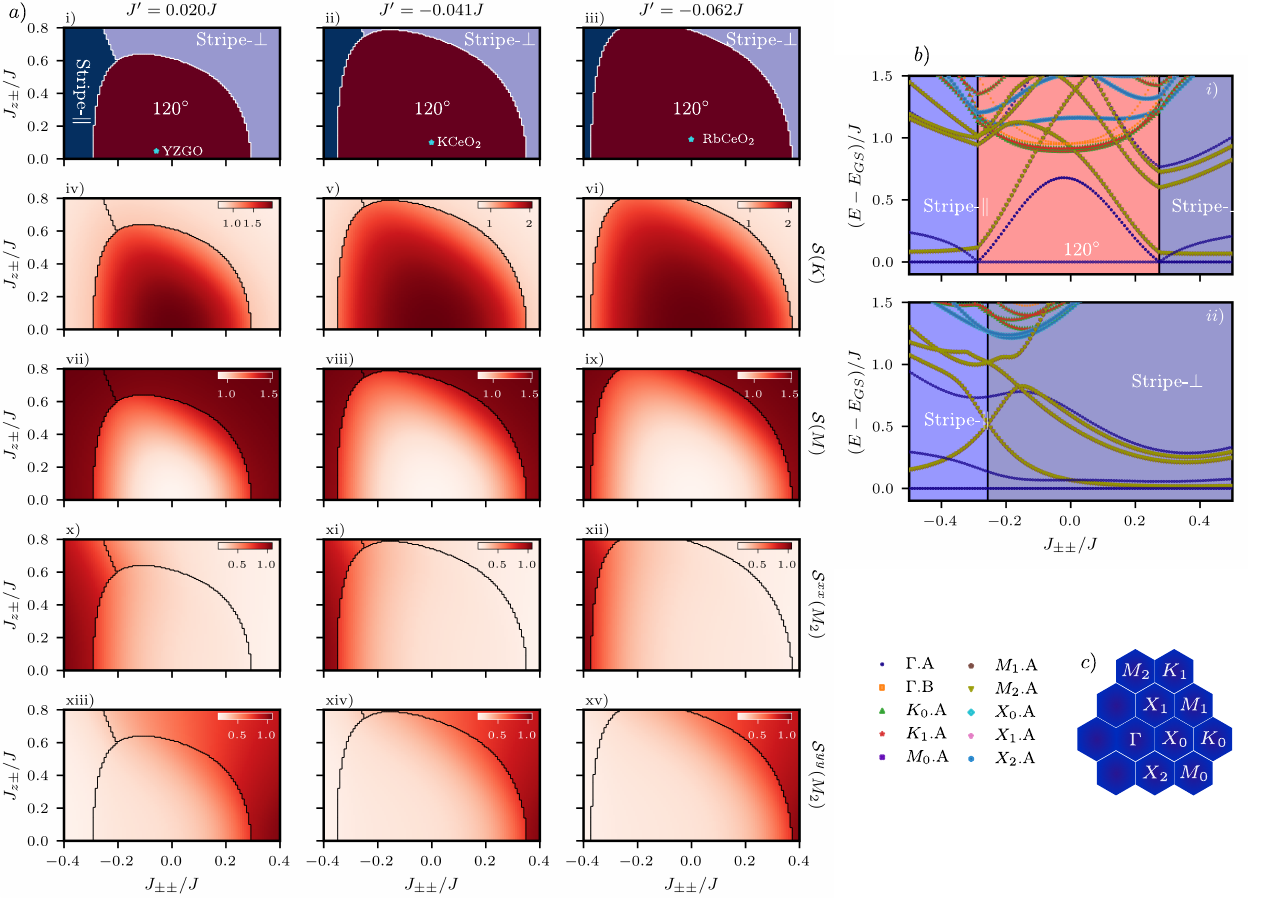}
    \caption{a)-Approximate phase diagram as a function of $J_{\pm\pm}/J$ and $J_{z\pm}/J$ for different values of $J^\prime/J$ [i)–iii)]. The blue cross marks the approximate coupling values of some of the compounds considered in this work (we are using the absolute value of $J_{z\pm}$ see the main text). $\mathcal{S}(\boldsymbol{k})$ is shown at the ordering vectors $K_0$ [iv–vi] and $M$ [vii–ix], as well as for the $x$ [x–xii] and $y$ [xiii–xv] spin components at the $M_2$ wave-vector. (b) Low-energy spectra resolved by symmetry sectors as a function of $J_{\pm\pm}/J$ for $J_{z\pm} = 0.20J$ [i)] and $J_{z\pm} = 0.69J$ [ii)], with $J^\prime = 0.02J$. (c) Sketch of the FBZ for the 12-site cluster.}
    \label{fig:12-site-results}
\end{figure*}

In this section, we discuss the phase diagram obtained for the 12-site cluster, whose results support those presented in the main text for the 32-site cluster. In contrast to the 32-site cluster, the 12-site cluster resolves the natural ordering vectors for both the stripe order and the $120^\circ$ AFM order, specifically the $M$ and $K$ wave-vectors, as depicted in Fig.~\ref{fig:12-site-results} $d)$. Moreover, the $X$ points are also resolved, which is important for the DSL phase.

\begin{figure*}[t]
    \centering
    \includegraphics{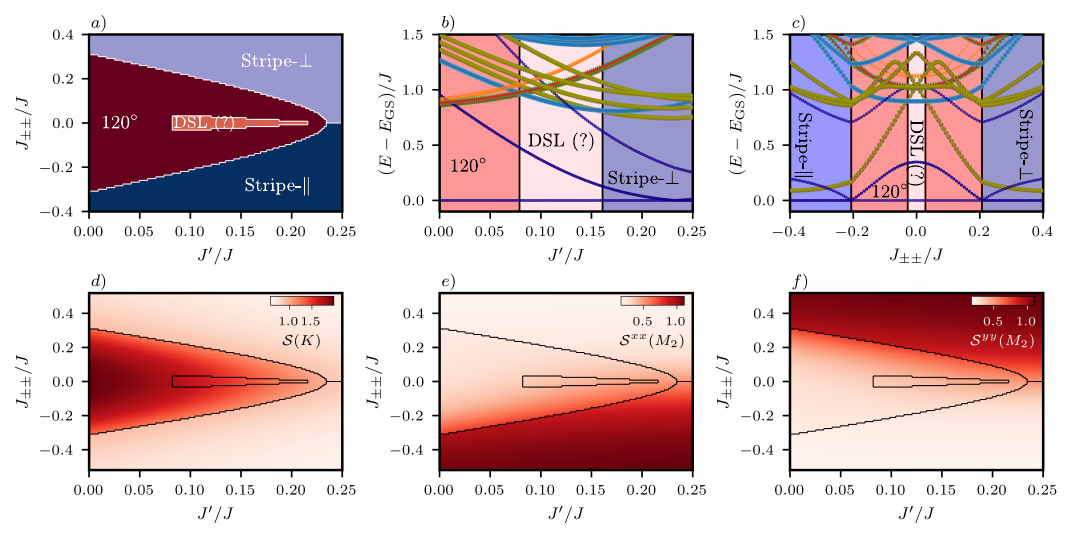}
    \caption{$a)$ Approximate phase diagram based on the structure of the quantum numbers of the low-energy spectra as a function of $J_{\pm\pm}/J$ and $J'/J$. The transition line between the two distinct stripe orders is obtained from the level crossing in the $M.\mathrm{A}$ irreps. $b)$ and $c)$ Low-energy spectra for $J_{\pm\pm} =0.0202J$ as a function of $J'/J$ $b)$ and for $J' = 0.1054J$ as a function of $J_{\pm\pm}/J$ $c)$. The color scheme labeling the different irreps is the same as in Fig.~\ref{fig:12-site-results}. $d)$--$f)$ Static spin structure factor for different as a function of $J_{\pm\pm}/J$ and $J'/J$ evaluated at momentum $K$ (panel $d)$) and $M_2$ (panels $e)$ and $f)$). Other parameters: $N = 12$ and $J_{z\pm} = 0$.}
    \label{fig:12_site_dsl}
\end{figure*}

We begin by studying the phase diagram as a function of $J_{\pm\pm}/J$ and $J_{z\pm}/J$ for fixed values of $J^\prime$ consistent with those obtained for the compounds considered in this work (see Tab.~III in the main text). The phase diagram is shown in panel a), sub-panels i), ii), and iii), corresponding to $J^\prime=0.020J$, $J^\prime=-0.041J$, and $J^\prime=-0.062J$, respectively. The approximate phase boundaries are estimated from the quantum numbers of the low-energy spectrum. In the $120^\circ$ phase, the GS is unique transforming under the $\Gamma.\text{A}$ irrep, with low-energy excitations in the $\Gamma.\text{B}$, $K_0.\text{A}$ and $K_1.\text{A}$ irreps. We also observe an exactly degenerate first excited state also transforming under $\Gamma.\text{A}$ irrep. These exact degeneracy is due to the richer symmetry group than the one considered in our ED code. Upon tuning either $J_{\pm\pm}/J$ or $J_{z\pm}/J$, we observe a closure of the energy gap between the GS energy and the first excited state within the $\Gamma.{\rm A}$ irrep. The exact point where this gap vanishes is identified as the transition into the stripe phase. This transition is further confirmed by the decrease in intensity in $\mathcal{S}(K)$ after the gap closes, as shown in Fig.~\ref{fig:12-site-results}~a), panels iv)–vi). In the phase diagrams shown in panel a), we have located some of the compounds considered in this work. Their positions are determined by the calculated values of \( J^\prime/J \), \( J_{\pm\pm}/J \), and the absolute value of \( J_{z\pm} \) (magnetic phases with \( J_{z\pm}<0 \) can be mapped to those with \( J_{z\pm}>0 \) by applying a \(\pi\)-rotation along the $z$ axis to all spins in the system). We find their locations to be consistent with previous calculations, namely within the \(120^\circ\) AFM state.

In the stripe-ordered phase, the ground state becomes doubly degenerate in $\Gamma.{\rm A}$, while the first excited states are degenerate and transform under the $M_0.{\rm A}$, $M_1.{\rm A}$, and $M_2.{\rm A}$ irreps. This also corroborates the onset of stripy magnetic order, as these quantum numbers match those predicted from the representation group theory calculations (see panel c) in Fig.~\ref{tab:stripy_ground-state_c2}). This is again corroborated by the static spin structure factors evaluated at the $M$ point, as shown in Fig.\ref{fig:12-site-results}~a), panels vii)–ix).

To distinguish between the stripe-$\parallel$ and stripe-$\perp$,we examine $\mathcal{S}^{xx}(\boldsymbol{k})$ and $\mathcal{S}^{yy}(\boldsymbol{k})$. Both exhibit peaks at the wave vector $M_2$, but in different regions of the phase diagram: the $xx$ peak identifies the stripe-$\parallel$ phase, while the $yy$ peak signals the stripe-$\perp$ phase. The phase diagram obtained from them is consistent with the transition line inferred from the energy-level crossing at the $M$ points. This crossing reflects the transition between the two stripe orders, which carry distinct quantum numbers in the low-energy sector (although both remain located at momentum $M$). \highlight{\sout{This crossing is clearly illustrated in Fig.~\ref{fig:12-site-results}c)} The energy level crossing is illustrated in Fig.~\ref{fig:12-site-results}b) in } panels i) and ii), where we show the energy spectra as functions of $J_{\pm\pm}/J$ for two representative values of $J_{z\pm}$: $J_{z\pm}\simeq 0.20J$ (i)) and $J_{z\pm}\simeq 0.69J$ (ii)).

We find that the $120^\circ$ order is suppressed as either $J_{\pm\pm}$ or $J_{z\pm}$ is increased, with negative values of $J^\prime/J$ enlarging this region in the phase diagram. By tuning $J_{\pm\pm}$ and/or $J_{z\pm}$, the GS instead exhibits stripe order. Similar to the 32-site cluster, we observe that for $J_{z\pm}=0$, negative $J_{\pm\pm}$ values favors the stripe-$\parallel$ phase, while positive values stabilize the stripe-$\perp$ phase. The stripe-$\perp$ phase is also supported for negative values of $J_{\pm\pm}/J$, when $J_{z\pm}\neq0$. For all these parameters, we do not find the first excited state (with a non-zero momentum) in the $X.\mathrm{A}$ irrep, which therefore does not indicate the presence of a DSL phase.

In Fig.~\ref{fig:12_site_dsl}, we set $J_{z\pm} = 0$ and study the phase diagram as a function of $J_{\pm\pm}/J$ and $J^\prime.J$. In panel~(a), the phase boundaries are determined in the same way as for Fig.~\ref{fig:12-site-results}. For $J'/J \lesssim 1\%$, the behavior matches that discussed previously: a $120^\circ$ AFM phase is stabilized for finite $J_{\pm\pm}/J$, which, upon increasing $|J_{\pm\pm}/J|$, gives way to a stripy phase. The nature of this phase depends on the sign of $J_{\pm\pm}/J$: for negative values it is stripe-$\parallel$, while for positive values it is stripe-$\perp$. This is corroborated by the static spin structure factors shown in panels~d) to f): in the $120^\circ$ phase we find strong peaks at both $K$ points, whereas in the stripe-$\parallel$ phase the $S^{xx}$ component exhibits a pronounced signal along the $M_2$ vector, and in the stripe-$\perp$ phase the same wave-vector is dominant in the $S^{yy}$ component. The transition between the two stripe phases is again identified by locating the level crossing between the ground state and the first excited state within the $M.\text{A}$ irreps.

Increasing the strength of NNN coupling, $J'/J$, leads to the appearance of first excited state with non-zero momentum at the the $X_1.A$ and $X_2.A$ irrep. This is clearly seen in panels~(b) and~(c), where we plot the low-energy spectrum as a function of $J'$ for $J_{\pm\pm} = 0.02J$ and as a function of $J_{\pm\pm}$ along the line $J' = 0.1054J$, respectively. As discussed in the main text, this behavior pins the possible DSL region. From panel~(a), we observe that the range of $J_{\pm\pm}/J$ supporting this region shrinks as $J'$ increases. For sufficiently large values of either coupling, the system transitions into a stripy ordered phase whose orientation again depends on the sign of $J_{\pm\pm}/J$ (negative for stripe-$\parallel$, positive for stripe-$\perp$). Exactly at $J_{\pm\pm} = 0$, the model recovers the SU(2) and U(1) spin-symmetries. These findings are consistent with the results for the 32-site cluster presented in the main text: while a possible DSL regime is identified, all \textit{ab initio} estimates of $J'/J$ lie below the required range.


%